\documentclass[%
 reprint,
superscriptaddress,
preprintnumbers,
nofootinbib,
 amsmath,amssymb,
 aps,
longbibliography
]{revtex4-1}

\usepackage{graphicx}
\usepackage{dcolumn}
\usepackage{bm}
\usepackage[normalem]{ulem}
\usepackage[
colorlinks=true, 
linkcolor=blue,
citecolor=blue,
urlcolor=blue
]{hyperref}
\usepackage{xcolor}

\newcommand{\dd}{\mathrm{d}}

\begin{document}

\preprint{TTK-21-53}
\preprint{DESY-21-216}

\title{The ups and downs of inelastic dark matter: \\ Electron recoils from terrestrial upscattering}

\author{Timon Emken}
\email{timon.emken@fysik.su.se}
\affiliation{
 The Oskar Klein Centre, Department of Physics, Department of Physics, Stockholm University, AlbaNova, SE-10691 Stockholm, Sweden
}%

\author{Jonas Frerick}
\email{jonas.frerick@desy.de}%
\affiliation{%
 Institute for Theoretical Particle Physics and Cosmology (TTK), RWTH Aachen University, 52056 Aachen, Germany
}%
\affiliation{%
Deutsches Elektronen-Synchrotron DESY, Notkestr. 85, 22607 Hamburg, Germany
}%

\author{Saniya Heeba}%
 \email{saniya.heeba@mcgill.ca}
\affiliation{%
 Institute for Theoretical Particle Physics and Cosmology (TTK), RWTH Aachen University, 52056 Aachen, Germany
}%
\affiliation{Department of Physics \& McGill Space Institute, McGill University, Montr\'eal, QC H3A 2T8, Canada}

 \author{Felix Kahlhoefer}%
 \email{kahlhoefer@physik.rwth-aachen.de}
\affiliation{%
 Institute for Theoretical Particle Physics and Cosmology (TTK), RWTH Aachen University, 52056 Aachen, Germany
}%

\date{\today}

\begin{abstract}
The growing interest in the interactions between dark matter particles and electrons has received a further boost by the observation of an excess in electron recoil events in the XENON1T experiment. Of particular interest are dark matter models in which the scattering process is inelastic, such that the ground state can upscatter into an excited state. The subsequent exothermic downscattering of such excited states on electrons can lead to observable signals in direct detection experiments and gives a good fit to the XENON1T excess. In this work, we study terrestrial upscattering, i.e.\ inelastic scattering of dark matter particles on nuclei in the Earth, as a plausible origin of such excited states. Using both analytical and Monte Carlo methods, we obtain detailed predictions of their density and velocity distribution. These results enable us to explore the time dependence of the flux of excited states resulting from the rotation of the Earth. For the case of XENON1T, we find the resulting daily modulation of the electron recoil signal to be at the level of 10\% with a strong dependence on the dark matter mass.
\end{abstract}

\maketitle

\section{Introduction}
\label{sec: introduction}

For many years, the expectation that dark matter (DM) particles should have a mass at the TeV scale and scatter predominantly on nuclei has guided the development of direct detection experiments~\cite{Lin:2019uvt}. Only  quite recently have strategies been developed to search for electron recoils as a signature of the scattering of DM particles at the GeV scale or below~\cite{Essig:2011nj}. While many experiments designed to look for nuclear recoils also have excellent sensitivity to electron recoils, there have been many ideas and proposals for new technologies looking specifically for the signatures of electron scattering, using for example CCDs~\cite{Graham:2012su,Essig:2015cda,DAMIC:2019dcn,SENSEI:2020dpa}, graphene~\cite{Hochberg:2016ntt,PTOLEMY:2018jst}, three-dimensional Dirac materials~\cite{Hochberg:2017wce,Coskuner:2019odd,Geilhufe:2019ndy}, superconductors~\cite{Hochberg:2015pha,Hochberg:2015fth,Hochberg:2021ymx}, polar targets~\cite{Knapen:2017ekk}, superconducting nanowires~\cite{Hochberg:2019cyy,Hochberg:2021yud,Chiles:2021gxk}, scintillators~\cite{Derenzo:2016fse,Blanco:2019lrf}, and more~\cite{Essig:2016crl,Bunting:2017net,Geilhufe:2018gry,Griffin:2019mvc,Kurinsky:2019pgb,Griffin:2020lgd,Kahn:2021ttr}.
In addition to technological advances, a number of phenomenological ideas on how to probe lower DM~masses were proposed, e.g.\ by looking for Migdal scatterings~\cite{Ibe:2017yqa,Dolan:2017xbu,Bell:2019egg,Baxter:2019pnz,Essig:2019xkx,Flambaum:2020xxo,Wang:2021oha}, or a high-energetic DM~population from solar reflection~\cite{An:2017ojc,Emken:2017hnp,Chen:2020gcl,Emken:2021lgc,An:2021qdl} or cosmic ray upscatterings~\cite{Bringmann:2018cvk,Ema:2018bih,Cappiello:2019qsw,Bondarenko:2019vrb}.
At the same time, the theoretical description of DM-electron scatterings in material is continuously getting extended and improved~\cite{Roberts:2016xfw,Trickle:2019nya,Catena:2019gfa,Trickle:2020oki,Gelmini:2020xir,Borah:2020smw,Griffin:2021znd,Catena:2021qsr,Hochberg:2021pkt,Knapen:2021run}.

As expected, these rapid developments have been accompanied by the observation of a number of experimental excesses for which no known background model exists. Most notable among these are an excess seen close to threshold across several experiments~\cite{Kurinsky:2020dpb,Du:2020ldo} and an excess at a few keV electron recoil energy reported by the XENON1T experiment~\cite{XENON:2020rca}. Unfortunately, neither of these excesses can be readily interpreted in terms of elastic scattering of DM particles off individual electrons. This has led to rapidly growing interest in scattering processes that are inelastic due to the excitation (or de-excitation) of internal modes of either the detector or the DM particle.

Indeed, it has been shown that the XENON1T excess can be well fitted in models of inelastic DM~\cite{Tucker-Smith:2001myb}, in which an excited state downscatters to its ground state and releases an energy comparable to the observed electron recoil energy, see e.g.\ Refs.~\cite{Bell:2020bes,Harigaya:2020ckz,Bloch:2020uzh,Baryakhtar:2020rwy,Choi:2020ysq,An:2020tcg,Chao:2020yro,He:2020wjs,He:2020sat,Borah:2020jzi,Baek:2021yos}. The origin of the population of excited states depends on the specific model under consideration. The cases most commonly considered are that the excited states are a cosmological relic (i.e.\ their lifetime exceeds the age of the universe) or that they are produced in astrophysical objects such as the Sun~\cite{Baryakhtar:2020rwy}.

In the present work, we consider terrestrial upscattering as an alternative mechanism to produce a population of excited states, which can subsequently create observable signals in direct detection experiments. In this set-up a DM particle in the ground state is excited by upscattering on an atom in the Earth and subsequently de-excited by downscattering in the detector. We note that a similar mechanism has been considered previously under the name of ``luminous DM''~\cite{Feldstein:2010su}, but this model assumes that the excited DM particles de-excite spontaneously (under the emission of a photon) rather than via downscattering~\cite{Eby:2019mgs}.

For the case of downscattering, it is essential to accurately calculate not only the fraction of excited particles, but also of their velocity distribution. To achieve this goal, we extend the analytical formalism for terrestrial elastic nuclear scattering from Kavanagh et al.~\cite{Kavanagh:2016pyr} to account for an inelastic splitting. We find that because of the inelasticity of the collision, DM particles may be slowed down considerably during upscattering, which enhances their density via the ``traffic jam'' effect~\cite{pospelov:2020}, but makes it necessary to account for the probability that the excited state decays before reaching the detector. We validate these findings using explicit Monte Carlo simulations similarly to Ref.~\cite{Emken:2017qmp}.

Applying our formalism to the XENON1T excess, we find that the observed signal can be fitted for DM masses of a few GeV, provided that the DM-nucleon cross section (responsible for upscattering) and the DM-electron cross section (responsible for downscattering) are of a similar magnitude. Furthermore, we calculate for the first time the modulation of the signal resulting from the daily rotation of the Earth and find that the effect may be large enough to be observable in future experiments aiming to confirm the excess.

The remainder of this work is structured as follows. In Sec.~\ref{sec: direct detection} we review the direct detection of electron recoils using noble gas targets with a specific focus on the case of inelastic scattering. Section~\ref{sec: upscattering formalism} then describes our new formalism for terrestrial upscattering and presents our calculation of the resulting flux of excited DM particles. In Sec.~\ref{sec: results} we combine both effects to calculate direct detection signals from the combination of terrestrial upscattering and subsequent downscattering. We use the XENON1T excess as an illustrative example to constrain the parameter space of the model and to predict the daily modulation of the signal. Section~\ref{sec: discussion} reviews some of the model-building challenges and complementary constraints for the scenario that we consider. Additional technical details are provided in appendix~\ref{app: simulations}.

\section{Direct detection of inelastic dark matter}
\label{sec: direct detection}

The basic idea of inelastic DM is that there is a mass splitting $\delta>0$ between the ground state $\chi$ with mass $m_\chi$ and the excited state $\chi^\ast$, where $\delta \ll m_\chi$ \cite{Tucker-Smith:2001myb,Tucker-Smith:2004mxa}. The couplings of these particles are off-diagonal, meaning that every scattering process must involve one ground state and one excited state. This allows for inelastic upscattering $\chi + X \to \chi^\ast + X$, where $X$ can for example be a nucleus, and for exothermic downscattering $\chi^\ast + X \to \chi + X$. While in the former process a part of the kinetic energy of the incoming DM particle is absorbed, in the latter process additional energy is released in the form of recoil energy of the outgoing particles. We begin our discussion by briefly reviewing the scattering kinematics and the resulting event rates in direct detection experiments for both nuclear and electron recoils for the case of inelastic and exothermic scattering. 

The differential event rate of nuclear recoils with respect to recoil energy $E_\mathrm{nr}$ is given by
\begin{equation}
\frac{\text{d}R}{\text{d}E_\mathrm{nr}} = 
\frac{\rho}{m_{N} m_\chi}  \int_{v>v_\text{min}} 
v f(\mathbf{v}) \frac{\text{d} \sigma}{\mathrm{d} E_\mathrm{nr}} \, \text{d}^3 v\; ,
\label{eq:dRdE}
\end{equation}
where $m_\chi$ and $m_{N}$ denote respectively the DM and target nucleus mass and $\rho$ and $f(\mathbf{v})$ are the DM density and velocity distribution in the laboratory frame. The differential scattering cross section can be written as
\begin{equation}
\label{eq:usualSI}
\frac{\text{d}\sigma}{\text{d} E_\mathrm{nr}} = C^2_\text{T} (A,Z) F(E_\mathrm{nr})^2 \frac{m_N \sigma_p}{2 \mu_{n}^2 v^2} \;,
\end{equation}
where $\mu_{p} = m_\chi m_p / (m_\chi + m_p)$ is the reduced DM-nucleon mass, $F(E_\mathrm{nr})$ is the nuclear form factor, $\sigma_p$ is the DM-proton scattering cross section and the function $C_\text{T}(A,Z)$ gives the scaling of the DM-nucleus cross section $\sigma$ with mass number $A$ and charge $Z$. In the following we will focus on the case that the DM couplings to SM particles are proportional to their charge, such that $C_\text{T}(A,Z) = Z$. Moreover, we assume that the mediator of the interaction is heavy compared to the momentum transfer, such that no additional form factor is needed to parametrize the momentum dependence of the scattering process itself.

Energy and momentum conservation are encoded in the minimum velocity $v_\text{min}$ required to produce a nuclear recoil of energy $E_\mathrm{nr}$:
\begin{equation}
v_\text{min} = \left|\frac{m_N \, E_\mathrm{nr}}{\mu_N} \pm \delta \right| \frac{1}{\sqrt{2 \, E_\mathrm{nr} \, m_N}} \; ,
\end{equation}
where $\mu_N$ is the reduced DM-\emph{nucleus} mass. The positive sign corresponds to upscattering, the negative sign to downscattering.

\begin{figure}[t!]
    \centering
    \includegraphics[width=\columnwidth]{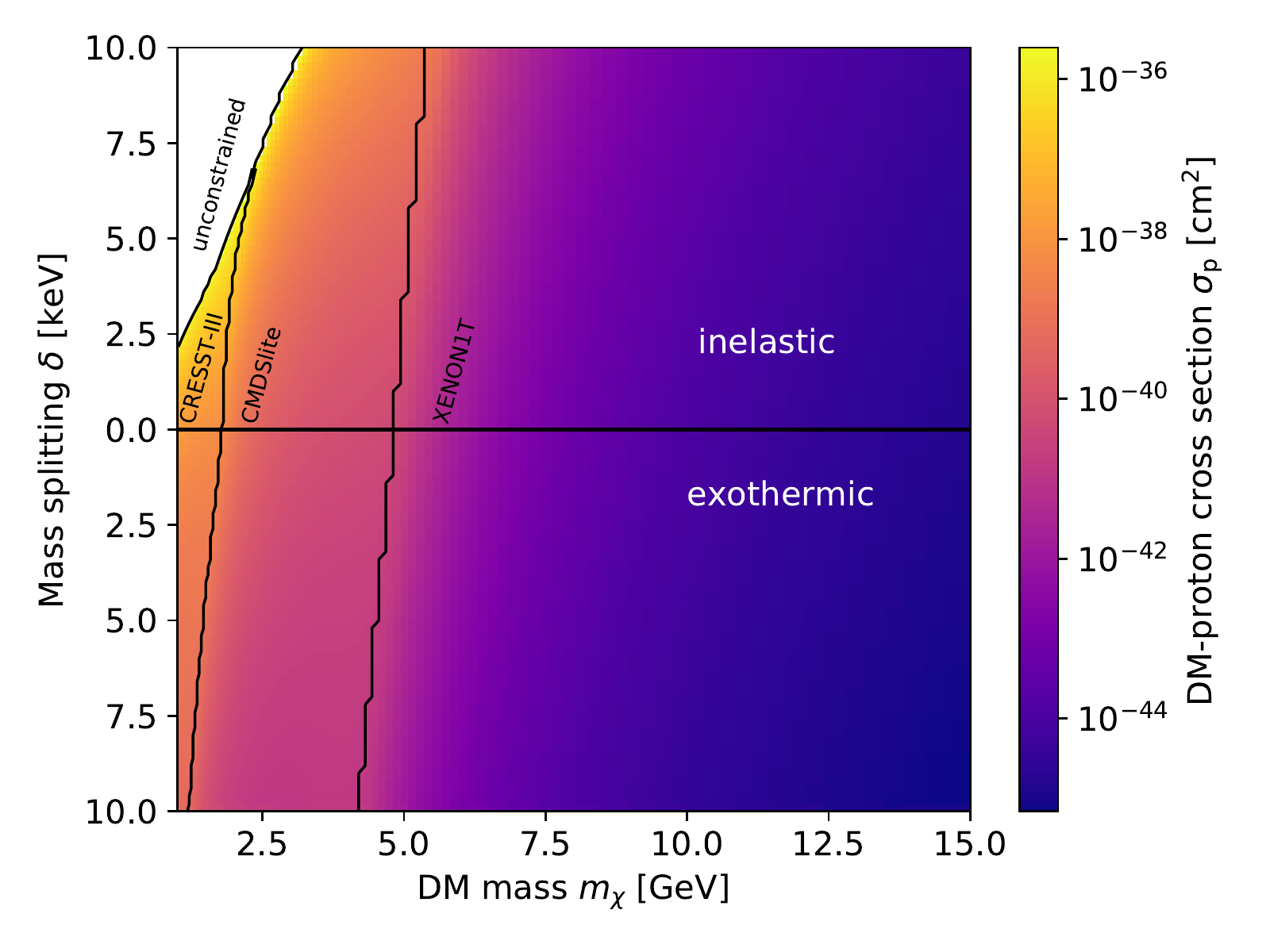}
    \caption{Upper bound at 90\% confidence level on the DM-nucleon scattering cross section $\sigma_{p}$ as a function of $m_\chi$ and $\delta$ for inelastic scattering (top half) and exothermic scattering (bottom half) under the assumption that the DM-nucleus scattering cross section is proportional to $Z^2$. The black lines separate the regions where the strongest constraint stems from CRESST-III, CDMSlite and XENON1T. In the top-left corner $\delta / \mu_N$ is so large that scattering is impossible and no constraint can be obtained.}
    \label{fig:inelastic_bounds}
\end{figure}

In the following, we will assume that almost all DM particles are in the ground state. For upscattering, the DM density is therefore given by the local DM density, $\rho = 0.4 \, \mathrm{GeV \, cm^{-3}}$~\cite{Catena:2009mf}, and the velocity distribution is given by the Standard Halo Model:
\begin{equation}
 f(\mathbf{v}) = f_\text{SHM}(\mathbf{v} + \mathbf{v}_\text{E}(t)) \; ,
 \label{eq:SHM}
\end{equation}
where $f_\text{SHM}(\mathbf{v})$ is a Maxwell-Boltzmann distribution with $v_0=220$~km\,s$^{-1}$ cut off at the escape velocity $v_{\rm{esc}}=544$~km\,s$^{-1}$, and $\mathbf{v}_\text{E}(t)$ is the velocity of the Earth relative to the Galactic rest frame~\cite{Evans:2018bqy}. Under these assumptions, upscattering is possible only if $v_\text{min} < v_\text{esc} + v_\text{E} \approx 2.5 \cdot 10^{-3} c$, which in turn requires $\delta/\mu_N \lesssim 3 \cdot 10^{-6}$. Thus, for DM particles in the GeV range, inelastic scattering is possible only if the mass splitting is in the keV range. This is illustrated in Fig.~\ref{fig:inelastic_bounds}, which shows the upper bounds from CRESST-III~\cite{CRESST:2019jnq}, CDMSlite~\cite{SuperCDMS:2015eex} and XENON1T~\cite{XENON:2018voc} on the DM-proton scattering cross section $\sigma_p$ as a function of $\delta$ and $m_\chi$. These constraints have been obtained using a modified version of \texttt{DDCalc}~\cite{GAMBITDarkMatterWorkgroup:2017fax,GAMBIT:2018eea}. 

Let us now turn to the case of electron scattering. In this case, we need to account for the binding energy $E_b$ of the electron. Denoting its final kinetic energy by $E_\mathrm{er} = k^{\prime \, 2} / (2 m_e)$ and the initial (final) DM velocity by $\mathbf{v}$ ($\mathbf{v}'$), energy conservation implies
\begin{align}
 \frac{1}{2} m_\chi v^2 - E_b & = \frac{1}{2} m_\chi v'^2 + E_\mathrm{er} \pm \delta \; ,
\end{align}
where we have neglected terms of higher order in $v$ and $\delta$ and used that the recoil energy of the nucleus is negligible for $m_N \gg m_\chi$~\cite{Baxter:2019pnz}. 
In the following we will focus on downscattering, corresponding to the negative sign in the first equation. Defining the momentum change of the DM particle by $\mathbf{q} = m_\chi (\mathbf{v} - \mathbf{v}')$ and the energy transfer $\Delta E_e = E_b + E_\mathrm{er}$ one then finds
\begin{equation}
 v_\text{min} = \left| \frac{\Delta E_e - \delta}{q} + \frac{q}{2 m_\chi}\right| \; .
\end{equation}
Moreover, using the upper bound on the DM velocity $v_\text{max}$ we can determine the range of allowed momentum transfer, which can be written as $q_\text{min} < q < q_\text{max}$ with
\begin{align}
q_\text{min} & =\text{sign}(\Delta E_e - \delta) m_\chi v_\text{max}\left(1-\sqrt{1-\frac{2(\Delta E_e - \delta)}{m_\chi v_\text{max}^2}}\right)\\
q_\text{max} & = m_\chi v_\text{max}\left(1+\sqrt{1-\frac{2(\Delta E_e-\delta)}{m_\chi v_\text{max}^2}}\right).
\end{align}
The differential rate of electron recoil events is then given by
\begin{align}
\frac{\mathrm{d}R_\text{ion}}{\mathrm{d}E_\mathrm{er}} = \frac{\rho}{m_\chi} \frac{\sigma_e}{8 E_\mathrm{er} \mu_e^2} \sum_{n,l} & \int_{q_\text{min}}^{q_\text{max}} q \mathrm{d}q\ |f_{n,l\rightarrow E_r}(q)|^2 \nonumber \\ & \times \int_{v>v_\text{min}} \mathrm{d}^3 v \frac{f^\ast(\mathbf{v})}{v} \label{eq:exorate} \; ,
\end{align}
where $f^\ast(\mathbf{v})$ denotes the velocity distribution of excited states (normalized such that $\rho \int \mathrm{d}^3 v f^\ast(\mathbf{v}) \equiv \rho^\ast$ yields the density of excited states), $\sigma_e$ is the DM-electron scattering cross section (which we again assume to be momentum-independent) and $\mu_e$ is the DM-electron reduced mass. The quantum numbers $n$ and $l$ denote the different atomic shells and the corresponding ionization form factors for a final state energy $E_\mathrm{er}$ are given by
\begin{equation}
f_{n,l\rightarrow E_\mathrm{er}}(q) = \frac{4 k^{\prime 3}}{(2\pi)^3}\sum_{l^\prime=0}^{\infty}\sum_{m=-l}^{l}\sum_{m^\prime=-l^\prime}^{l^\prime}|f_{1\rightarrow 2}(q)|^2 \; . 
\end{equation}

To calculate these form factors for Xenon, we use \texttt{DarkARC} \cite{Catena:2019gfa} which employs Rothaan-Hartree-Fock orbitals \cite{BUNGE1993113} for the initial state and solves the Schr\"{o}dinger equation for continuum states of a hydrogen-like potential with adjusted charge for the final state \cite{Catena:2019gfa,bethe2012quantum}. We have checked that this approach agrees well with the one from Ref.~\cite{Essig:2011nj}.

\begin{figure}[t!]
    \centering
    \includegraphics[width=\columnwidth]{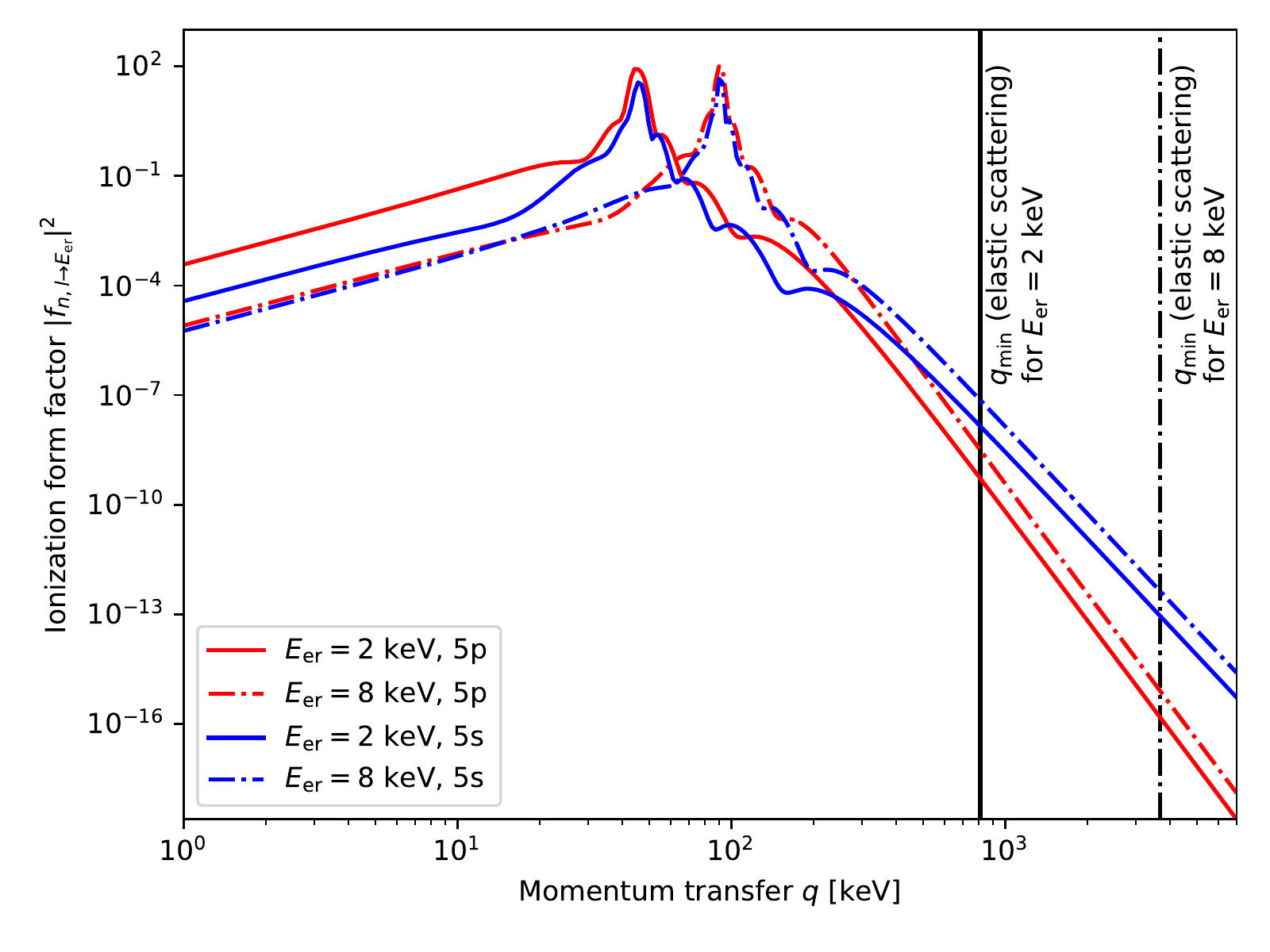}
    \caption{Ionization form factors of the 5p and 5s shell for different recoil energies. For elastic scattering, the momentum transfer must satisfy $q > q_\text{min}$ as indicated by the solid (dashed) vertical line for $E_\mathrm{er} = 2 \, \mathrm{keV}$ ($E_\mathrm{er} = 8 \, \mathrm{keV}$). For exothermic scattering, on the other hand, the entire range of momentum transfer can contribute, leading to a significant enhancement of the signal.}
    \label{fig:ionisationff}
\end{figure}

Figure~\ref{fig:ionisationff} shows the ionization form factors for the 5p and 5s shells and different electron recoil energies in the keV range. We find that these form factors are strongly peaked at momentum transfer $q \sim \sqrt{2 m_e E_\mathrm{er}}$. These momentum transfers are however tiny compared to the values of $q_\text{min}$ obtained for elastic scattering ($\delta = 0$), as indicated by the vertical lines. For exothermic scattering, on the other hand, much smaller values of $q_\text{min}$ are possible if $\delta \approx \Delta E_e$. If this is the case the integration over $q$ leads to a strong enhancement of the scattering rate. If $\delta$ is large compared to the initial energy of the electron, we therefore find that the differential event rate for exothermic scattering will be peaked at $E_\mathrm{er} \approx \delta$.

In the following we will be most interested in electron recoil energies in the range 2--3 keV, corresponding to the excess observed in the XENON1T experiment. This consideration fixes the mass splitting $\delta$ to the same range, which in turn implies $m_\chi \gtrsim 1 \, \mathrm{GeV}$ in order for upscattering on nuclei to be kinematically allowed.\footnote{Upscattering on electrons is also possible in principle, but the required momentum transfer is so large that the ionization form factor is heavily suppressed. For the cross sections that we will be interested in, this effect is therefore completely negligible.} A more detailed analysis of the parameter space will be performed in Sec.~\ref{sec: results}. First we however need to take a closer look at terrestrial upscattering in order to calculate the presently unknown density $\rho^\ast$ of excited states and their speed distribution $f^\ast(v) \equiv \int \mathrm{d}\Omega_v v^2 f^\ast(\mathbf{v})$. 

\section{Terrestrial upscatterings}
\label{sec: upscattering formalism}

In the context of exothermic and luminous~DM, the origin of the excited states that pass through our detector is a central question.
For a long enough mean lifetime~$\tau$, there might be a primordial population that originates from the thermal bath of the early Universe and survives until the present time when it can trigger our detectors~\cite{Baryakhtar:2020rwy, An:2020tcg,CarrilloGonzalez:2021lxm,Fitzpatrick:2021cij,Finkbeiner:2008gw}. 
For shorter lifetimes, these particles will have decayed by now, and in order for us to be able to detect exothermic and luminous~DM, we rely on local mechanisms to generate a detectable amount of excited DM~states inside our solar system, either from the Sun or from inside the Earth~\cite{Baryakhtar:2020rwy}.
In this paper, we focus on terrestrial upscattering which is the most important source of excited DM~states if the excited states created inside the Sun decay before reaching the Earth, i.e. if $v\tau\ll 1$~AU.

In the context of daily modulations due to elastic DM-nuclear scatterings, Kavanagh et al. have developed a general analytic framework to describe the impact of Earth scatterings on the DM~distribution inside a detector~\cite{Kavanagh:2016pyr}.
While its validity is limited to the single-scattering regime, and the impact of multiple elastic scatterings typically require Monte Carlo simulations~\cite{Collar:1992qc,Collar:1993ss,Hasenbalg:1997hs,Emken:2017qmp,Kavanagh:2020cvn}, the formalism is ideal to describe upscatterings of inelastic DM.
For this purpose, we extend the formalism by Kavanagh et al.\ in two major ways.
We need to account for
\begin{itemize}
    \item[(a)] the modified kinematics of inelastic scatterings, and\\
    \item[(b)] the possibility that an excited state created deep inside the Earth might decay before reaching the Earth's surface.
\end{itemize}  

The main result of this section is an analytic expression of the speed distribution~$f^*(v)$ of upscattered DM~particles through any detector on Earth, presented in Eq.~\eqref{eq:master final}, which allows us to compute the expected electron recoil event rates and modulation signature of exothermic and luminuous~DM.
Finally, the density~$\rho^*$ of excited states is a crucial parameter for direct detection and is encapsulated in $f^*(v)$ by its relative normalization:
\begin{align}
    \rho^* = \rho\int \mathrm{d}v \; f^*(v)\, . \label{eq: upscattered state density}
\end{align}
\begin{figure}[t!]
    \centering
    \includegraphics[width=0.48\columnwidth,clip,trim=100 0 100 0]{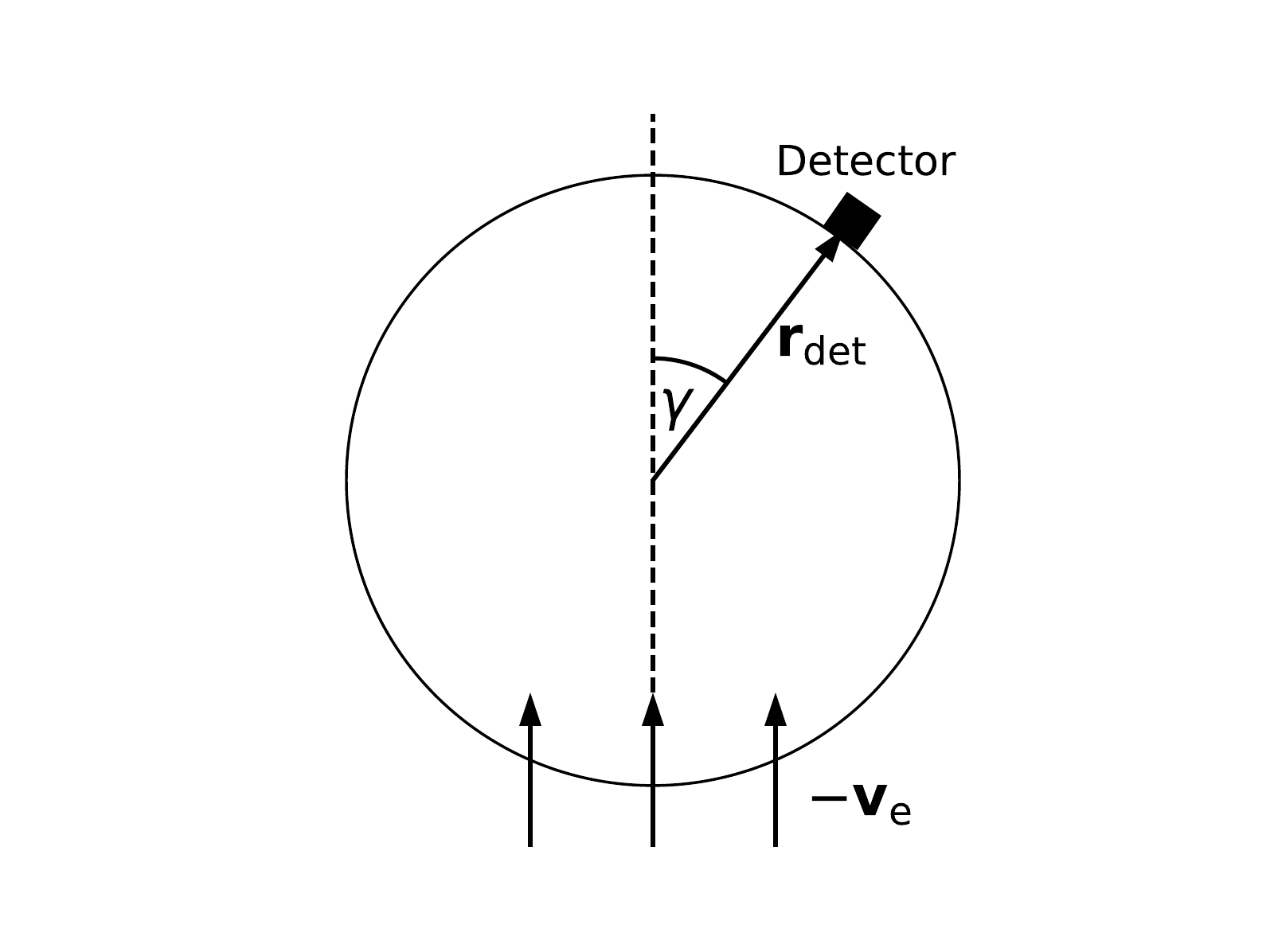}
    \includegraphics[width=0.48\columnwidth,clip,trim=100 0 100 0]{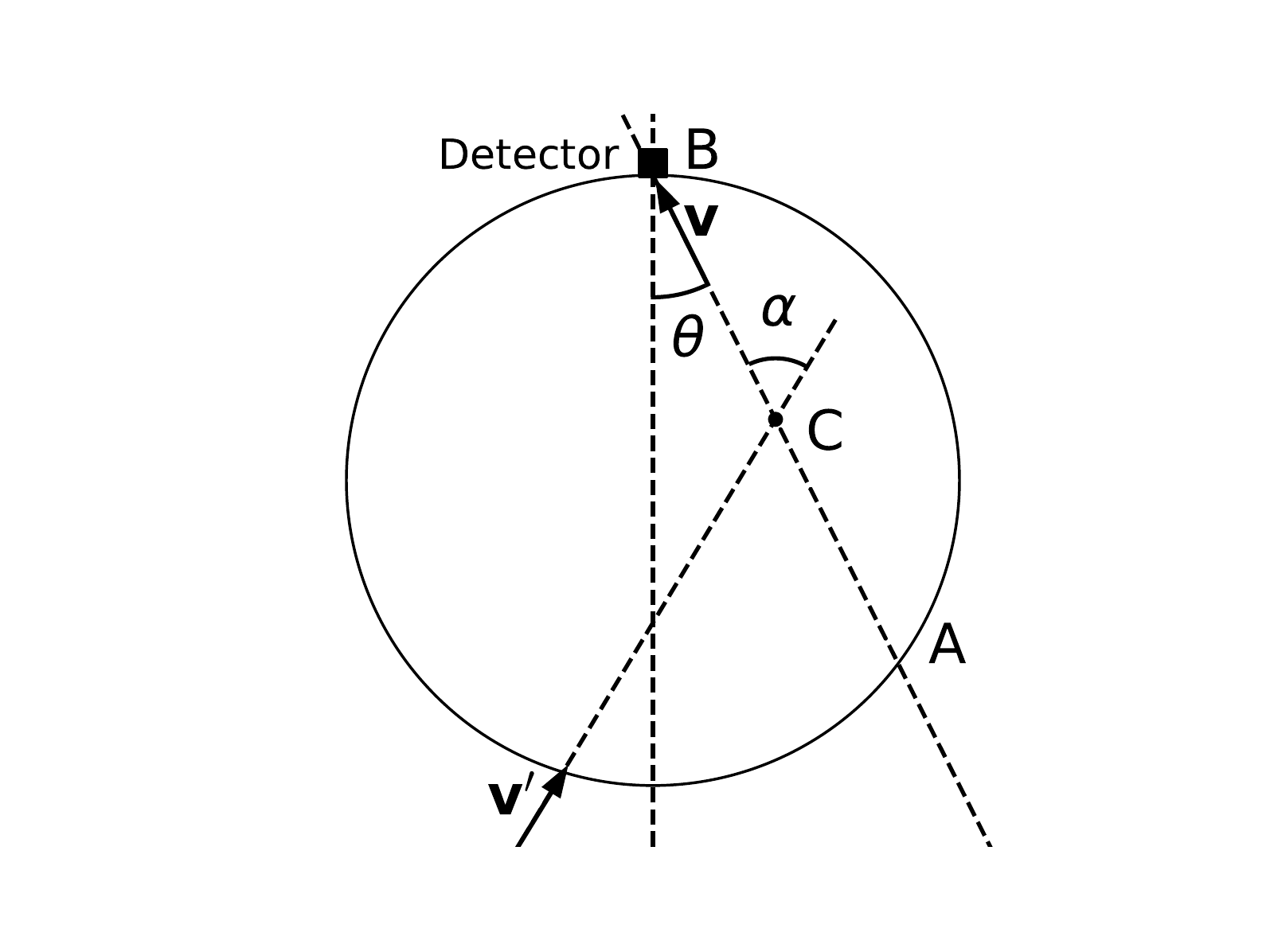}
    \caption{Definition of the various angles occurring in the calculation of terrestrial upscattering. Figures adapted from Fig.~1 and~2b of Ref.~\cite{Kavanagh:2016pyr}.}
    \label{fig: angles}
\end{figure}

The final results obtained in this section depend on the assumed velocity distribution of the incoming DM particles (see above), the particle physics properties of DM and the position of the detector $\mathbf{r}_\text{det}$. The latter dependence can be simplified by exploiting the system's axial symmetry around the direction of the Earth's velocity in the galactic rest frame.
As shown in the left panel of Fig.~\ref{fig: angles} the polar angle~$\gamma$ relative to this symmetry axis is defined by
\begin{align}
    \cos\gamma = -\frac{\mathbf{v}_\mathrm{E}\cdot \mathbf{r}_\mathrm{det}}{v_\mathrm{E} r_\mathrm{E}}\, , \label{eq: isodetection angle}
\end{align}
where $r_\mathrm{E}$ denotes the Earth's radius. This angle serves as a measure of the detector's location on Earth relative to the DM~wind, with~$\gamma = 0^\circ(180^\circ)$ corresponding to the DM~wind passing through the detector from below (above). 

\subsection{Contribution from one direction}
\label{ss: contribution from one point}

Every point inside the Earth acts as a source of excited~DM upscattered by terrestrial nuclei.
First, we focus on excited states that arrive at a detector from a certain direction defined by the line between a point~A on the Earth's surface and the detector's location~B, characterized by the angle~$\theta$ as sketched in the right panel of Fig.~\ref{fig: angles}.
Furthermore, we assume a DM particle from the galactic halo with initial velocity~$\mathbf{v}^\prime$ to potentially upscatter at a point~C along the line~AB.
The probability of a DM~particle to scatter within an infinitesimal interaction region around~C of length~$\dd l$ (along AB) and surface~$\dd \mathbf{S}$ (perpendicular to AB) is given by
\begin{align}
\dd p_\mathrm{scat}=\frac{\dd l}{\lambda(r)\cos\alpha}\, ,\label{eq: dpscat}
\end{align}
where~$\lambda(r)$ is the mean free path, which depends on the density of scattering targets and hence on the distance~$r$ between point~C and the Earth's center, and~$\alpha$ is the scattering angle in the Earth's rest frame, as shown in Fig.~\ref{fig: angles}.

The rate of halo DM~particles of velocity~$\mathbf{v}^\prime$ entering the interaction region, upscattering on a terrestrial nucleus, and ending up with a final velocity~$\mathbf{v}$ towards the detector is given by
\begin{align}
\underbrace{\left[f_0(\mathbf{v^\prime})\mathbf{v^\prime}\cdot \dd\mathbf{S} \dd^3v^\prime\right]}_{\text{entering rate}}\times\underbrace{\left[\dd p_\mathrm{scat} P(\mathbf{v^\prime}\rightarrow \mathbf{v})\dd^3v\right]}_{\text{probability to scatter to $\mathbf{v}$}}\, .
\end{align}
This can be equated to the rate of upscattered DM~particles leaving the infinitesimal interaction region towards the detector,
\begin{align}
f^*(\mathbf{v},\mathbf{v}^\prime,\mathbf{r}_C) \mathbf{v}\cdot \dd\mathbf{S} \dd^3v\, ,
\end{align}
which provides us with an expression for the contributions of the interaction point~C to the upscattered DM~distribution~$f^*(\mathbf{v})$.
Using  $\mathbf{v}\cdot \dd\mathbf{S}=v\,\dd S$ and $\mathbf{v^\prime}\cdot \dd\mathbf{S}=v \,\dd S\cos\alpha$, we find
\begin{align}
f^*(\mathbf{v},\mathbf{v}^\prime,\mathbf{r}_C)=\frac{\dd l}{\lambda(r)}\frac{v^\prime}{v}f_0(\mathbf{v^\prime})P(\mathbf{v^\prime}\rightarrow \mathbf{v})\dd^3v^\prime\label{eq:master1}\, .
\end{align}
In order to obtain all contributions for a given velocity~$\mathbf{v}$ or equivalently a direction, we need to integrate over all interaction points along the line~AB.
In doing so, we need to account for the fact that the~$\chi^*$ particles are unstable with a mean lifetime of~$\tau$,
\begin{align}
& f^*(\mathbf{v},\mathbf{v}^\prime)\nonumber\\
& \quad = \int_\mathrm{AB}\frac{\dd l}{\lambda(r)}\underbrace{\exp\left(-\frac{l}{ v\tau}\right)}_\text{decays}\frac{v^\prime}{v}f_0(\mathbf{v^\prime})P(\mathbf{v^\prime}\rightarrow \mathbf{v})\dd^3v^\prime\, . \label{eq: AB integral}
\end{align}
The exponential weight factor describes the particles that decay before reaching the detector's location.
It depletes the density of upscattered states, in particular of slow ones.

Following the steps of~Kavanagh et al., we assume a single nuclear target in order not to clutter the notation with additional indices. A generalization to multiple targets is trivial, and we will restore the target index in the very end.
Due to~$\lambda^{-1}(r)=n(r)\sigma$, where~$n(r)$ is the target number density, and~$\sigma$ is the total upscattering cross-section, we can isolate the only factors of Eq.~\eqref{eq: AB integral} that depend on the position inside the Earth, and we absorb the integral into an effective Earth-crossing distance~$d_{\mathrm{eff}}(\cos\theta)$,
\begin{align}
     d_{\mathrm{eff}}(\cos\theta)&\equiv\int_\text{AB}\dd l\, \frac{n(r)}{\bar{n}} \exp\left(-\frac{l}{v\tau}\right)\, .
\end{align}
Here, we defined an averaged number density of target nucleus~$i$,
\begin{align}
    \bar{n} \equiv \frac{1}{r_\mathrm{E}} \int_0^{r_\mathrm{E}} \dd r\, n(r)\, .
\end{align}

Next, we change the variable of integration to~$r$.
For a given value of~$r$, the distance~$l$ to the detector is given by
\begin{align}
    l &= r_\mathrm{E}\cos\theta\pm\sqrt{r^2-r_\mathrm{E}^2\sin^2\theta}\, , 
\label{eq:pathrad}
\end{align}
where by construction $\cos \theta \geq 0$.
We can therefore write the integral along AB as
\begin{widetext}
    \begin{align}
        \bar{n} d_{\mathrm{eff}}(\cos\theta) &= \exp\left(-\frac{r_\mathrm{E}\cos\theta}{v\tau}\right) \int_{r_\mathrm{E}\sin\theta}^{r_\mathrm{E}}\dd r\;n(r)\frac{r}{\sqrt{r^2-r_\mathrm{E}^2\sin^2\theta}}\left[\ \exp\left(\tfrac{\sqrt{r^2-r_\mathrm{E}^2\sin^2\theta}}{v\tau}\right)+ \exp\left(-\tfrac{\sqrt{r^2-r_\mathrm{E}^2\sin^2\theta}}{v\tau}\right)\right]\\ \label{eq:denspm}
	&=2 v\tau\exp\left(-\frac{r_\mathrm{E}\cos\theta}{v\tau}\right)\int_{r_\mathrm{E}\sin\theta}^{r_\mathrm{E}}\dd r\; n(r)\frac{\dd}{\dd r}\sinh\left(\frac{\sqrt{r^2-r_\mathrm{E}^2\sin^2\theta}}{v\tau}\right)
\end{align}
We assume that the mass density inside the Earth's mantle and core is constant respectively.
\begin{align}
    n(r) = \begin{cases}
    n_c\, , \quad&\text{ for }   r < r_\mathrm{core}\, ,\\
    n_m\, , \quad&\text{ for }r_\mathrm{core} \leq r \leq r_\mathrm{E}\, ,
    \end{cases}
\end{align}
We list the various elements included in our analysis and their respective abundances in the mantle and core in table~\ref{tab:el}. The core radius is taken to be $r_\text{core} = 3500\,\mathrm{km}$. Core and mantle are found to contribute 32\% and 68\% to the total mass of Earth, respectively. 

\begin{table*}[t]
\center
\caption{Fractions of the total mass in core and mantle. The values are taken from Ref.~\cite{Kavanagh:2016pyr} apart from Ni, which was found in Ref.~\cite{Lundberg_2004}. 
}
 \begin{tabular}{ p{4cm} p{1.2cm} p{1.2cm} p{1.2cm} p{1.2cm} p{1.2cm} p{1.2cm} p{1.2cm} p{1.2cm} p{1.2cm} p{1.5cm} } 
 \hline
 \hline
Element & O &  Si &  Mg & Fe & Ca & Na & S & Al & Ni & total \\
\hline
Mass in GeV & 14.9 & 26.1 & 22.3 & 52.1 & 37.2 & 21.4 & 29.8 & 25.1 & 58.7 & \\
Relative abundance mantle & 0.4400 & 0.2100 & 0.2280 & 0.0626 &  0.0253 & 0.0027  & 0.0003 & 0.0235 & 0 & 0.9924 \\
Relative abundance core & 0 & 0.060 & 0 & 0.855 & 0 & 0 & 0.019 & 0 & 0.052 & 0.986 \\
\hline
 \hline
\end{tabular}
\label{tab:el}
\end{table*}

Depending on whether the AB~line crosses through the Earth's core (i.e. whether $r_\mathrm{E}\sin\theta < r_\mathrm{core}$), there are two possible results,
\begin{align}
\label{eq:decinc}
    \bar{n} d_\mathrm{eff}(\cos\theta) &=\begin{cases}
    2v\tau\exp\left(-\frac{r_\mathrm{E}\cos\theta}{\tau v}\right)\Big[n_c\sinh\Big(\frac{\sqrt{r_\mathrm{core}^2-r_\mathrm{E}^2\sin^2\theta}}{v\tau}\Big)&\\
        \qquad\qquad\qquad\qquad\quad +n_m\Big(\sinh\Big(\frac{r_\mathrm{E}\cos\theta}{v\tau}\Big)-\sinh\Big(\frac{\sqrt{r_\mathrm{core}^2-r_\mathrm{E}^2\sin^2\theta}}{v\tau}\Big)\Big)\Big]\, ,&\text{ if } r_\mathrm{E}\sin\theta<r_\mathrm{core}\, ,\\
        2v\tau\exp\left(-\frac{r_\mathrm{E}\cos\theta}{\tau v}\right)n_m\sinh\left(\frac{r_\mathrm{E}\cos\theta}{\tau v}\right)\, ,&\text{ otherwise.}
    \end{cases}
    \end{align}
\end{widetext}
This relation allows us to rewrite Eq.~\eqref{eq: AB integral} as
\begin{equation}
f^*(\mathbf{v},\mathbf{v}^\prime)=\sigma\bar{n} d_\mathrm{eff}(\cos\theta)\frac{v^\prime}{v}f_0(\mathbf{v^\prime})P(\mathbf{v^\prime}\rightarrow \mathbf{v})\dd^3v^\prime\label{eq:master2}\, .
\end{equation}

The last piece of this expression that we need to evaluate is the probability~$P(\mathbf{v^\prime}\rightarrow \mathbf{v})$ to upscatter from initial velocity~$\mathbf{v}^\prime$ to final velocity~$\mathbf{v}$.
In order to do so, we will need further knowledge of the kinematics of the scattering process.

\subsection{Probability and kinematics for upscatterings}
\label{ss: kinematics}
Due to the azimuthal symmetry of the system, the probability $P(\mathbf{v^\prime}\rightarrow \mathbf{v})$ only depends on the scattering angle~$\alpha$ in the Earth's rest frame.
We can therefore parametrize the probability as
\begin{align}
P(\mathbf{v^\prime}\rightarrow \mathbf{v})=\frac{1}{2\pi v^2}\delta(v-\kappa^{-1}(v^\prime,\alpha)) P(\cos\alpha)\, , \label{eq: P(v'->v)}
\end{align}
where $\kappa(v,\alpha)$ is the kinematic relation yielding the initial state speed~$v^\prime$ in terms of~$v$ and~$\alpha$, the function~$\kappa^{-1}(v^\prime,\alpha)$ is the corresponding inverse, and~$P(\cos\alpha)$ is the probability to scatter with scattering angle~$\alpha$.
The kinematic functions~$\kappa$ and~$\kappa^{-1}$ can be obtained from energy and momentum conservation,
\begin{align}
    \frac{1}{2}m_\chi v^{\prime 2} &= \frac{1}{2}m_\chi v^2 +\frac{1}{2}m_N v_N^2  +\delta\, ,\\
    m_\chi \mathbf{v}^\prime &= m_\chi \mathbf{v} + m_N \mathbf{v}_i\, ,
\end{align}
where we use~$\delta\ll m_\chi$.
Using~$\mathbf{v}^\prime\cdot\mathbf{v}=v^\prime v \cos\alpha$, we obtain
\begin{align}
\label{eq: kappas}
    \kappa^{-1}_\pm(v^\prime,\alpha) &= v^\prime \frac{ \cos\alpha\pm\sqrt{\tfrac{m_N^2}{m_\chi^2}-\sin^2\alpha-\frac{2\delta m_N(m_N+m_\chi)}{m_\chi^3 v^{\prime 2}}}}{1+m_N/m_\chi} \,
 , \\
 \kappa_\pm(v,\alpha) &= v\frac{ \cos\alpha\mp\sqrt{\tfrac{m_N^2}{m_\chi^2}-\sin^2\alpha+\frac{2\delta m_N(m_N-m_\chi)}{m_\chi^3 v^2}}}{1 - m_N/m_\chi}\, . 
\end{align}
The physical solutions are found by demanding~$\kappa^{-1}_\pm(v^\prime,\alpha)>0$ and $\kappa_\pm(v,\alpha)>0$.
Depending on the sign in $\kappa^{-1}_\pm(v^\prime,\alpha)$ and $\kappa_\pm(v,\alpha)$ as well as the scattering angle~$\alpha$, this can be translated in conditions on~$v^\prime$. To simplify our calculations, we will from now on assume that $m_\chi < m_N$.
In this case, we find
\begin{align}
    v^\prime &\geq \sqrt{\frac{2\delta m_N^2 }{\mu_N (m_N^2-m_\chi^2\sin^2\alpha)}}\geq\sqrt{\frac{2\delta}{\mu_N}}\, ,
\end{align}
which is the kinetic threshold for the upscattering process.
Moreover, we find that for $\kappa_\pm(v,\alpha)$ only the `+' solution is physical, which is why we simply write~$\kappa(v,\alpha)\equiv\kappa_+(v,\alpha)$ in the following. For $\kappa^{-1}_\pm(v^\prime, \alpha)$ we include both solutions (if they are both physical) but find that the `+' solution gives the dominant contribution. 

The second ingredient of Eq.~\eqref{eq: P(v'->v)} is the probability~$P(\cos\alpha)$, which can be related to the probability for a given scattering angle~$\alpha_\mathrm{cms}$ in the center-of-mass system~(CMS),
\begin{align}
    P(\cos\alpha) = P(\cos\alpha_\mathrm{cms}) \frac{\dd \cos\alpha_\mathrm{cms}}{\dd \cos \alpha}\, . \label{eq: P(cos a)}
\end{align}
Throughout this study, we assume isotropic contact interactions, i.e.~$P(\cos\alpha_\mathrm{cms}) = \frac{1}{2}$.
Next we express~$\cos\alpha_\mathrm{cms}$ in terms of~$\cos\alpha$ and~$v^\prime$ in the Earth's rest frame,
\begin{align}
    \cos\alpha_\mathrm{cms} &= \frac{\mathbf{v}^\prime_\mathrm{cms}\cdot\mathbf{v}_\mathrm{cms}}{v^\prime_\mathrm{cms}v_\mathrm{cms}}\, .
\end{align}
Using
\begin{align}
    \mathbf{v}^\prime_\mathrm{cms} = \frac{\mu_N}{m_\chi}\mathbf{v}^\prime \, , & & 
    \mathbf{v}_\mathrm{cms} = \mathbf{v}-\frac{\mu_N}{m_N}\mathbf{v}^\prime\, ,
\end{align}
we find
\begin{align}
    \cos\alpha_\mathrm{cms} &=\left.\frac{v\cos\alpha-\frac{\mu_N}{m_N}v^\prime}{\sqrt{v^2+\frac{\mu_N^2}{m_N^2}v^{\prime 2}-2\frac{\mu_N}{m_N}v v^\prime \cos\alpha}} \right|_{v=\kappa^{-1}_\pm(v^\prime,\alpha)}  \nonumber \\ &=\left.\frac{v\cos\alpha-\frac{\mu_N}{m_N}v^\prime}{\sqrt{\frac{\mu_N^2}{m_\chi^2}v^{\prime 2}-2\frac{\mu_N}{m_\chi^2}\delta}}\right|_{v=\kappa^{-1}_\pm(v^\prime,\alpha)}\, ,
\end{align}
where we have used energy and momentum conservation in the CMS in the second step. We can now evaluate the derivative in Eq.~\eqref{eq: P(cos a)},
\begin{align}
    \frac{\dd \cos\alpha_\mathrm{cms}}{\dd \cos \alpha} &= \left.\frac{\frac{\dd v}{\dd \cos\alpha}\cos\alpha+v}{\sqrt{\frac{\mu_N^2}{m_\chi^2}v^{\prime 2}-2\frac{\mu_N}{m_\chi^2}\delta}}\right|_{v=\kappa^{-1}_\pm(v^\prime,\alpha)}\end{align}
with
    \begin{align}
    &\frac{\dd \kappa^{-1}_\pm(v^\prime,\alpha)}{\dd \cos\alpha} \nonumber \\
    &\quad = \frac{\mu_N v^\prime}{m_N} 
    \Bigg(1\pm\frac{\cos\alpha}{\sqrt{\tfrac{m_N^2}{m_\chi^2}-\sin^2\alpha-\frac{2\delta m_N^2}{\mu_N m_\chi^2 v^{\prime3}}}}\Bigg) \, .
\end{align}

\begin{figure}[t!]
    \centering
    \includegraphics[width=\columnwidth]{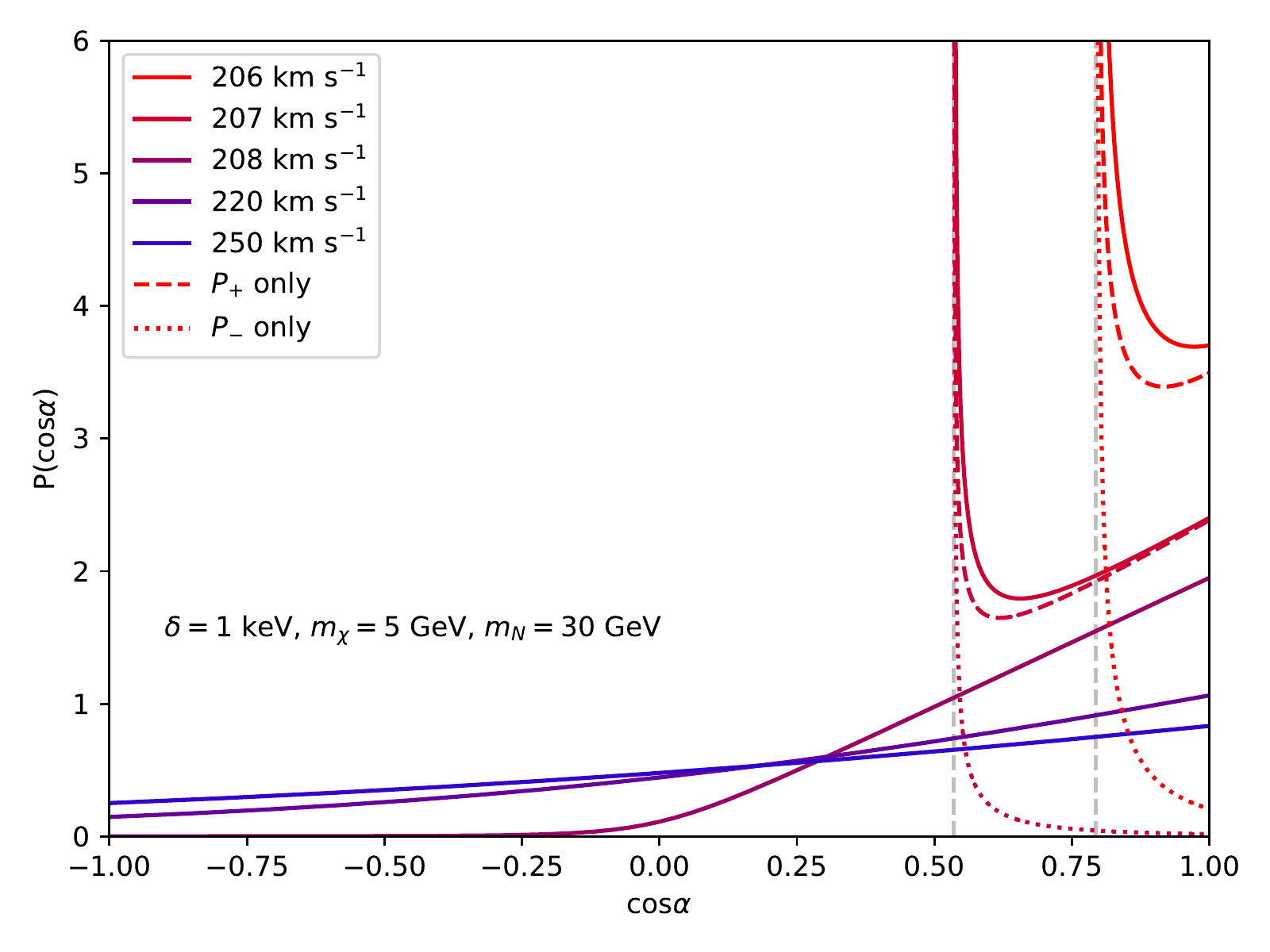}
    \caption{Distribution of the scattering angle $\alpha$ in terms of the scattering probability $P(\cos \alpha)$ for different velocities $v^\prime$ of the incoming DM particle (solid lines). Each curve is normalized to unity. For the two smallest velocities considered only large values of $\cos \alpha$ are physical (as indicated by the vertical grey lines). In these cases the total scattering probability receives two separate contributions from the two solutions of Eq.~(\ref{eq: kappas}), called $P_+$ and $P_-$, which are indicated by the dashed and dotted lines, respectively.}
    \label{fig:pscat}
\end{figure}

These results enable us to evaluate~$P(\cos\alpha)$ for all physical solutions. An example is shown in Fig.~\ref{fig:pscat} for the case $\delta = 1\,\mathrm{keV}$, $m_\chi = 5 \, \mathrm{GeV}$ and $m_N = 30 \, \mathrm{GeV}$. Physical solutions are found for $\tfrac{1}{2}\mu_N v^2 > \delta$ corresponding to $v > 205 \, \mathrm{km\,s^{-1}}$. We find that for large velocities of the incoming DM particle the scattering becomes nearly isotropic. For the smallest velocities considered, on the other hand, almost all of the kinetic energy in the CMS is required for upscattering, such that the outgoing particles are almost stationary in this frame. Their velocities in the laboratory frame are then dominated by the relative velocity between the two frames, such that only a finite range of scattering angles in the forward direction is physical. In these cases the scattering probability is obtained by summing over the two physical solutions: $P(\cos \alpha) = P_+(\cos \alpha) + P_-(\cos \alpha)$, as indicated by the dashed and dotted lines. For even smaller velocities, scattering is completely forbidden.

We can use these scattering probabilities to evaluate Eq.~\eqref{eq: P(v'->v)} and Eq.~\eqref{eq:master2}, leading to
\begin{align}
    f^*(\mathbf{v},\mathbf{v}^\prime)=& \frac{1}{2\pi}\sigma\bar{n} d_\mathrm{eff}(\cos\theta)\frac{v^\prime}{v^3}f_0(\mathbf{v^\prime}) \nonumber \\ &\times\sum_\pm \delta(v-\kappa^{-1}_\pm(v^\prime,\alpha))P_\pm(\cos\alpha)\dd^3v^\prime\label{eq:master3}\, .
\end{align}
Again we sum over both solutions of Eq.~\eqref{eq: kappas} if they are both physical.

\subsection{Summing over initial velocities, directions, and targets}
\label{ss: all contributions}

We continue by integrating Eq.~\eqref{eq:master3} over the initial velocities~$\mathbf{v}^\prime$ of the DM~particles in spherical coordinates~$(v^\prime,\theta^\prime,\phi^\prime)$.
Performing the integral over the initial speed~$v^\prime$ fixes its value through the~$\delta$-distribution,
\begin{align}
    \delta(v-\kappa_\pm^{-1}(v^\prime,\alpha))= & \left| \frac{\dd \kappa_\pm^{-1}(v^\prime,\alpha)}{\dd v^\prime}\right|^{-1}_{v^\prime = \kappa(v,\alpha)} \delta(v^\prime-\kappa(v,\alpha))\, ,
\end{align}
with
\begin{align}
    & \frac{\dd\kappa_\pm^{-1}(v^\prime,\alpha)}{\dd v^\prime} = \frac{\tfrac{(m_N^2-m_\chi^2\sin^2\alpha)v^\prime}{\sqrt{(m_N^2-m_\chi^2\sin^2\alpha)v^{\prime 2}-2\delta m_N^2/\mu_N}}\pm m_\chi\cos\alpha}{m_N+m_\chi}\, ,
\end{align}
which leaves us with
\begin{align}
    f^*(\mathbf{v})=& \sum_{\pm} \int_0^{2\pi}\dd\phi^\prime\int_{-1}^{1}\dd\cos\theta^\prime \nonumber \\
    & \times \frac{\sigma\bar{n} d_\mathrm{eff}(\cos\theta)}{2\pi} \left| \frac{\dd \kappa_\pm^{-1}(v^\prime,\alpha)}{\dd v^\prime}\right|^{-1} \nonumber \\
    & \times \left.\frac{v^{\prime 3}}{v^3}f_0(v^\prime,\cos\theta^\prime,\phi^\prime)P_\pm(\cos\alpha)\right|_{v^\prime = \kappa(v,\alpha)}\label{eq:master4}\, .
\end{align}
The speed distribution is then obtained via
\begin{align}
    f^*(v) &= v^2 \int_{0}^{2\pi}\dd\phi\int_{0}^{1}\cos\theta\;f^*(\mathbf{v})\, ,
\end{align}
where the integration bounds of~$\cos\theta$ reflect the fact that the upscattered states pass through the detector from below.

At this point we should note the relation between the spherical coordinates of~$\mathbf{v}^\prime,\mathbf{v}$ and the scattering angle~$\alpha$,
\begin{align}
    \cos\alpha = \sin\theta\sin\theta^\prime\cos(\phi-\phi^\prime)+\cos\theta\cos\theta^\prime\, .
\end{align}
As we integrate over all values of~$\phi$, we can eliminate the dependency of~$\cos\alpha$ on~$\phi^\prime$ by a shift.
The only remaining part depending on~$\phi^\prime$ is the initial DM~velocity distribution, and we perform the integral over~$\phi^\prime$ separately.
To denote this, we omit the~$\phi^\prime$ argument, i.e.~$f_0(v^\prime,\cos\theta^\prime)\equiv\int_0^{2\pi} \dd \phi^\prime\,f_0(v^\prime,\cos\theta^\prime,\phi^\prime)$.

\begin{figure}[t!]
    \centering
    \includegraphics[width=\columnwidth]{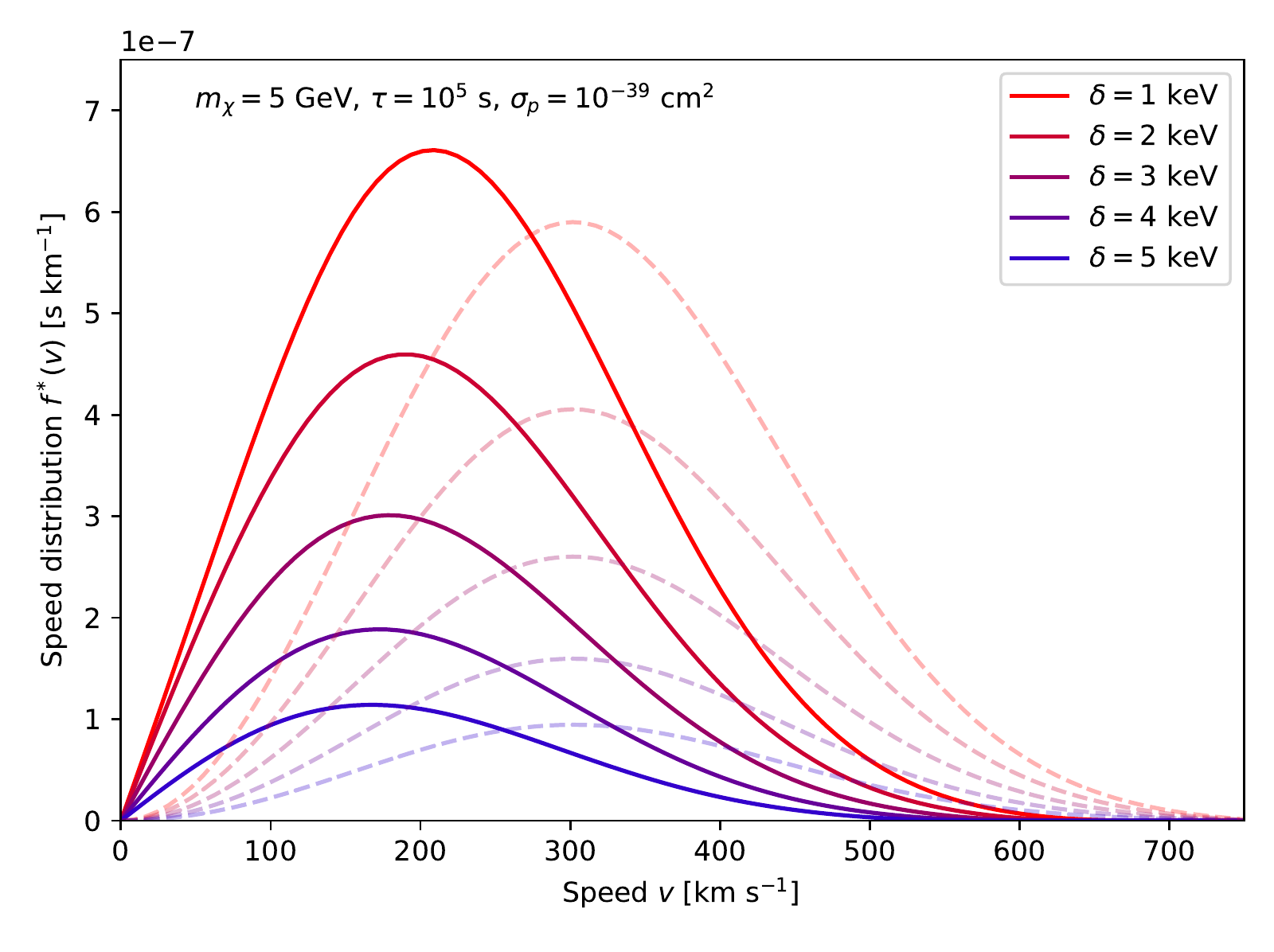}
    \includegraphics[width=\columnwidth]{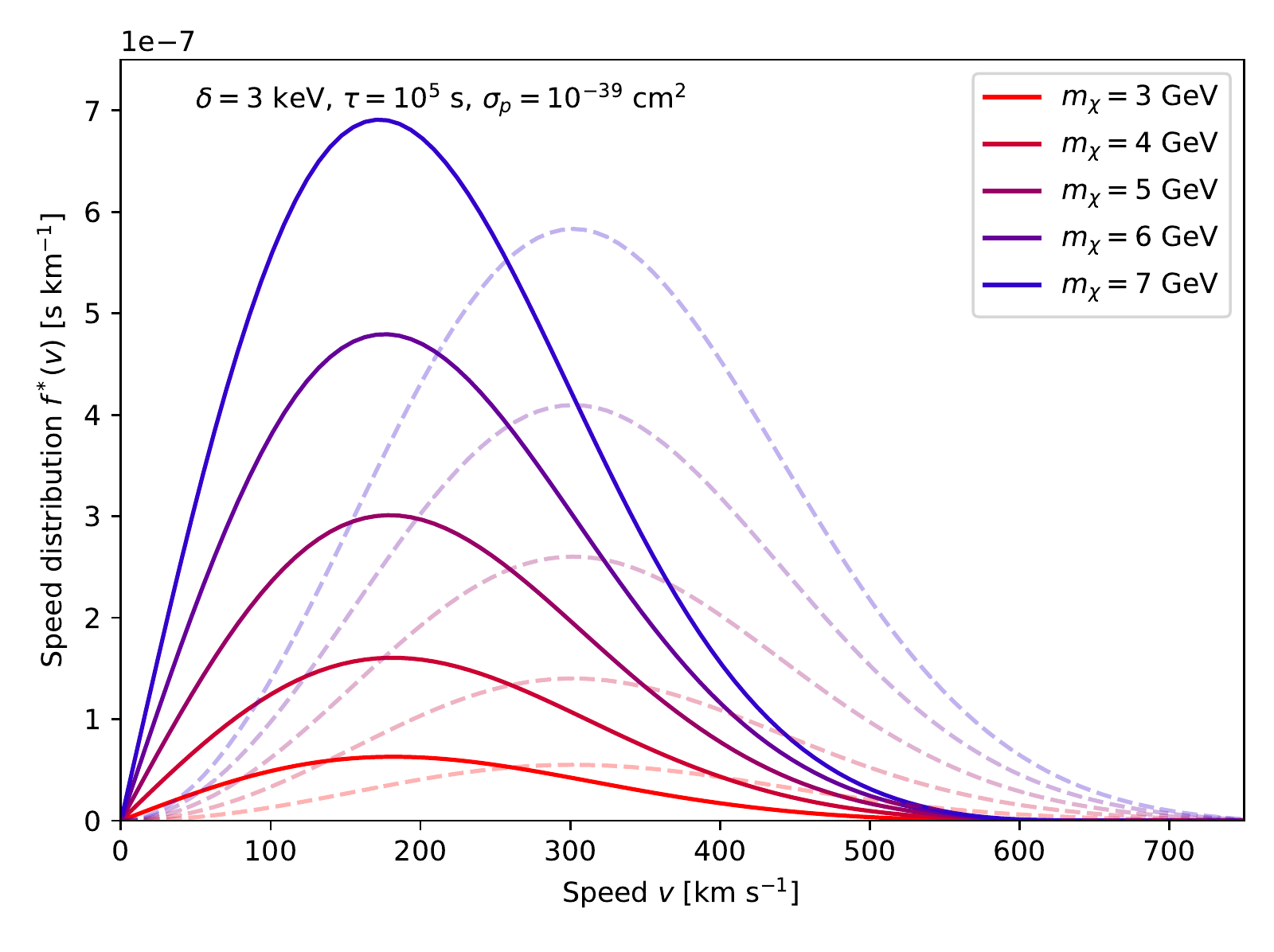}
    \includegraphics[width=\columnwidth]{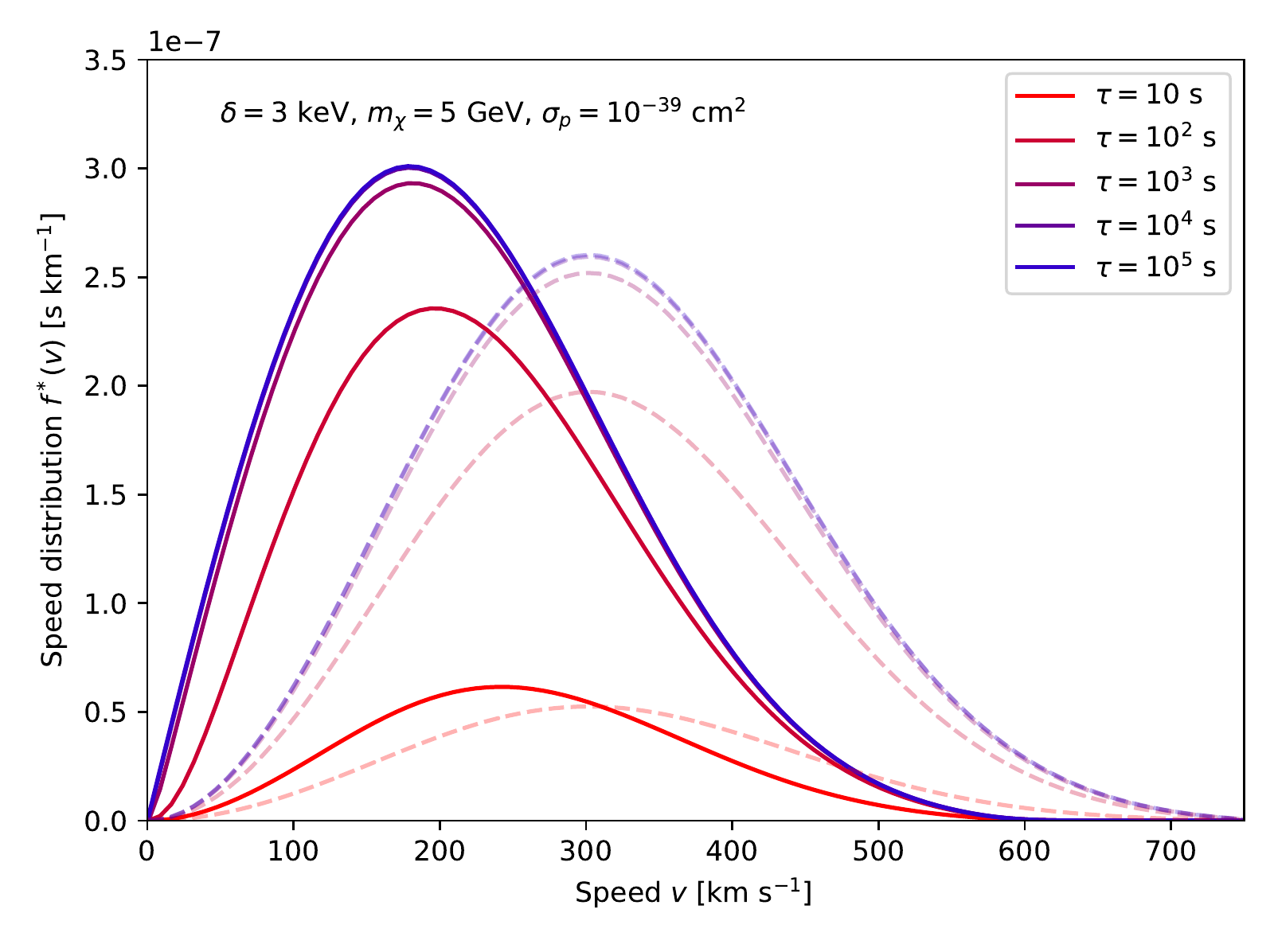}
    \caption{Speed distribution of the excited fraction for different values of $\delta$ (top), $m_\chi$ (center) and $\tau$ (bottom). 
    For the sake of comparison, the dashed lines show the speed distribution of the incoming DM particles in the Standard Halo Model, cf.\ Eq.~(\ref{eq:SHM}), normalized to match the fractional density of excited states.}
    \label{fig: speed distribution}
\end{figure}

So far, we have assumed that the Earth consists of a single nucleus species.
By summing over all nuclear targets present in the Earth's mantle and core, we obtain the final expression for the speed distribution of upscattered DM states,
\begin{align}
    f^*(v) = & \sum_{\pm,i} \int_{0}^{1}\dd\cos\theta \int_{0}^{2\pi}\dd\phi\int_{-1}^{1}\dd\cos\theta^\prime \nonumber \\ & \times \frac{\sigma_i\bar{n}_id_{\mathrm{eff},i}(\cos\theta)}{2\pi}\left| \frac{\dd \kappa_{\pm,i}^{-1}(v^\prime,\alpha)}{\dd v^\prime}\right|^{-1} \nonumber \\ & \times \left.\frac{v^{\prime 3}}{v}f_0(v^\prime,\cos\theta^\prime)P_{\pm,i}(\cos\alpha)\right|_{v^\prime = \kappa_i(v,\alpha)}\, . \label{eq:master final}
\end{align}
Examples for speed distributions are shown in Fig.~\ref{fig: speed distribution} for various values of the mass splitting~$\delta$, DM mass~$m_\chi$, and mean lifetime~$\tau$.

We would like to draw attention to one particular feature of Eq.~(\ref{eq:master final}), which is the explicit factor $v^{-1}$ that appears in the final result. This factor can be understood as the so-called ``traffic jam'' effect, i.e. an enhancement of the density as the velocity decreases~\cite{pospelov:2020}. This enhancement is particularly significant in our case, as the inelastic nature of the scattering process leads to a loss of kinetic energy and allows for very small values of the final speed $v$.

We have confirmed the validity of our analytic formalism resulting in Eq.~\eqref{eq:master final} by describing the same process using Monte Carlo simulations of DM~particles as they traverse through the Earth's bulk mass and get upscattered by terrestrial nuclei. For details on this crucial consistency check, we refer to App.~\ref{app: simulations}.
In summary, we find very good agreement between the two independent approaches.

We emphasize that throughout the derivation, we have assumed the single scattering regime.
In particular, Eq.~\eqref{eq:master final} does not account for the possibility that the upscattered particle scatters down before reaching the detector (as opposed to decaying).
Assuming that the probability to up- and downscatter are comparable, this is justified for all parameters assumed in this study, as the upscattered fraction~$\rho^*/\rho$, and thereby also the upscattering probability, will always fall well below unity.

\section{Results}
\label{sec: results}

Let us now combine the various calculations discussed in the previous two sections in order to obtain the event rate for electron recoils from terrestrial upscattering on nuclei followed by downscattering on electrons in the experiment. Before turning our attention to the XENON1T excess, we discuss a few general features of the signal.

First of all, we note that in our set-up the fraction of excited DM particles at a given experiment is proportional to the DM-proton scattering cross section $\sigma_p$. In the following, we will furthermore make the assumption that the lifetime of the excited state is large enough that spontaneous de-excitation inside the Earth is negligible, which corresponds to $\tau \gg 100 \, \mathrm{s}$. In this case the entire Earth contributes to upscattering and the fraction of excited states becomes to good approximation independent of $\tau$ (see the bottom panel of Fig.~\ref{fig: speed distribution} and Sec.~\ref{sec: discussion} for further discussion).

\begin{figure}[t!]
    \centering
    \includegraphics[width=\columnwidth]{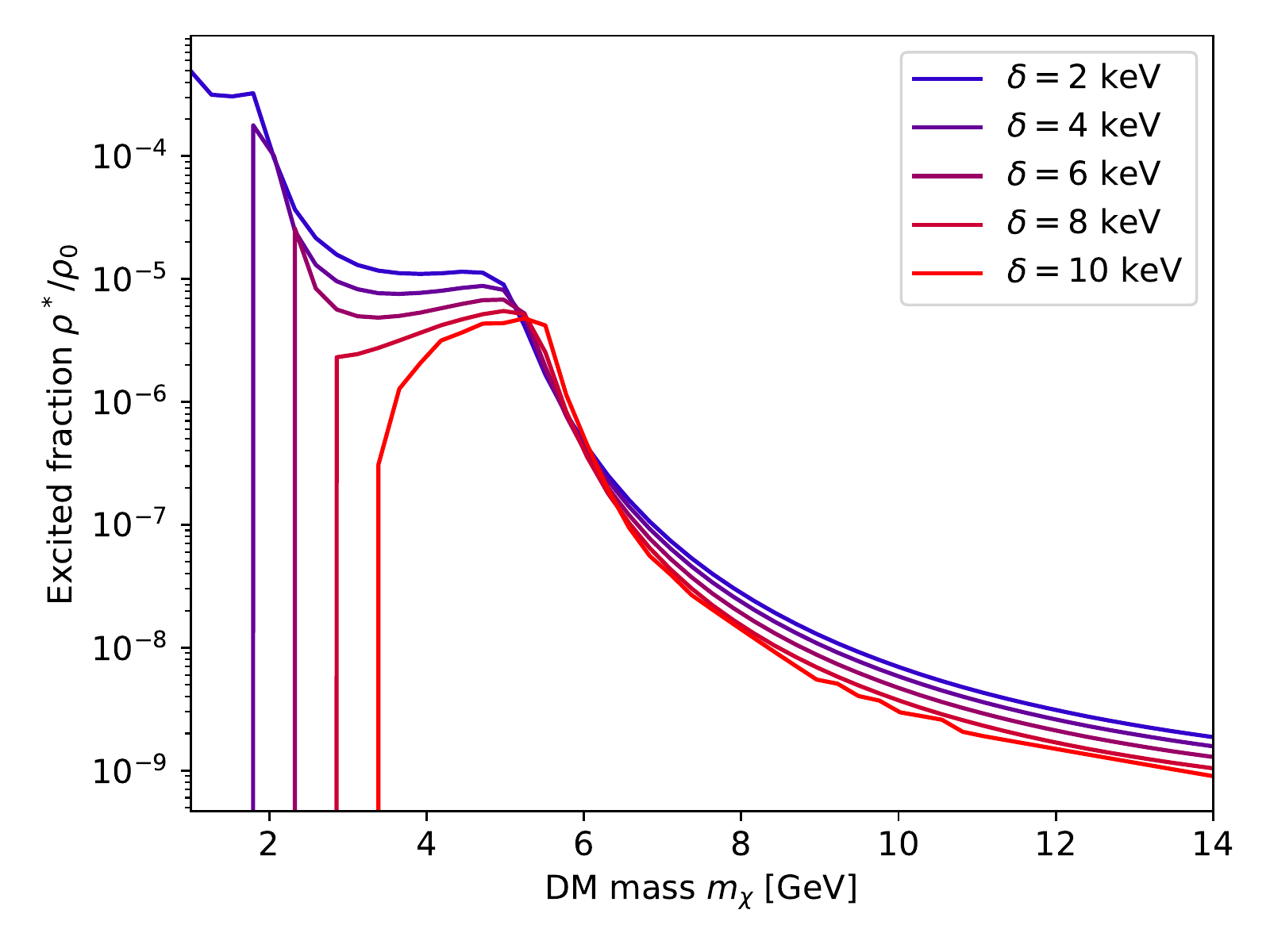}
    \caption{Upper bound at 90\% confidence level on the fraction of excited DM particles from terrestrial upscattering as a function of $m_\chi$ for different values of $\delta$.}
    \label{fig:excited_fraction}
\end{figure}

The usual searches for nuclear recoils in direct detection experiments (see Fig.~\ref{fig:inelastic_bounds}) imply an upper bound on the fraction of excited DM particles at a given experiment as a function of the DM mass and the mass splitting $\delta$. Figure~\ref{fig:excited_fraction} shows this bound at the (average) position of the XENON1T experiment in the Gran Sasso laboratory ($\cos \gamma = -0.5$) for different values of $\delta$. We find that as long as the DM particle is light enough for upscattering to be unconstrained by the XENON1T experiment, the fraction of excited states can be in the range $10^{-6} \text{--} 10^{-4}$ for $\delta$ of the order of a few keV. For larger values of $\delta$ only heavier DM particles can experience upscattering and the overall upscattering probability is suppressed. 

The probability of downscattering on the electrons in a given experiment is then proportional to $\sigma_e$. An electron recoil signal therefore probes the effective cross section $\sigma_\text{eff} \equiv \sqrt{\sigma_p \sigma_e}$ as a function of the DM mass $m_\chi$ and $\delta$, which determine the probability for upscattering. By fitting both the magnitude and shape of a given signal we can then infer all three parameters. 

\subsection{The XENON1T excess}
\label{ss: xenon1t}

Let us now turn our attention to the XENON1T excess,  which is located at electron recoil energies of about $3 \, \mathrm{keV}$. It has been shown that both a mono-energetic recoil spectrum (as expected for example from the absorption of a bosonic DM particle~\cite{Takahashi:2020bpq,Alonso-Alvarez:2020cdv,Athron:2020maw}) and the slightly broader spectrum expected for exothermic DM-electron scattering can give a good fit to the excess \cite{Bloch:2020uzh}.  The reason is that the predicted recoil spectrum is broadened by the energy resolution of the detector, which is given by
\begin{equation}
 \sigma(E) = a \cdot \sqrt{E} + b \cdot E
\end{equation}
with $a = 0.31 \sqrt{\mathrm{keV}}$ and $b = 0.0037$~\cite{XENON:2020rca}. The event rate in a given bin $[E_i, E_{i+1}]$ is therefore found to be
\begin{align}
 R_i = \int_{E_i}^{E_{i+1}} & \mathrm{d}E_\mathrm{er} \frac{\mathrm{d}R}{\mathrm{d}E_\mathrm{er}}(E_\mathrm{er}) \xi(E_\mathrm{er}) \nonumber \\ & \times \frac{1}{2}\left[\text{erf}\left(\tfrac{E_{i+1} - E_\mathrm{er}}{\sqrt{2}\sigma(E_\mathrm{er})}\right) - \text{erf}\left(\tfrac{E_{i} - E_\mathrm{er}}{\sqrt{2}\sigma(E_\mathrm{er})}\right)\right] \; ,
\end{align}
where $\xi(E_\mathrm{er})$ denotes the detector efficiency.

\begin{figure*}[t!]
    \centering
    \includegraphics[width=\columnwidth]{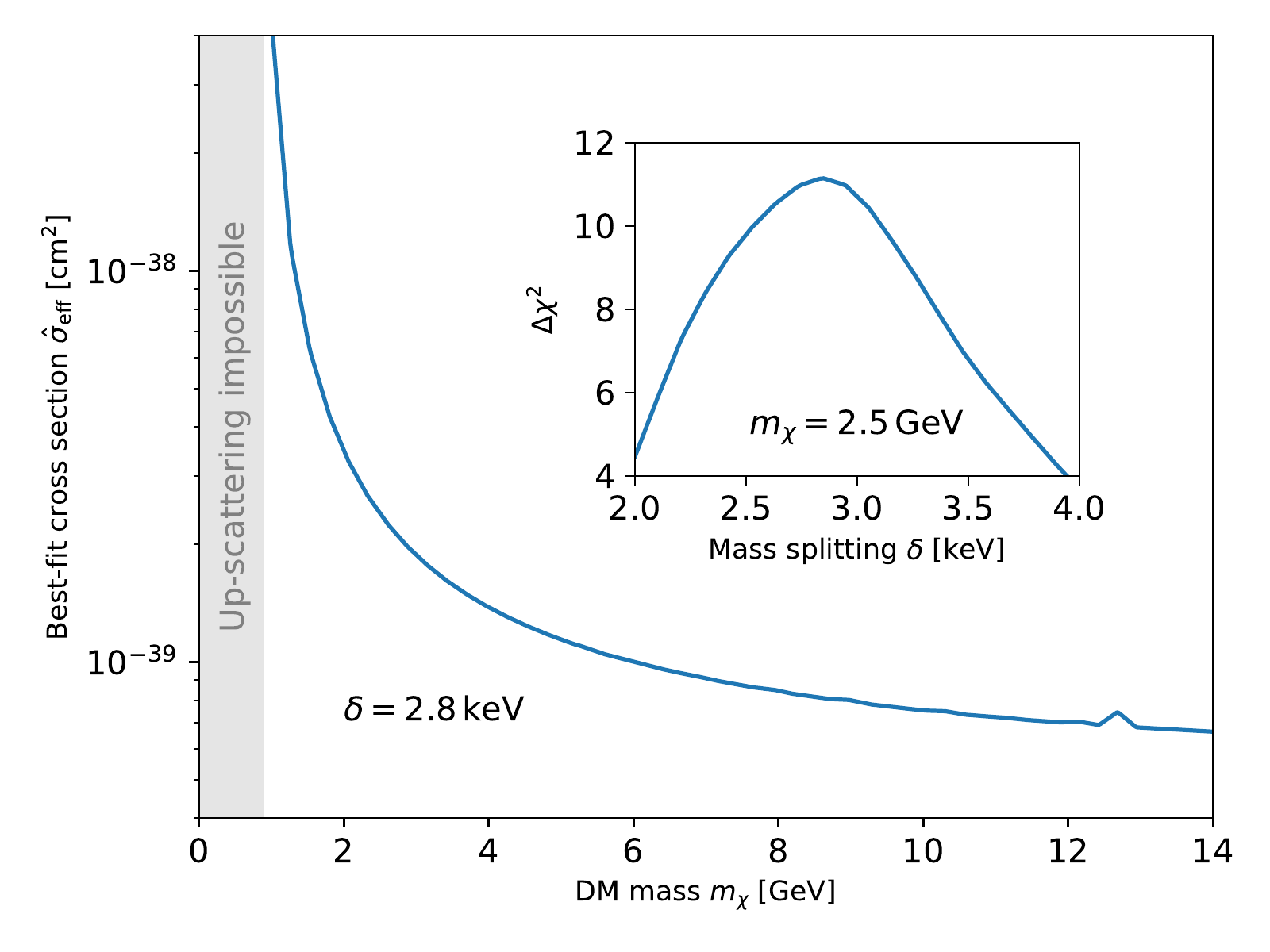}\hfill
    \includegraphics[width=\columnwidth]{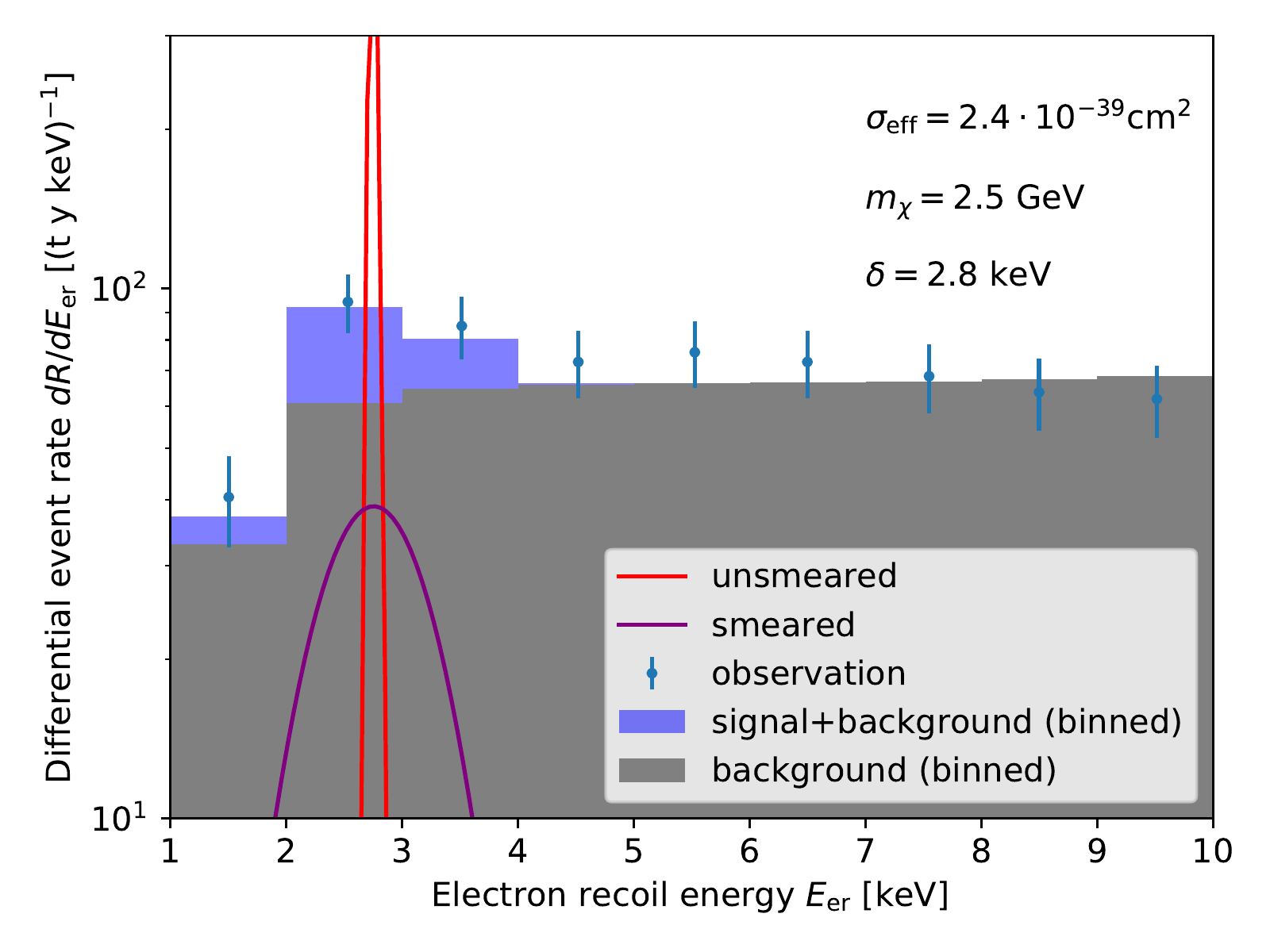}
    \caption{\textit{Left:} Value of the best-fit effective cross section $\hat{\sigma}_\text{eff}$ as a function of $m_\chi$ for $\delta = 2.8 \, \mathrm{keV}$. The value of $\delta$ is chosen to maximize the signal preference $\Delta \chi^2$ (see inset). \textit{Right:} Comparison of the background expectation and the predicted signal for $m_\chi = 2.5\,\mathrm{GeV}$, $\delta = 2.8\,\mathrm{keV}$ and $\sigma_\text{eff} = 2.4 \cdot 10^{-39} \, \mathrm{cm^2}$ with the event rates observed by the XENON1T experiment.}
    \label{fig:preferred_regions}
\end{figure*}

To determine the parameter regions of our model compatible with the XENON1T excess, we consider a $\chi^2$ test statistic:
\begin{equation}
    \chi^2 = \sum_{i=1}^4 \frac{(R_{i,\text{pred}} - R_{i,\text{obs}})^2}{\Delta_i^2} \; ,
\end{equation}
where $R_{i,\text{pred}}$ is the sum of the DM signal and the expected background in each bin (as given in Ref.~\cite{XENON:2020rca}) and the observed event rates as well as their uncertainties $\Delta_i$ from Poisson noise are also taken from Ref.~\cite{XENON:2020rca}. For the purpose of parameter estimation it is sufficient to include the first four bins, beyond which the DM signal is expected to vanish (for $\delta \sim 3 \, \mathrm{keV}$). For each combination of $m_\chi$ and $\delta$ we can then find the best-fit value of the effective cross section, $\hat{\sigma}_\text{eff}$, and calculate the preference over the background-only hypothesis via $\Delta \chi^2 \equiv \chi^2(\sigma_\text{eff} = 0) - \chi^2(\hat{\sigma}_\text{eff})$.

The left panel of Fig.~\ref{fig:preferred_regions} shows the best-fit value of the effective cross section as a function of $m_\chi$ for $\delta = 2.8\,\mathrm{keV}$, which is found to maximize the value of $\Delta \chi^2$ (see inset). We find that terrestrial upscattering followed by exothermic downscattering on electrons can fit the XENON1T excess across the whole range of DM masses that we consider down to the kinematic limit where upscattering becomes forbidden. As an example we show in the right panel of Fig.~\ref{fig:preferred_regions} the parameter point $m_\chi = 2.5\,\mathrm{GeV}$ and $\delta = 2.8\,\mathrm{keV}$, for which $\hat{\sigma}_\text{eff} = 2.4 \cdot 10^{-39} \, \mathrm{cm^2}$ and $\Delta \chi^2 = 11.1$. We will use this parameter point as a benchmark value in the following.

\subsection{Daily modulations}
\label{ss: daily modulations}

Should the XENON1T excess be confirmed by future direct detection experiments, the central question will be how to disentangle the various possible explanations. A key strategy to answering this question is to study the time-dependence of the signal. Indeed it is well known that many direct detection signals exhibit an annual modulation resulting from the motion of the Earth around the Sun. In the present case, there turns out to be an even more promising signature, namely a daily modulation resulting from the rotation of the Earth.

The origin of the daily modulation lies in the anisotropy of the DM velocity distribution arriving on Earth. Indeed, the motion of the Sun (and hence the Earth) through the Milky Way leads to a so-called ``DM wind'' from the direction of Cygnus, meaning that DM particles arriving from this direction are on average faster than those from other directions. Due to the Earth's rotation, the orientation of the DM wind in the laboratory frame changes over the course of a day, leading to modulating signals in any experiment sensitive to the direction of the incoming DM particles~\cite{Bozorgnia:2011tk,Hochberg:2016ntt,Griffin:2018bjn,Coskuner:2019odd,Geilhufe:2019ndy,Blanco:2021hlm}. In our case, the experiment itself is not sensitive to the direction of the incoming DM particles, but the incoming flux of excited states varies over the course of the day (see also Refs.~\cite{Kouvaris:2015xga,Kavanagh:2016pyr,Eby:2019mgs}).

If the rotation axis of the Earth points in the $z$ direction, we can parametrize the detector position and the direction of the DM wind via
\begin{align}
    \frac{\mathbf{r}_\text{det}}{r_\mathrm{E}} & = \left(\cos \theta_l \cos \omega t, \cos \theta_l \sin \omega t, \sin \theta_l\right) \, ,\\
    \frac{\mathbf{v}_\mathrm{E}}{v_\mathrm{E}} & = \left(\sin \beta, 0, \cos \beta\right) \; ,
\end{align}
where $\omega = 2\pi \, \mathrm{day^{-1}}$, $\theta_l$ denotes the latitude of the detector and $\beta = 42.8^\circ$ is the angle between the Earth's rotation axis and the DM wind. Here we have chosen the time coordinate in such a way that at $t = 0$ the detector position lies in the plane of the DM wind and the Earth's rotation axis.

In the laboratory frame, the direction of the DM wind is given by the angle
\begin{equation}
    \cos \gamma \equiv - \frac{\mathbf{v}_\mathrm{E} \cdot \mathbf{r}_\text{det}}{v_\mathrm{E} r_\mathrm{E}} = -\left(\cos \theta_l \cos \omega t \sin \beta + \sin \theta_l \cos \beta\right) \; .
\end{equation}
The latitude of the XENON1T experiment is $\theta_l = 42.5^\circ$, such that $\gamma$ varies between about $-1$ and 0. In other words, at $t = 0$ the DM wind comes almost directly from above, while at $t = 12 \, \mathrm{hours}$ it points almost exactly sideways. This means that at $t = 0$ the DM particles would have to upscatter at a large angle ($> 90^\circ$) in order to contribute to the downscattering signal, which suppresses the contribution from slow DM particles (see Fig.~\ref{fig:pscat}). As a result, we expect the event rate to exhibit a minimum at $t = 0$ and a maximum at $t = 12 \, \mathrm{hours}$, where smaller upscattering angles are sufficient.

\begin{figure}[t!]
    \centering
    \includegraphics[width=\columnwidth]{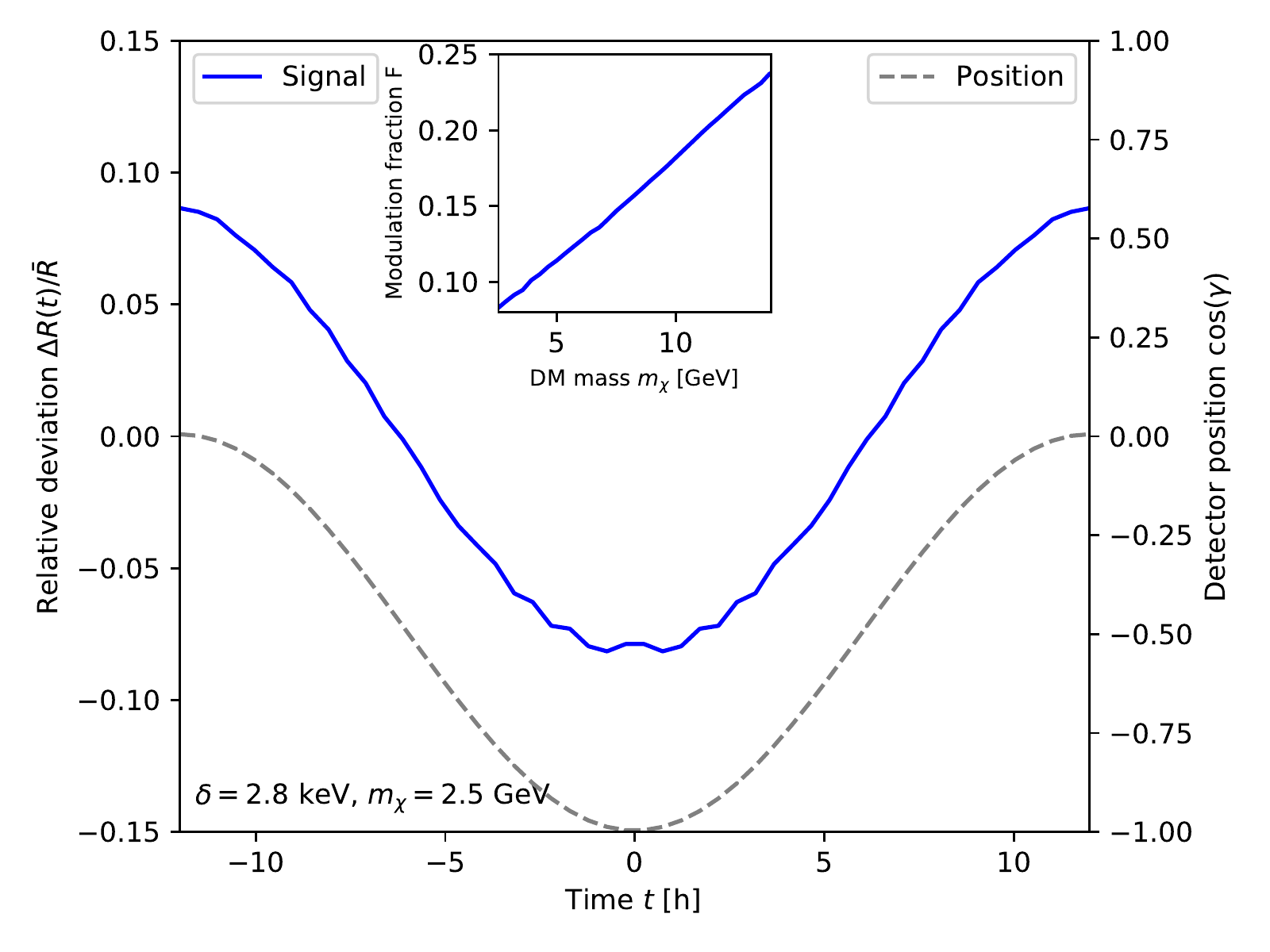}
    \caption{Time dependence of the predicted signal for $m_\chi = 2.5 \, \mathrm{GeV}$ and $\delta = 2.8 \, \mathrm{keV}$ (solid blue line, left $y$-axis) and of the detector position (dashed gray line, right $y$-axis). As shown in the inset, the modulation fraction increases with increasing DM mass (keeping $\delta$ fixed).}
    \label{fig:modulation_example}
\end{figure}

This expectation is confirmed in Fig.~\ref{fig:modulation_example}, which shows the time dependence of $\cos \gamma$ (dashed gray line, right $y$-axis) and of the signal for $\delta = 2.8 \, \mathrm{keV}$ and $m_\chi = 2.5 \, \mathrm{GeV}$ (blue line, left $y$-axis). To characterize the modulation we define the minimum and maximum rate $R_\text{min,max}$, the average rate $\bar{R} = (R_\text{min} + R_\text{max})/2$, the modulation $\Delta R(t) = R(t) - \bar{R}$ and the modulation fraction $F = (R_\text{max} - R_\text{min}) / (R_\text{max} + R_\text{min})$. In the present case, the modulation fraction is approximately $8\%$ for $m_\chi = 2.5 \, \mathrm{keV}$, which is too small to be observable with the number of electron recoil events seen by XENON1T but may be observable in future experiments.\footnote{For concreteness, Ref.~\cite{Geilhufe:2019ndy} proposes a simple hypothesis test, for which the significance of modulation is approximately $F \sqrt{N_s}$ standard deviations, where $N_s$ is the total number of signal events and backgrounds are assumed to be negligible. Hence, to observe daily modulation with $3\sigma$ significance would require $N_s \approx 900$ ($N_s \approx 225$) for $F = 10\%$ ($F = 20\%$), corresponding to about a factor of 20 (factor of 5) more statistics than XENON1T.}

We note that the modulation fraction grows rapidly with increasing DM mass, as a result of the upscattering process (in the Earth frame) being increasingly peaked in the forward direction. For example, for $m_\chi = 14 \, \mathrm{GeV}$ the daily modulation is already at the level of 24\%. Hence, a precise determination of the modulation fraction may not only enable us to distinguish between the mechanism proposed here and alternative explanations of the XENON1T excess, but also to determine the DM mass, which has almost no effect on the energy dependence of the electron recoil spectrum. 

\section{Discussion}
\label{sec: discussion}

In the previous section we have established exothermic downscattering on electrons as a viable explanation of the XENON1T excess and inelastic upscattering in the Earth as an interesting possibility to create excited states with a time-dependent density. So far, we have phrased our analysis in terms of the derived parameters $\tau$ and $\sigma_\text{eff}$. In this section we discuss how these quantities may be obtained from a more fundamental theory and what corresponding model-building challenges to expect.

\subsection{Decays of the excited states}

As mentioned above, for our results we have assumed that $\tau \gg 100 \, \mathrm{s}$, such that the entire Earth contributes to upscattering. In principle, also smaller values of $\tau$ could be considered, but doing so would imply larger cross sections in order to achieve the same signal strength. In terms of upper bounds on $\tau$ our calculations assume that there are no other sources of excited states apart from terrestrial upscattering. In particular, we do not consider upscattering in the Sun, which has been studied previously in Ref.~\cite{Baryakhtar:2020rwy}. This is a good approximation as long as $\tau \lesssim 10^5 \, \mathrm{s}$, such that any excited states produced in the Sun would decay before reaching the Earth. For larger values of $\tau$ the Sun may contribute significantly to the flux of excited DM particles on Earth, because its core temperature is large enough to efficiently excite DM particles through electron scattering.

So far we have not specified the mechanism through which the excited state decays. Within the SM the two possible decay modes are $\chi^\ast \to \chi \gamma$ and $\chi^\ast \to \chi \nu \bar{\nu}$. However, for the latter process the available phase space is so small that it will be very challenging to achieve sufficiently small lifetimes, $\tau \lesssim 10^5 \, \mathrm{s}$. In the former case, the excited state can decay for example via an inelastic magnetic dipole moment~\cite{Chang:2010en}
\begin{equation}
 \mathcal{L} = \mu_\chi F^{\mu\nu} \bar{\chi} \sigma_{\mu\nu} \chi^\ast + \text{h.c.}
\end{equation}
The corresponding lifetime is given by $\tau = 4\pi (\mu_\chi^2 \delta^3)$, which is in the desired range for $\mu_\chi \sim 10^{-8} \mu_B$ with $\mu_B$ being the Bohr magneton.\footnote{We note that for values of $\mu_\chi$ in this range there may be an additional contribution to upscattering via long-range interactions, which is not covered by the formalism presented in this work.}

\begin{figure}[t!]
    \centering
    \includegraphics[width=\columnwidth,clip]{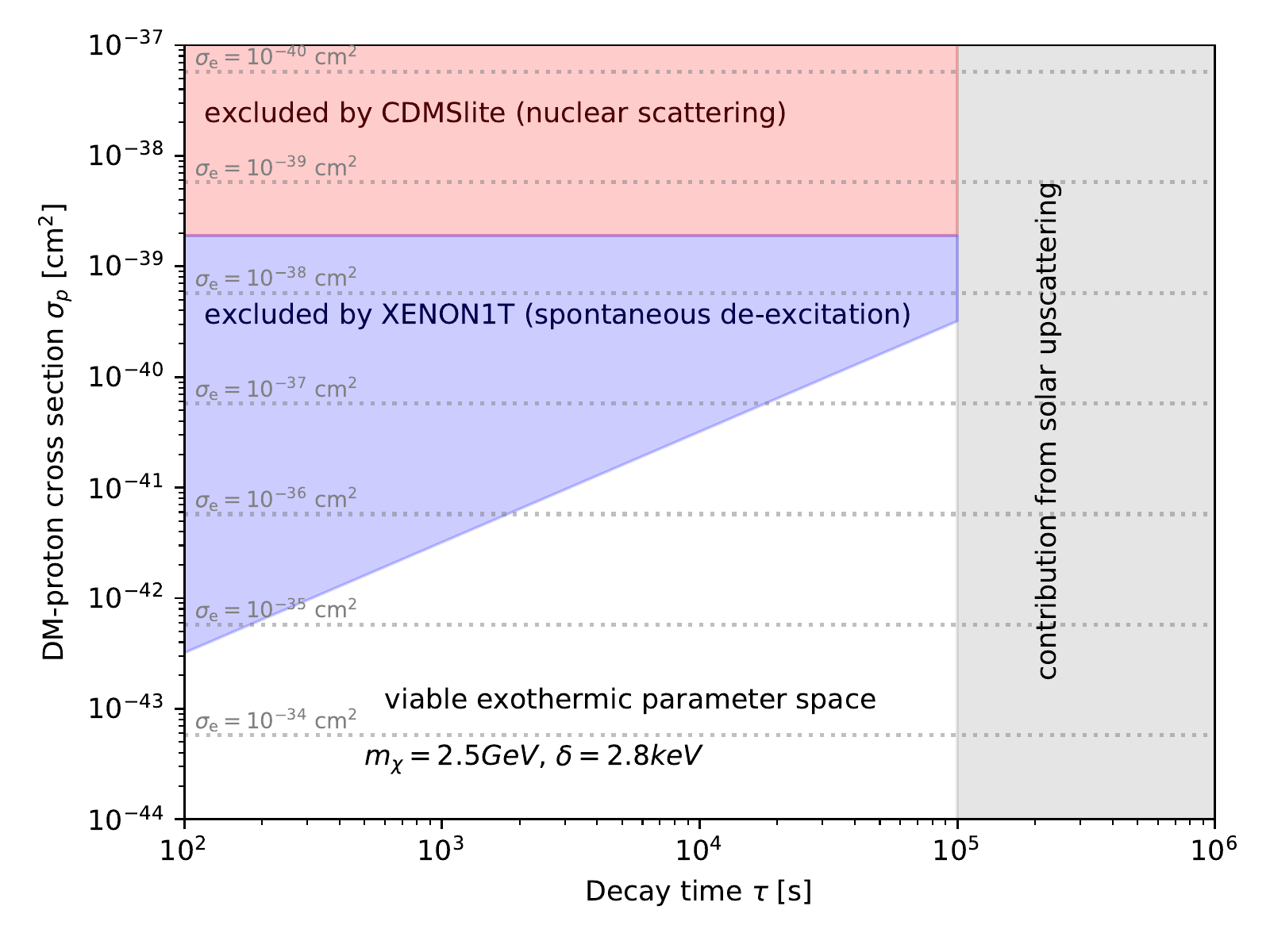}
    \caption{Available parameter space for terrestrial upscattering followed by exothermic downscattering on electrons in the $\tau$--$\sigma_p$ parameter plane under the assumption that the lifetime is determined by the decay $\chi^\ast \to \chi + \gamma$. For $m_\chi = 2.5 \, \mathrm{GeV}$ and $\delta = 2.8 \, \mathrm{keV}$ we find that the conventional bound on nuclear scattering (red shaded region) is significantly weaker than the one on spontaneous de-excitation (blue shaded region). For $\tau > 10^5 \, \mathrm{s}$ the contribution from solar upscattering can no longer be neglected.}
    \label{fig:luminous_constraint}
\end{figure}

However, if the excited states decay into the ground state and a mono-energetic photon, there are two additional constraints that need to be considered. First of all, spontaneous de-excitation in the detector followed by the absorption of the resulting photon constitutes an additional source of electron recoils with a rate given by
\begin{equation}
R = \frac{\rho^\ast V_\text{det}}{\tau m_\chi} \; ,
\end{equation}
where $V_\text{det}$ is the active volume of the detector.
Requiring that the resulting signal does not violate experimental constraints leads to an upper bound on $\sigma_p$ as a function of $\tau$, which can be significantly stronger than the from Fig.~\ref{fig:inelastic_bounds}.\footnote{We note in passing that it is also possible to fit the XENON1T excess with terrestrial upscattering followed by spontaneous de-excitation in the detector, as in models of luminous DM~\cite{Bell:2020bes}. For $\tau \gg 100 \, \mathrm{s}$ we find that the daily modulation is very similar to the case of exothermic downscattering, making it very difficult to distinguish between these two possibilities experimentally.} This is illustrated in Fig.~\ref{fig:luminous_constraint}, which shows the available parameter space as a function of $\sigma_p$ and $\tau$ for $m_\chi = 2.5 \, \mathrm{GeV}$ under the assumption that the excited states decay via $\chi^\ast \to \chi \gamma$. The dashed gray lines show the values of the downscattering cross-section, $\sigma_e$ required to fit the excess for $\delta=2.8\,\mathrm{keV}$.

A second set of constraints originate from searches for x-ray lines, which are sensitive to the upscattering and subsequent de-excitation of DM particles in the DM halo of the Milky Way or of another galaxy or galaxy cluster~\cite{Finkbeiner:2014sja,DEramo:2016gqz}. However, the expected magnitude of this signal depends on the detailed distribution of SM particles in the DM halo as well as on the cross section of inelastic DM self-scattering. A calculation of these model-dependent constraints is beyond the scope of the present work.

To conclude this discussion, we note that it is also possible for the excited state to decay invisibly, provided there exists another light boson with sub-keV mass beyond the SM. Indeed, such a light particle may be directly related to the mechanism that generates the DM mass splitting (see Ref.~\cite{Duerr:2020muu} for a similar discussion in the context of accelerator searches). We leave a more detailed analysis of such models as well as their cosmological viability to future work.

\subsection{Cross section hierarchy}

So far we have only considered the effective cross section $\sigma_\text{eff} = \sqrt{\sigma_e \sigma_p}$, which is required to be of the order of $10^{-39} \, \mathrm{cm^2}$ in order to fit the XENON1T excess (see Fig.~\ref{fig:preferred_regions}). However, for the range of $m_\chi$ and $\delta$ that we are interested in, experimental bounds on DM-nucleus scattering require $\sigma_p \lesssim 10^{-39} \, \mathrm{cm^2}$, which implies the hierarchy $\sigma_e > \sigma_p$. While this seems like an innocuous requirement at first sight, it actually turns out to be quite challenging from the model-building point of view to realize such a hierarchy. The reason is that, assuming the same process mediates both DM-nucleon and DM-electron scattering, one would expect
\begin{equation}
 \frac{\sigma_e}{\sigma_p} = \frac{\mu_e^2}{\mu_p^2} \approx \frac{m_e^2}{m_p^2} < 10^{-6} \; . 
\end{equation}
Hence, to achieve $\sigma_e > \sigma_p$ it is necessary for DM to couple much more strongly to electrons than to nucleons.

A conceivable solution might be to consider a leptophilic vector mediator with no tree-level couplings to quarks, but even the couplings induced at the one-loop level~\cite{DEramo:2016gos,DEramo:2017zqw} would spoil the desired hierarchy. For a leptophilic scalar mediator, on the other hand, DM-quark interactions only arise at the two-loop level~\cite{Kopp:2009et} making it possible to achieve $\sigma_e > \sigma_p$. A more detailed investigation of the related model-building challenges is beyond the scope of the present work.

We emphasize that the problem is exacerbated for heavier DM masses. Already for $m_\chi = 10 \, \mathrm{GeV}$ the XENON1T results for DM-nucleus scattering imply $\sigma_p \lesssim 10^{-44} \, \mathrm{cm^2}$, which would correspond to an implausibly large DM-electron cross section of $\sigma_e \sim 10^{-34} \, \mathrm{cm^2}$. Our set-up therefore clearly favors DM masses close to the threshold for upscattering ($m_\chi \approx 2 \text{--} 3 \, \mathrm{GeV})$. Intriguingly, experimental constraints on $\sigma_p$ in this mass range are expected to improve considerably in coming years (see e.g.\ Ref.~\cite{SuperCDMS:2016wui}), such that it may be possible to observe both nuclear and electron recoils originating from inelastic upscattering and downscattering, respectively. 

\section{Conclusions}
\label{sec: conclusions}

Searches for electron recoils originating from the scattering of DM particles are an exciting avenue to extend the sensitivity of direct detection experiments to smaller DM masses. Great advances have been made in recent years both in terms of addressing the technological challenges and in terms of improving the theoretical framework needed for accurate signal predictions. At the same time a wealth of models of light DM have been proposed that allow for a consistent cosmological history and make testable predictions for near-future experiments.

The XENON1T excess in electron recoil events offers an ideal test case to apply these recent developments and identify the viable explanations as well as the required calculational methods. One particularly interesting example is the exothermic downscattering of a sub-dominant population of excited DM states, for which the ionization probability is greatly enhanced and an excellent fit to the observed shape of the excess is obtained. However, the origin of these excited states is often left unspecified and may be difficult to calculate in detail.

In this work we have addressed this question by studying a specific mechanism for producing a small fraction of excited states, namely inelastic upscattering on nuclei in the Earth. For this purpose, we have extended previous analyses of terrestrial scattering to include the kinematics of inelastic collisions, the decay probability of the excited state and the ``traffic jam'' effect, which enhances the density of low-velocity particles. We have validated our analytical results using explicit Monte Carlo simulations of the scattering processes.

As the central result, we obtain the density $\rho^\ast$ and velocity distribution $f^\ast(\mathbf{v})$ of excited states as a function of the model parameters and the position of the detector on the Earth's surface. Since this position changes (relative to the DM wind) over the course of each day, the resulting flux of excited states exhibits a characteristic daily modulation, that may be used to identify the origin of the signal and determine the DM mass.

We find that the XENON1T excess can be fitted for a wide range of DM masses, provided the mass splitting $\delta$ is comparable to the typical electron recoil energy. However, the requirement of a sufficiently large fraction of excited states points towards DM masses in the range 1--5 GeV. In this mass range, the modulation fraction is found to be of the order of 10\%, which is too small to be detectable with current data but a promising target for future measurements.

Given experimental upper bounds on the DM-proton scattering cross section (in particular if the excited state decays under the emission of a photon), the required DM-electron cross sections are quite large. This finding points towards rather specific underlying models, in which the DM particle couples much more strongly to leptons than to baryons. It will be exciting to explore these implications further, should the XENON1T excess be confirmed by the next generation of experiments.

\acknowledgments

We thank Riccardo Catena and Jan Conrad for initiating this project, Tongyan Lin and Katelin Schutz for discussions, Bradley J. Kavanagh for valuable feedback on this manuscript, and Diego Redigolo for clarifications regarding Ref.~\cite{Bloch:2020uzh}. This  work  is  funded  by  the  Deutsche Forschungsgemeinschaft (DFG) through the Emmy Noether Grant No.\ KA 4662/1-1. TE was supported by the Knut and Alice Wallenberg Foundation (PI, Jan Conrad). SH acknowledges the support of the Natural Sciences and Engineering Research Council of Canada (NSERC), SAPIN-2021-00034. TE thanks the Theoretical Subatomic Physics group at Chalmers University of Technology for its hospitality.

\appendix

\section{Verification of analytic formalism via Monte Carlo simulations}
\label{app: simulations}

To validate our analytical formalism, we can use Monte Carlo (MC) simulations as an independent approach to this problem. These simulations are powerful tools that can be applied far beyond the single scattering approximation \cite{Emken:2017qmp,Emken:2021lgc} but for our purposes it is sufficient to remain within this assumption to test the analytical results. The concept is fairly simple: First, we draw initial states from the DM distribution in the halo. In doing so, we note a subtlety pointed out in \cite{Emken:2017qmp} which adds an intermediate step between the initialization of the particle and the scattering event: To ensure spatial homogeneity, we randomly position the particle on a disk of radius $r_E$ perpendicular to the initial velocity. Assuming the latter to point in the $z$ direction, we can therefore write the initial position as
\begin{equation}
\mathbf{r} = r_\mathrm{ini} \mathbf{\hat{e}}_z + \sqrt{\xi} r_\mathrm{E} \left(\cos(\phi)\mathbf{\hat{e}}_x+\sin(\phi)\mathbf{\hat{e}}_y\right) \; ,
\end{equation}
where $r_\mathrm{ini} > r_\mathrm{E}$ is an arbitrary distance and the parameters $\xi$ and $\phi$ are uniformly distributed on $[0,1]$ and $[0,2\pi]$, respectively.

The next step is to calculate the particle's path through the Earth according to its initial position and direction in order to account for the density model of the Earth, c.f.\ Eq.~(\ref{eq:decinc}). For a given path we randomly determine the distance that the particle travels before scattering. This distance is distributed as
\begin{equation}
P(d) = \Lambda\exp\left(-\frac{d}{\Lambda}\right) \; ,
\label{eq:freepath}
\end{equation}
with the mean free path $\Lambda=\sum_i (n_i \sigma_i)^{-1}$. Here, $n_i$ and $\sigma_i$ denote the element specific number density and interaction cross section, respectively.\footnote{Note that in practice, we will consider each element separately and only once we have derived the speed distribution, we sum up all contributions.} We consider two scenarios:
\begin{enumerate}
 \item The particle passes only through the mantle. In this case, we draw a free path from the distribution in Eq.~\ref{eq:freepath} and check if the point of scattering lies within the Earth, in which case we can continue with the simulation.
 \item If the particle passes through the core, we check each segment of the path (``mantle-core-mantle'') for a scattering event.
\end{enumerate}

Once we have established the point of scattering, we can calculate the outgoing velocity of the particle via the kinematic relation 
\begin{equation}
\mathbf{v}=\frac{\sqrt{m_a^2 v^{\prime 2}-\frac{2\delta}{m_\chi}m_a(m_a+m_\chi)}\mathbf{n}+m_\chi \mathbf{v^\prime}}{m_a+m_\chi}
\; .
\end{equation}
The isotropy of scattering in the center of mass frame is represented by the isotropically distributed unit vector~$\mathbf{n}$. Combining the point of exit with the final velocity, we have successfully completed one simulated event. If we are interested in the inclusion of decays as well we can follow a similar approach as for the free path. We can translate the decay time $\tau$ into an effective mean free path $\Lambda=v\tau$, taking into account the speed of the particle. Then, we determine the point where the particle decays and select only those particles which survive until they leave the Earth.

In the following, we will outline how to extract the speed distribution from the MC data. Our procedure closely follows Ref.~\cite{Emken:2017qmp}, deviating only at one major step. The basic idea is that the remaining symmetry of the initial velocity distribution around the DM wind allows us to sort the simulated particles into bins of $\cos \gamma$ according to where they leave the Earth after scattering. For a given detector position, the particles in the corresponding bin can be used to infer the reduced flux $\Phi/v$ through the surface, where
\begin{equation}
 \Phi(\mathbf{v}) = n^\ast f(\mathbf{v}) v \cos \theta \label{eq:flux}
\end{equation}
with $\theta$ being the angle between the particle trajectory and the surface and $n^\ast$ being the number density of excited states. The reason that the simulation yields the reduced flux rather than the flux itself is that all simulated trajectories are counted equally, irrespective of how long it takes the particle to reach the detector.

Since we want to infer the velocity distribution $f(\mathbf{v})$ rather than the flux, we need to reweight each event according to Eq.~(\ref{eq:flux}). In this context it is essential to account for the ``DM traffic jam'' effect~\cite{pospelov:2020}, i.e.\ the fact that a loss of velocity leads to an increase in density proportional to $v' / v$.\footnote{This effect is negligible for elastic scattering of light DM and was therefore not considered in Ref.~\cite{Emken:2017qmp}. In our case, however, $v$ can be very different from $v'$ and the effect leads to a significant enhancement of the contribution from slow particles.} Hence, the appropriate weighting factor for each particle is given by
\begin{equation}
 w_i = \frac{v_i'}{v_i \cos \theta_{i}} \; .
\end{equation}
Using these weights, the shape of the speed distribution is correctly reproduced.

\begin{figure}[t!]
    \centering
    \includegraphics[width=\columnwidth]{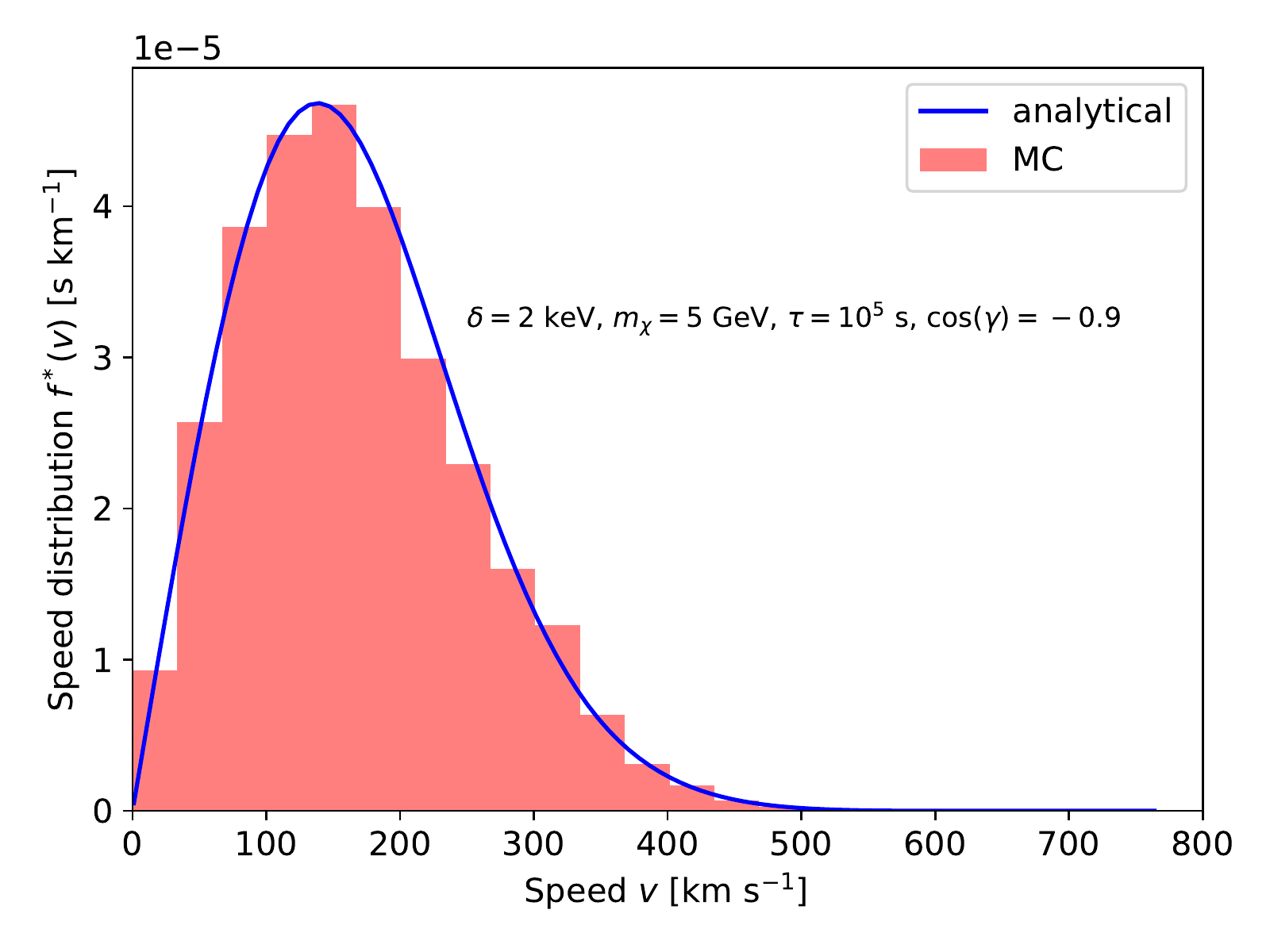}
    \caption{Comparison of the speed distribution obtained from the analytical approach (solid line) and the Monte Carlo simulation (histogram).}
    \label{fig:mcvsan}
\end{figure}

To obtain also the correct normalization, we need to determine the fraction of excited DM particles. For this purpose we can compare the MC data set to the case of an ``empty Earth'', i.e.\ the absence of scattering events, for which the number of outgoing particles in a given bin of $\cos \gamma$ can be calculated analytically. Including this normalization factor, we obtain the histogram shown in Fig.~\ref{fig:mcvsan}. We find excellent agreement between the analytical approach and the MC simulation for a wide range of parameter choices, validating the calculations in Sec.~\ref{sec: upscattering formalism} for the single scattering regime. Since the MC simulations are significantly more time-consuming, we use the analytical calculation for our main results.

\newpage 

\bibliography{references_paper}

\begin{thebibliography}{106}%
\makeatletter
\providecommand \@ifxundefined [1]{%
 \@ifx{#1\undefined}
}%
\providecommand \@ifnum [1]{%
 \ifnum #1\expandafter \@firstoftwo
 \else \expandafter \@secondoftwo
 \fi
}%
\providecommand \@ifx [1]{%
 \ifx #1\expandafter \@firstoftwo
 \else \expandafter \@secondoftwo
 \fi
}%
\providecommand \natexlab [1]{#1}%
\providecommand \enquote  [1]{``#1''}%
\providecommand \bibnamefont  [1]{#1}%
\providecommand \bibfnamefont [1]{#1}%
\providecommand \citenamefont [1]{#1}%
\providecommand \href@noop [0]{\@secondoftwo}%
\providecommand \href [0]{\begingroup \@sanitize@url \@href}%
\providecommand \@href[1]{\@@startlink{#1}\@@href}%
\providecommand \@@href[1]{\endgroup#1\@@endlink}%
\providecommand \@sanitize@url [0]{\catcode `\\12\catcode `\$12\catcode
  `\&12\catcode `\#12\catcode `\^12\catcode `\_12\catcode `\%12\relax}%
\providecommand \@@startlink[1]{}%
\providecommand \@@endlink[0]{}%
\providecommand \url  [0]{\begingroup\@sanitize@url \@url }%
\providecommand \@url [1]{\endgroup\@href {#1}{\urlprefix }}%
\providecommand \urlprefix  [0]{URL }%
\providecommand \Eprint [0]{\href }%
\providecommand \doibase [0]{http://dx.doi.org/}%
\providecommand \selectlanguage [0]{\@gobble}%
\providecommand \bibinfo  [0]{\@secondoftwo}%
\providecommand \bibfield  [0]{\@secondoftwo}%
\providecommand \translation [1]{[#1]}%
\providecommand \BibitemOpen [0]{}%
\providecommand \bibitemStop [0]{}%
\providecommand \bibitemNoStop [0]{.\EOS\space}%
\providecommand \EOS [0]{\spacefactor3000\relax}%
\providecommand \BibitemShut  [1]{\csname bibitem#1\endcsname}%
\let\auto@bib@innerbib\@empty
\bibitem [{\citenamefont {Lin}(2019)}]{Lin:2019uvt}%
  \BibitemOpen
  \bibfield  {author} {\bibinfo {author} {\bibfnamefont {Tongyan}\ \bibnamefont
  {Lin}},\ }\bibfield  {title} {\enquote {\bibinfo {title} {{Dark matter models
  and direct detection}},}\ }\href {\doibase 10.22323/1.333.0009} {\bibfield
  {journal} {\bibinfo  {journal} {PoS}\ }\textbf {\bibinfo {volume} {333}},\
  \bibinfo {pages} {009} (\bibinfo {year} {2019})},\ \Eprint
  {http://arxiv.org/abs/1904.07915} {arXiv:1904.07915 [hep-ph]} \BibitemShut
  {NoStop}%
\bibitem [{\citenamefont {Essig}\ \emph {et~al.}(2012)\citenamefont {Essig},
  \citenamefont {Mardon},\ and\ \citenamefont {Volansky}}]{Essig:2011nj}%
  \BibitemOpen
  \bibfield  {author} {\bibinfo {author} {\bibfnamefont {Rouven}\ \bibnamefont
  {Essig}}, \bibinfo {author} {\bibfnamefont {Jeremy}\ \bibnamefont {Mardon}},
  \ and\ \bibinfo {author} {\bibfnamefont {Tomer}\ \bibnamefont {Volansky}},\
  }\bibfield  {title} {\enquote {\bibinfo {title} {{Direct Detection of Sub-GeV
  Dark Matter}},}\ }\href {\doibase 10.1103/PhysRevD.85.076007} {\bibfield
  {journal} {\bibinfo  {journal} {Phys. Rev. D}\ }\textbf {\bibinfo {volume}
  {85}},\ \bibinfo {pages} {076007} (\bibinfo {year} {2012})},\ \Eprint
  {http://arxiv.org/abs/1108.5383} {arXiv:1108.5383 [hep-ph]} \BibitemShut
  {NoStop}%
\bibitem [{\citenamefont {Graham}\ \emph {et~al.}(2012)\citenamefont {Graham},
  \citenamefont {Kaplan}, \citenamefont {Rajendran},\ and\ \citenamefont
  {Walters}}]{Graham:2012su}%
  \BibitemOpen
  \bibfield  {author} {\bibinfo {author} {\bibfnamefont {Peter~W.}\
  \bibnamefont {Graham}}, \bibinfo {author} {\bibfnamefont {David~E.}\
  \bibnamefont {Kaplan}}, \bibinfo {author} {\bibfnamefont {Surjeet}\
  \bibnamefont {Rajendran}}, \ and\ \bibinfo {author} {\bibfnamefont
  {Matthew~T.}\ \bibnamefont {Walters}},\ }\bibfield  {title} {\enquote
  {\bibinfo {title} {{Semiconductor Probes of Light Dark Matter}},}\ }\href
  {\doibase 10.1016/j.dark.2012.09.001} {\bibfield  {journal} {\bibinfo
  {journal} {Phys. Dark Univ.}\ }\textbf {\bibinfo {volume} {1}},\ \bibinfo
  {pages} {32--49} (\bibinfo {year} {2012})},\ \Eprint
  {http://arxiv.org/abs/1203.2531} {arXiv:1203.2531 [hep-ph]} \BibitemShut
  {NoStop}%
\bibitem [{\citenamefont {Essig}\ \emph {et~al.}(2016)\citenamefont {Essig},
  \citenamefont {Fernandez-Serra}, \citenamefont {Mardon}, \citenamefont
  {Soto}, \citenamefont {Volansky},\ and\ \citenamefont {Yu}}]{Essig:2015cda}%
  \BibitemOpen
  \bibfield  {author} {\bibinfo {author} {\bibfnamefont {Rouven}\ \bibnamefont
  {Essig}}, \bibinfo {author} {\bibfnamefont {Marivi}\ \bibnamefont
  {Fernandez-Serra}}, \bibinfo {author} {\bibfnamefont {Jeremy}\ \bibnamefont
  {Mardon}}, \bibinfo {author} {\bibfnamefont {Adrian}\ \bibnamefont {Soto}},
  \bibinfo {author} {\bibfnamefont {Tomer}\ \bibnamefont {Volansky}}, \ and\
  \bibinfo {author} {\bibfnamefont {Tien-Tien}\ \bibnamefont {Yu}},\ }\bibfield
   {title} {\enquote {\bibinfo {title} {{Direct Detection of sub-GeV Dark
  Matter with Semiconductor Targets}},}\ }\href {\doibase
  10.1007/JHEP05(2016)046} {\bibfield  {journal} {\bibinfo  {journal} {JHEP}\
  }\textbf {\bibinfo {volume} {05}},\ \bibinfo {pages} {046} (\bibinfo {year}
  {2016})},\ \Eprint {http://arxiv.org/abs/1509.01598} {arXiv:1509.01598
  [hep-ph]} \BibitemShut {NoStop}%
\bibitem [{\citenamefont {Aguilar-Arevalo}\ \emph {et~al.}(2019)\citenamefont
  {Aguilar-Arevalo} \emph {et~al.}}]{DAMIC:2019dcn}%
  \BibitemOpen
  \bibfield  {author} {\bibinfo {author} {\bibfnamefont {A.}~\bibnamefont
  {Aguilar-Arevalo}} \emph {et~al.} (\bibinfo {collaboration} {DAMIC}),\
  }\bibfield  {title} {\enquote {\bibinfo {title} {{Constraints on Light Dark
  Matter Particles Interacting with Electrons from DAMIC at SNOLAB}},}\ }\href
  {\doibase 10.1103/PhysRevLett.123.181802} {\bibfield  {journal} {\bibinfo
  {journal} {Phys. Rev. Lett.}\ }\textbf {\bibinfo {volume} {123}},\ \bibinfo
  {pages} {181802} (\bibinfo {year} {2019})},\ \Eprint
  {http://arxiv.org/abs/1907.12628} {arXiv:1907.12628 [astro-ph.CO]}
  \BibitemShut {NoStop}%
\bibitem [{\citenamefont {Barak}\ \emph {et~al.}(2020)\citenamefont {Barak}
  \emph {et~al.}}]{SENSEI:2020dpa}%
  \BibitemOpen
  \bibfield  {author} {\bibinfo {author} {\bibfnamefont {Liron}\ \bibnamefont
  {Barak}} \emph {et~al.} (\bibinfo {collaboration} {SENSEI}),\ }\bibfield
  {title} {\enquote {\bibinfo {title} {{SENSEI: Direct-Detection Results on
  sub-GeV Dark Matter from a New Skipper-CCD}},}\ }\href {\doibase
  10.1103/PhysRevLett.125.171802} {\bibfield  {journal} {\bibinfo  {journal}
  {Phys. Rev. Lett.}\ }\textbf {\bibinfo {volume} {125}},\ \bibinfo {pages}
  {171802} (\bibinfo {year} {2020})},\ \Eprint
  {http://arxiv.org/abs/2004.11378} {arXiv:2004.11378 [astro-ph.CO]}
  \BibitemShut {NoStop}%
\bibitem [{\citenamefont {Hochberg}\ \emph {et~al.}(2017)\citenamefont
  {Hochberg}, \citenamefont {Kahn}, \citenamefont {Lisanti}, \citenamefont
  {Tully},\ and\ \citenamefont {Zurek}}]{Hochberg:2016ntt}%
  \BibitemOpen
  \bibfield  {author} {\bibinfo {author} {\bibfnamefont {Yonit}\ \bibnamefont
  {Hochberg}}, \bibinfo {author} {\bibfnamefont {Yonatan}\ \bibnamefont
  {Kahn}}, \bibinfo {author} {\bibfnamefont {Mariangela}\ \bibnamefont
  {Lisanti}}, \bibinfo {author} {\bibfnamefont {Christopher~G.}\ \bibnamefont
  {Tully}}, \ and\ \bibinfo {author} {\bibfnamefont {Kathryn~M.}\ \bibnamefont
  {Zurek}},\ }\bibfield  {title} {\enquote {\bibinfo {title} {{Directional
  detection of dark matter with two-dimensional targets}},}\ }\href {\doibase
  10.1016/j.physletb.2017.06.051} {\bibfield  {journal} {\bibinfo  {journal}
  {Phys. Lett. B}\ }\textbf {\bibinfo {volume} {772}},\ \bibinfo {pages}
  {239--246} (\bibinfo {year} {2017})},\ \Eprint
  {http://arxiv.org/abs/1606.08849} {arXiv:1606.08849 [hep-ph]} \BibitemShut
  {NoStop}%
\bibitem [{\citenamefont {Baracchini}\ \emph {et~al.}(2018)\citenamefont
  {Baracchini} \emph {et~al.}}]{PTOLEMY:2018jst}%
  \BibitemOpen
  \bibfield  {author} {\bibinfo {author} {\bibfnamefont {E.}~\bibnamefont
  {Baracchini}} \emph {et~al.} (\bibinfo {collaboration} {PTOLEMY}),\
  }\bibfield  {title} {\enquote {\bibinfo {title} {{PTOLEMY: A Proposal for
  Thermal Relic Detection of Massive Neutrinos and Directional Detection of MeV
  Dark Matter}},}\ }\href@noop {} {\  (\bibinfo {year} {2018})},\ \Eprint
  {http://arxiv.org/abs/1808.01892} {arXiv:1808.01892 [physics.ins-det]}
  \BibitemShut {NoStop}%
\bibitem [{\citenamefont {Hochberg}\ \emph {et~al.}(2018)\citenamefont
  {Hochberg}, \citenamefont {Kahn}, \citenamefont {Lisanti}, \citenamefont
  {Zurek}, \citenamefont {Grushin}, \citenamefont {Ilan}, \citenamefont
  {Griffin}, \citenamefont {Liu}, \citenamefont {Weber},\ and\ \citenamefont
  {Neaton}}]{Hochberg:2017wce}%
  \BibitemOpen
  \bibfield  {author} {\bibinfo {author} {\bibfnamefont {Yonit}\ \bibnamefont
  {Hochberg}}, \bibinfo {author} {\bibfnamefont {Yonatan}\ \bibnamefont
  {Kahn}}, \bibinfo {author} {\bibfnamefont {Mariangela}\ \bibnamefont
  {Lisanti}}, \bibinfo {author} {\bibfnamefont {Kathryn~M.}\ \bibnamefont
  {Zurek}}, \bibinfo {author} {\bibfnamefont {Adolfo~G.}\ \bibnamefont
  {Grushin}}, \bibinfo {author} {\bibfnamefont {Roni}\ \bibnamefont {Ilan}},
  \bibinfo {author} {\bibfnamefont {Sin\'ead~M.}\ \bibnamefont {Griffin}},
  \bibinfo {author} {\bibfnamefont {Zhen-Fei}\ \bibnamefont {Liu}}, \bibinfo
  {author} {\bibfnamefont {Sophie~F.}\ \bibnamefont {Weber}}, \ and\ \bibinfo
  {author} {\bibfnamefont {Jeffrey~B.}\ \bibnamefont {Neaton}},\ }\bibfield
  {title} {\enquote {\bibinfo {title} {{Detection of sub-MeV Dark Matter with
  Three-Dimensional Dirac Materials}},}\ }\href {\doibase
  10.1103/PhysRevD.97.015004} {\bibfield  {journal} {\bibinfo  {journal} {Phys.
  Rev. D}\ }\textbf {\bibinfo {volume} {97}},\ \bibinfo {pages} {015004}
  (\bibinfo {year} {2018})},\ \Eprint {http://arxiv.org/abs/1708.08929}
  {arXiv:1708.08929 [hep-ph]} \BibitemShut {NoStop}%
\bibitem [{\citenamefont {Coskuner}\ \emph {et~al.}(2021)\citenamefont
  {Coskuner}, \citenamefont {Mitridate}, \citenamefont {Olivares},\ and\
  \citenamefont {Zurek}}]{Coskuner:2019odd}%
  \BibitemOpen
  \bibfield  {author} {\bibinfo {author} {\bibfnamefont {Ahmet}\ \bibnamefont
  {Coskuner}}, \bibinfo {author} {\bibfnamefont {Andrea}\ \bibnamefont
  {Mitridate}}, \bibinfo {author} {\bibfnamefont {Andres}\ \bibnamefont
  {Olivares}}, \ and\ \bibinfo {author} {\bibfnamefont {Kathryn~M.}\
  \bibnamefont {Zurek}},\ }\bibfield  {title} {\enquote {\bibinfo {title}
  {{Directional Dark Matter Detection in Anisotropic Dirac Materials}},}\
  }\href {\doibase 10.1103/PhysRevD.103.016006} {\bibfield  {journal} {\bibinfo
   {journal} {Phys. Rev. D}\ }\textbf {\bibinfo {volume} {103}},\ \bibinfo
  {pages} {016006} (\bibinfo {year} {2021})},\ \Eprint
  {http://arxiv.org/abs/1909.09170} {arXiv:1909.09170 [hep-ph]} \BibitemShut
  {NoStop}%
\bibitem [{\citenamefont {Geilhufe}\ \emph {et~al.}(2020)\citenamefont
  {Geilhufe}, \citenamefont {Kahlhoefer},\ and\ \citenamefont
  {Winkler}}]{Geilhufe:2019ndy}%
  \BibitemOpen
  \bibfield  {author} {\bibinfo {author} {\bibfnamefont {R.~Matthias}\
  \bibnamefont {Geilhufe}}, \bibinfo {author} {\bibfnamefont {Felix}\
  \bibnamefont {Kahlhoefer}}, \ and\ \bibinfo {author} {\bibfnamefont
  {Martin~Wolfgang}\ \bibnamefont {Winkler}},\ }\bibfield  {title} {\enquote
  {\bibinfo {title} {{Dirac Materials for Sub-MeV Dark Matter Detection: New
  Targets and Improved Formalism}},}\ }\href {\doibase
  10.1103/PhysRevD.101.055005} {\bibfield  {journal} {\bibinfo  {journal}
  {Phys. Rev. D}\ }\textbf {\bibinfo {volume} {101}},\ \bibinfo {pages}
  {055005} (\bibinfo {year} {2020})},\ \Eprint
  {http://arxiv.org/abs/1910.02091} {arXiv:1910.02091 [hep-ph]} \BibitemShut
  {NoStop}%
\bibitem [{\citenamefont {Hochberg}\ \emph
  {et~al.}(2016{\natexlab{a}})\citenamefont {Hochberg}, \citenamefont {Zhao},\
  and\ \citenamefont {Zurek}}]{Hochberg:2015pha}%
  \BibitemOpen
  \bibfield  {author} {\bibinfo {author} {\bibfnamefont {Yonit}\ \bibnamefont
  {Hochberg}}, \bibinfo {author} {\bibfnamefont {Yue}\ \bibnamefont {Zhao}}, \
  and\ \bibinfo {author} {\bibfnamefont {Kathryn~M.}\ \bibnamefont {Zurek}},\
  }\bibfield  {title} {\enquote {\bibinfo {title} {{Superconducting Detectors
  for Superlight Dark Matter}},}\ }\href {\doibase
  10.1103/PhysRevLett.116.011301} {\bibfield  {journal} {\bibinfo  {journal}
  {Phys. Rev. Lett.}\ }\textbf {\bibinfo {volume} {116}},\ \bibinfo {pages}
  {011301} (\bibinfo {year} {2016}{\natexlab{a}})},\ \Eprint
  {http://arxiv.org/abs/1504.07237} {arXiv:1504.07237 [hep-ph]} \BibitemShut
  {NoStop}%
\bibitem [{\citenamefont {Hochberg}\ \emph
  {et~al.}(2016{\natexlab{b}})\citenamefont {Hochberg}, \citenamefont {Pyle},
  \citenamefont {Zhao},\ and\ \citenamefont {Zurek}}]{Hochberg:2015fth}%
  \BibitemOpen
  \bibfield  {author} {\bibinfo {author} {\bibfnamefont {Yonit}\ \bibnamefont
  {Hochberg}}, \bibinfo {author} {\bibfnamefont {Matt}\ \bibnamefont {Pyle}},
  \bibinfo {author} {\bibfnamefont {Yue}\ \bibnamefont {Zhao}}, \ and\ \bibinfo
  {author} {\bibfnamefont {Kathryn~M.}\ \bibnamefont {Zurek}},\ }\bibfield
  {title} {\enquote {\bibinfo {title} {{Detecting Superlight Dark Matter with
  Fermi-Degenerate Materials}},}\ }\href {\doibase 10.1007/JHEP08(2016)057}
  {\bibfield  {journal} {\bibinfo  {journal} {JHEP}\ }\textbf {\bibinfo
  {volume} {08}},\ \bibinfo {pages} {057} (\bibinfo {year}
  {2016}{\natexlab{b}})},\ \Eprint {http://arxiv.org/abs/1512.04533}
  {arXiv:1512.04533 [hep-ph]} \BibitemShut {NoStop}%
\bibitem [{\citenamefont {Hochberg}\ \emph
  {et~al.}(2021{\natexlab{a}})\citenamefont {Hochberg}, \citenamefont {Kramer},
  \citenamefont {Kurinsky},\ and\ \citenamefont {Lehmann}}]{Hochberg:2021ymx}%
  \BibitemOpen
  \bibfield  {author} {\bibinfo {author} {\bibfnamefont {Yonit}\ \bibnamefont
  {Hochberg}}, \bibinfo {author} {\bibfnamefont {Eric~David}\ \bibnamefont
  {Kramer}}, \bibinfo {author} {\bibfnamefont {Noah}\ \bibnamefont {Kurinsky}},
  \ and\ \bibinfo {author} {\bibfnamefont {Benjamin~V.}\ \bibnamefont
  {Lehmann}},\ }\bibfield  {title} {\enquote {\bibinfo {title} {{Directional
  Detection of Light Dark Matter in Superconductors}},}\ }\href@noop {} {\
  (\bibinfo {year} {2021}{\natexlab{a}})},\ \Eprint
  {http://arxiv.org/abs/2109.04473} {arXiv:2109.04473 [hep-ph]} \BibitemShut
  {NoStop}%
\bibitem [{\citenamefont {Knapen}\ \emph {et~al.}(2018)\citenamefont {Knapen},
  \citenamefont {Lin}, \citenamefont {Pyle},\ and\ \citenamefont
  {Zurek}}]{Knapen:2017ekk}%
  \BibitemOpen
  \bibfield  {author} {\bibinfo {author} {\bibfnamefont {Simon}\ \bibnamefont
  {Knapen}}, \bibinfo {author} {\bibfnamefont {Tongyan}\ \bibnamefont {Lin}},
  \bibinfo {author} {\bibfnamefont {Matt}\ \bibnamefont {Pyle}}, \ and\
  \bibinfo {author} {\bibfnamefont {Kathryn~M.}\ \bibnamefont {Zurek}},\
  }\bibfield  {title} {\enquote {\bibinfo {title} {{Detection of Light Dark
  Matter With Optical Phonons in Polar Materials}},}\ }\href {\doibase
  10.1016/j.physletb.2018.08.064} {\bibfield  {journal} {\bibinfo  {journal}
  {Phys. Lett. B}\ }\textbf {\bibinfo {volume} {785}},\ \bibinfo {pages}
  {386--390} (\bibinfo {year} {2018})},\ \Eprint
  {http://arxiv.org/abs/1712.06598} {arXiv:1712.06598 [hep-ph]} \BibitemShut
  {NoStop}%
\bibitem [{\citenamefont {Hochberg}\ \emph {et~al.}(2019)\citenamefont
  {Hochberg}, \citenamefont {Charaev}, \citenamefont {Nam}, \citenamefont
  {Verma}, \citenamefont {Colangelo},\ and\ \citenamefont
  {Berggren}}]{Hochberg:2019cyy}%
  \BibitemOpen
  \bibfield  {author} {\bibinfo {author} {\bibfnamefont {Yonit}\ \bibnamefont
  {Hochberg}}, \bibinfo {author} {\bibfnamefont {Ilya}\ \bibnamefont
  {Charaev}}, \bibinfo {author} {\bibfnamefont {Sae-Woo}\ \bibnamefont {Nam}},
  \bibinfo {author} {\bibfnamefont {Varun}\ \bibnamefont {Verma}}, \bibinfo
  {author} {\bibfnamefont {Marco}\ \bibnamefont {Colangelo}}, \ and\ \bibinfo
  {author} {\bibfnamefont {Karl~K.}\ \bibnamefont {Berggren}},\ }\bibfield
  {title} {\enquote {\bibinfo {title} {{Detecting Sub-GeV Dark Matter with
  Superconducting Nanowires}},}\ }\href {\doibase
  10.1103/PhysRevLett.123.151802} {\bibfield  {journal} {\bibinfo  {journal}
  {Phys. Rev. Lett.}\ }\textbf {\bibinfo {volume} {123}},\ \bibinfo {pages}
  {151802} (\bibinfo {year} {2019})},\ \Eprint
  {http://arxiv.org/abs/1903.05101} {arXiv:1903.05101 [hep-ph]} \BibitemShut
  {NoStop}%
\bibitem [{\citenamefont {Hochberg}\ \emph
  {et~al.}(2021{\natexlab{b}})\citenamefont {Hochberg}, \citenamefont
  {Lehmann}, \citenamefont {Charaev}, \citenamefont {Chiles}, \citenamefont
  {Nam},\ and\ \citenamefont {Berggren}}]{Hochberg:2021yud}%
  \BibitemOpen
  \bibfield  {author} {\bibinfo {author} {\bibfnamefont {Yonit}\ \bibnamefont
  {Hochberg}}, \bibinfo {author} {\bibfnamefont {Benjamin~V.}\ \bibnamefont
  {Lehmann}}, \bibinfo {author} {\bibfnamefont {Ilya}\ \bibnamefont {Charaev}},
  \bibinfo {author} {\bibfnamefont {Jeff}\ \bibnamefont {Chiles}}, \bibinfo
  {author} {\bibfnamefont {Sae~Woo}\ \bibnamefont {Nam}}, \ and\ \bibinfo
  {author} {\bibfnamefont {Karl~K.}\ \bibnamefont {Berggren}},\ }\bibfield
  {title} {\enquote {\bibinfo {title} {{New Constraints on Dark Matter from
  Superconducting Nanowires}},}\ }\href@noop {} {\  (\bibinfo {year}
  {2021}{\natexlab{b}})},\ \Eprint {http://arxiv.org/abs/2110.01586}
  {arXiv:2110.01586 [hep-ph]} \BibitemShut {NoStop}%
\bibitem [{\citenamefont {Chiles}\ \emph {et~al.}(2021)\citenamefont {Chiles}
  \emph {et~al.}}]{Chiles:2021gxk}%
  \BibitemOpen
  \bibfield  {author} {\bibinfo {author} {\bibfnamefont {Jeff}\ \bibnamefont
  {Chiles}} \emph {et~al.},\ }\bibfield  {title} {\enquote {\bibinfo {title}
  {{First Constraints on Dark Photon Dark Matter with Superconducting Nanowire
  Detectors in an Optical Haloscope}},}\ }\href@noop {} {\  (\bibinfo {year}
  {2021})},\ \Eprint {http://arxiv.org/abs/2110.01582} {arXiv:2110.01582
  [hep-ex]} \BibitemShut {NoStop}%
\bibitem [{\citenamefont {Derenzo}\ \emph {et~al.}(2017)\citenamefont
  {Derenzo}, \citenamefont {Essig}, \citenamefont {Massari}, \citenamefont
  {Soto},\ and\ \citenamefont {Yu}}]{Derenzo:2016fse}%
  \BibitemOpen
  \bibfield  {author} {\bibinfo {author} {\bibfnamefont {Stephen}\ \bibnamefont
  {Derenzo}}, \bibinfo {author} {\bibfnamefont {Rouven}\ \bibnamefont {Essig}},
  \bibinfo {author} {\bibfnamefont {Andrea}\ \bibnamefont {Massari}}, \bibinfo
  {author} {\bibfnamefont {Adr\'\i{}an}\ \bibnamefont {Soto}}, \ and\ \bibinfo
  {author} {\bibfnamefont {Tien-Tien}\ \bibnamefont {Yu}},\ }\bibfield  {title}
  {\enquote {\bibinfo {title} {{Direct Detection of sub-GeV Dark Matter with
  Scintillating Targets}},}\ }\href {\doibase 10.1103/PhysRevD.96.016026}
  {\bibfield  {journal} {\bibinfo  {journal} {Phys. Rev. D}\ }\textbf {\bibinfo
  {volume} {96}},\ \bibinfo {pages} {016026} (\bibinfo {year} {2017})},\
  \Eprint {http://arxiv.org/abs/1607.01009} {arXiv:1607.01009 [hep-ph]}
  \BibitemShut {NoStop}%
\bibitem [{\citenamefont {Blanco}\ \emph {et~al.}(2020)\citenamefont {Blanco},
  \citenamefont {Collar}, \citenamefont {Kahn},\ and\ \citenamefont
  {Lillard}}]{Blanco:2019lrf}%
  \BibitemOpen
  \bibfield  {author} {\bibinfo {author} {\bibfnamefont {Carlos}\ \bibnamefont
  {Blanco}}, \bibinfo {author} {\bibfnamefont {J.~I.}\ \bibnamefont {Collar}},
  \bibinfo {author} {\bibfnamefont {Yonatan}\ \bibnamefont {Kahn}}, \ and\
  \bibinfo {author} {\bibfnamefont {Benjamin}\ \bibnamefont {Lillard}},\
  }\bibfield  {title} {\enquote {\bibinfo {title} {{Dark Matter-Electron
  Scattering from Aromatic Organic Targets}},}\ }\href {\doibase
  10.1103/PhysRevD.101.056001} {\bibfield  {journal} {\bibinfo  {journal}
  {Phys. Rev. D}\ }\textbf {\bibinfo {volume} {101}},\ \bibinfo {pages}
  {056001} (\bibinfo {year} {2020})},\ \Eprint
  {http://arxiv.org/abs/1912.02822} {arXiv:1912.02822 [hep-ph]} \BibitemShut
  {NoStop}%
\bibitem [{\citenamefont {Essig}\ \emph {et~al.}(2017)\citenamefont {Essig},
  \citenamefont {Mardon}, \citenamefont {Slone},\ and\ \citenamefont
  {Volansky}}]{Essig:2016crl}%
  \BibitemOpen
  \bibfield  {author} {\bibinfo {author} {\bibfnamefont {Rouven}\ \bibnamefont
  {Essig}}, \bibinfo {author} {\bibfnamefont {Jeremy}\ \bibnamefont {Mardon}},
  \bibinfo {author} {\bibfnamefont {Oren}\ \bibnamefont {Slone}}, \ and\
  \bibinfo {author} {\bibfnamefont {Tomer}\ \bibnamefont {Volansky}},\
  }\bibfield  {title} {\enquote {\bibinfo {title} {{Detection of sub-GeV Dark
  Matter and Solar Neutrinos via Chemical-Bond Breaking}},}\ }\href {\doibase
  10.1103/PhysRevD.95.056011} {\bibfield  {journal} {\bibinfo  {journal} {Phys.
  Rev. D}\ }\textbf {\bibinfo {volume} {95}},\ \bibinfo {pages} {056011}
  (\bibinfo {year} {2017})},\ \Eprint {http://arxiv.org/abs/1608.02940}
  {arXiv:1608.02940 [hep-ph]} \BibitemShut {NoStop}%
\bibitem [{\citenamefont {Bunting}\ \emph {et~al.}(2017)\citenamefont
  {Bunting}, \citenamefont {Gratta}, \citenamefont {Melia},\ and\ \citenamefont
  {Rajendran}}]{Bunting:2017net}%
  \BibitemOpen
  \bibfield  {author} {\bibinfo {author} {\bibfnamefont {Philip~C.}\
  \bibnamefont {Bunting}}, \bibinfo {author} {\bibfnamefont {Giorgio}\
  \bibnamefont {Gratta}}, \bibinfo {author} {\bibfnamefont {Tom}\ \bibnamefont
  {Melia}}, \ and\ \bibinfo {author} {\bibfnamefont {Surjeet}\ \bibnamefont
  {Rajendran}},\ }\bibfield  {title} {\enquote {\bibinfo {title} {{Magnetic
  Bubble Chambers and Sub-GeV Dark Matter Direct Detection}},}\ }\href
  {\doibase 10.1103/PhysRevD.95.095001} {\bibfield  {journal} {\bibinfo
  {journal} {Phys. Rev. D}\ }\textbf {\bibinfo {volume} {95}},\ \bibinfo
  {pages} {095001} (\bibinfo {year} {2017})},\ \Eprint
  {http://arxiv.org/abs/1701.06566} {arXiv:1701.06566 [hep-ph]} \BibitemShut
  {NoStop}%
\bibitem [{\citenamefont {Geilhufe}\ \emph {et~al.}(2018)\citenamefont
  {Geilhufe}, \citenamefont {Olsthoorn}, \citenamefont {Ferella}, \citenamefont
  {Koski}, \citenamefont {Kahlhoefer}, \citenamefont {Conrad},\ and\
  \citenamefont {Balatsky}}]{Geilhufe:2018gry}%
  \BibitemOpen
  \bibfield  {author} {\bibinfo {author} {\bibfnamefont {R.~Matthias}\
  \bibnamefont {Geilhufe}}, \bibinfo {author} {\bibfnamefont {Bart}\
  \bibnamefont {Olsthoorn}}, \bibinfo {author} {\bibfnamefont {Alfredo}\
  \bibnamefont {Ferella}}, \bibinfo {author} {\bibfnamefont {Timo}\
  \bibnamefont {Koski}}, \bibinfo {author} {\bibfnamefont {Felix}\ \bibnamefont
  {Kahlhoefer}}, \bibinfo {author} {\bibfnamefont {Jan}\ \bibnamefont
  {Conrad}}, \ and\ \bibinfo {author} {\bibfnamefont {Alexander~V.}\
  \bibnamefont {Balatsky}},\ }\bibfield  {title} {\enquote {\bibinfo {title}
  {{Materials Informatics for Dark Matter Detection}},}\ }\href {\doibase
  10.1002/pssr.201800293} {\bibfield  {journal} {\bibinfo  {journal} {Phys.
  Status Solidi RRL}\ }\textbf {\bibinfo {volume} {12}},\ \bibinfo {pages}
  {1800293} (\bibinfo {year} {2018})},\ \Eprint
  {http://arxiv.org/abs/1806.06040} {arXiv:1806.06040 [cond-mat.mtrl-sci]}
  \BibitemShut {NoStop}%
\bibitem [{\citenamefont {Griffin}\ \emph {et~al.}(2020)\citenamefont
  {Griffin}, \citenamefont {Inzani}, \citenamefont {Trickle}, \citenamefont
  {Zhang},\ and\ \citenamefont {Zurek}}]{Griffin:2019mvc}%
  \BibitemOpen
  \bibfield  {author} {\bibinfo {author} {\bibfnamefont {Sin\'ead~M.}\
  \bibnamefont {Griffin}}, \bibinfo {author} {\bibfnamefont {Katherine}\
  \bibnamefont {Inzani}}, \bibinfo {author} {\bibfnamefont {Tanner}\
  \bibnamefont {Trickle}}, \bibinfo {author} {\bibfnamefont {Zhengkang}\
  \bibnamefont {Zhang}}, \ and\ \bibinfo {author} {\bibfnamefont {Kathryn~M.}\
  \bibnamefont {Zurek}},\ }\bibfield  {title} {\enquote {\bibinfo {title}
  {{Multichannel direct detection of light dark matter: Target comparison}},}\
  }\href {\doibase 10.1103/PhysRevD.101.055004} {\bibfield  {journal} {\bibinfo
   {journal} {Phys. Rev. D}\ }\textbf {\bibinfo {volume} {101}},\ \bibinfo
  {pages} {055004} (\bibinfo {year} {2020})},\ \Eprint
  {http://arxiv.org/abs/1910.10716} {arXiv:1910.10716 [hep-ph]} \BibitemShut
  {NoStop}%
\bibitem [{\citenamefont {Kurinsky}\ \emph {et~al.}(2019)\citenamefont
  {Kurinsky}, \citenamefont {Yu}, \citenamefont {Hochberg},\ and\ \citenamefont
  {Cabrera}}]{Kurinsky:2019pgb}%
  \BibitemOpen
  \bibfield  {author} {\bibinfo {author} {\bibfnamefont {Noah~Alexander}\
  \bibnamefont {Kurinsky}}, \bibinfo {author} {\bibfnamefont {To~Chin}\
  \bibnamefont {Yu}}, \bibinfo {author} {\bibfnamefont {Yonit}\ \bibnamefont
  {Hochberg}}, \ and\ \bibinfo {author} {\bibfnamefont {Blas}\ \bibnamefont
  {Cabrera}},\ }\bibfield  {title} {\enquote {\bibinfo {title} {{Diamond
  Detectors for Direct Detection of Sub-GeV Dark Matter}},}\ }\href {\doibase
  10.1103/PhysRevD.99.123005} {\bibfield  {journal} {\bibinfo  {journal} {Phys.
  Rev. D}\ }\textbf {\bibinfo {volume} {99}},\ \bibinfo {pages} {123005}
  (\bibinfo {year} {2019})},\ \Eprint {http://arxiv.org/abs/1901.07569}
  {arXiv:1901.07569 [hep-ex]} \BibitemShut {NoStop}%
\bibitem [{\citenamefont {Griffin}\ \emph
  {et~al.}(2021{\natexlab{a}})\citenamefont {Griffin}, \citenamefont
  {Hochberg}, \citenamefont {Inzani}, \citenamefont {Kurinsky}, \citenamefont
  {Lin},\ and\ \citenamefont {Chin}}]{Griffin:2020lgd}%
  \BibitemOpen
  \bibfield  {author} {\bibinfo {author} {\bibfnamefont {Sin\'ead~M.}\
  \bibnamefont {Griffin}}, \bibinfo {author} {\bibfnamefont {Yonit}\
  \bibnamefont {Hochberg}}, \bibinfo {author} {\bibfnamefont {Katherine}\
  \bibnamefont {Inzani}}, \bibinfo {author} {\bibfnamefont {Noah}\ \bibnamefont
  {Kurinsky}}, \bibinfo {author} {\bibfnamefont {Tongyan}\ \bibnamefont {Lin}},
  \ and\ \bibinfo {author} {\bibfnamefont {To}~\bibnamefont {Chin}},\
  }\bibfield  {title} {\enquote {\bibinfo {title} {{Silicon carbide detectors
  for sub-GeV dark matter}},}\ }\href {\doibase 10.1103/PhysRevD.103.075002}
  {\bibfield  {journal} {\bibinfo  {journal} {Phys. Rev. D}\ }\textbf {\bibinfo
  {volume} {103}},\ \bibinfo {pages} {075002} (\bibinfo {year}
  {2021}{\natexlab{a}})},\ \Eprint {http://arxiv.org/abs/2008.08560}
  {arXiv:2008.08560 [hep-ph]} \BibitemShut {NoStop}%
\bibitem [{\citenamefont {Kahn}\ and\ \citenamefont
  {Lin}(2021)}]{Kahn:2021ttr}%
  \BibitemOpen
  \bibfield  {author} {\bibinfo {author} {\bibfnamefont {Yonatan}\ \bibnamefont
  {Kahn}}\ and\ \bibinfo {author} {\bibfnamefont {Tongyan}\ \bibnamefont
  {Lin}},\ }\bibfield  {title} {\enquote {\bibinfo {title} {{Searches for light
  dark matter using condensed matter systems}},}\ }\href@noop {} {\  (\bibinfo
  {year} {2021})},\ \Eprint {http://arxiv.org/abs/2108.03239} {arXiv:2108.03239
  [hep-ph]} \BibitemShut {NoStop}%
\bibitem [{\citenamefont {Ibe}\ \emph {et~al.}(2018)\citenamefont {Ibe},
  \citenamefont {Nakano}, \citenamefont {Shoji},\ and\ \citenamefont
  {Suzuki}}]{Ibe:2017yqa}%
  \BibitemOpen
  \bibfield  {author} {\bibinfo {author} {\bibfnamefont {Masahiro}\
  \bibnamefont {Ibe}}, \bibinfo {author} {\bibfnamefont {Wakutaka}\
  \bibnamefont {Nakano}}, \bibinfo {author} {\bibfnamefont {Yutaro}\
  \bibnamefont {Shoji}}, \ and\ \bibinfo {author} {\bibfnamefont {Kazumine}\
  \bibnamefont {Suzuki}},\ }\bibfield  {title} {\enquote {\bibinfo {title}
  {{Migdal Effect in Dark Matter Direct Detection Experiments}},}\ }\href
  {\doibase 10.1007/JHEP03(2018)194} {\bibfield  {journal} {\bibinfo  {journal}
  {JHEP}\ }\textbf {\bibinfo {volume} {03}},\ \bibinfo {pages} {194} (\bibinfo
  {year} {2018})},\ \Eprint {http://arxiv.org/abs/1707.07258} {arXiv:1707.07258
  [hep-ph]} \BibitemShut {NoStop}%
\bibitem [{\citenamefont {Dolan}\ \emph {et~al.}(2018)\citenamefont {Dolan},
  \citenamefont {Kahlhoefer},\ and\ \citenamefont {McCabe}}]{Dolan:2017xbu}%
  \BibitemOpen
  \bibfield  {author} {\bibinfo {author} {\bibfnamefont {Matthew~J.}\
  \bibnamefont {Dolan}}, \bibinfo {author} {\bibfnamefont {Felix}\ \bibnamefont
  {Kahlhoefer}}, \ and\ \bibinfo {author} {\bibfnamefont {Christopher}\
  \bibnamefont {McCabe}},\ }\bibfield  {title} {\enquote {\bibinfo {title}
  {{Directly detecting sub-GeV dark matter with electrons from nuclear
  scattering}},}\ }\href {\doibase 10.1103/PhysRevLett.121.101801} {\bibfield
  {journal} {\bibinfo  {journal} {Phys. Rev. Lett.}\ }\textbf {\bibinfo
  {volume} {121}},\ \bibinfo {pages} {101801} (\bibinfo {year} {2018})},\
  \Eprint {http://arxiv.org/abs/1711.09906} {arXiv:1711.09906 [hep-ph]}
  \BibitemShut {NoStop}%
\bibitem [{\citenamefont {Bell}\ \emph
  {et~al.}(2020{\natexlab{a}})\citenamefont {Bell}, \citenamefont {Dent},
  \citenamefont {Newstead}, \citenamefont {Sabharwal},\ and\ \citenamefont
  {Weiler}}]{Bell:2019egg}%
  \BibitemOpen
  \bibfield  {author} {\bibinfo {author} {\bibfnamefont {Nicole~F.}\
  \bibnamefont {Bell}}, \bibinfo {author} {\bibfnamefont {James~B.}\
  \bibnamefont {Dent}}, \bibinfo {author} {\bibfnamefont {Jayden~L.}\
  \bibnamefont {Newstead}}, \bibinfo {author} {\bibfnamefont {Subir}\
  \bibnamefont {Sabharwal}}, \ and\ \bibinfo {author} {\bibfnamefont
  {Thomas~J.}\ \bibnamefont {Weiler}},\ }\bibfield  {title} {\enquote {\bibinfo
  {title} {{Migdal effect and photon bremsstrahlung in effective field theories
  of dark matter direct detection and coherent elastic neutrino-nucleus
  scattering}},}\ }\href {\doibase 10.1103/PhysRevD.101.015012} {\bibfield
  {journal} {\bibinfo  {journal} {Phys. Rev. D}\ }\textbf {\bibinfo {volume}
  {101}},\ \bibinfo {pages} {015012} (\bibinfo {year} {2020}{\natexlab{a}})},\
  \Eprint {http://arxiv.org/abs/1905.00046} {arXiv:1905.00046 [hep-ph]}
  \BibitemShut {NoStop}%
\bibitem [{\citenamefont {Baxter}\ \emph {et~al.}(2020)\citenamefont {Baxter},
  \citenamefont {Kahn},\ and\ \citenamefont {Krnjaic}}]{Baxter:2019pnz}%
  \BibitemOpen
  \bibfield  {author} {\bibinfo {author} {\bibfnamefont {Daniel}\ \bibnamefont
  {Baxter}}, \bibinfo {author} {\bibfnamefont {Yonatan}\ \bibnamefont {Kahn}},
  \ and\ \bibinfo {author} {\bibfnamefont {Gordan}\ \bibnamefont {Krnjaic}},\
  }\bibfield  {title} {\enquote {\bibinfo {title} {{Electron Ionization via
  Dark Matter-Electron Scattering and the Migdal Effect}},}\ }\href {\doibase
  10.1103/PhysRevD.101.076014} {\bibfield  {journal} {\bibinfo  {journal}
  {Phys. Rev. D}\ }\textbf {\bibinfo {volume} {101}},\ \bibinfo {pages}
  {076014} (\bibinfo {year} {2020})},\ \Eprint
  {http://arxiv.org/abs/1908.00012} {arXiv:1908.00012 [hep-ph]} \BibitemShut
  {NoStop}%
\bibitem [{\citenamefont {Essig}\ \emph {et~al.}(2020)\citenamefont {Essig},
  \citenamefont {Pradler}, \citenamefont {Sholapurkar},\ and\ \citenamefont
  {Yu}}]{Essig:2019xkx}%
  \BibitemOpen
  \bibfield  {author} {\bibinfo {author} {\bibfnamefont {Rouven}\ \bibnamefont
  {Essig}}, \bibinfo {author} {\bibfnamefont {Josef}\ \bibnamefont {Pradler}},
  \bibinfo {author} {\bibfnamefont {Mukul}\ \bibnamefont {Sholapurkar}}, \ and\
  \bibinfo {author} {\bibfnamefont {Tien-Tien}\ \bibnamefont {Yu}},\ }\bibfield
   {title} {\enquote {\bibinfo {title} {{Relation between the Migdal Effect and
  Dark Matter-Electron Scattering in Isolated Atoms and Semiconductors}},}\
  }\href {\doibase 10.1103/PhysRevLett.124.021801} {\bibfield  {journal}
  {\bibinfo  {journal} {Phys. Rev. Lett.}\ }\textbf {\bibinfo {volume} {124}},\
  \bibinfo {pages} {021801} (\bibinfo {year} {2020})},\ \Eprint
  {http://arxiv.org/abs/1908.10881} {arXiv:1908.10881 [hep-ph]} \BibitemShut
  {NoStop}%
\bibitem [{\citenamefont {Flambaum}\ \emph {et~al.}(2020)\citenamefont
  {Flambaum}, \citenamefont {Su}, \citenamefont {Wu},\ and\ \citenamefont
  {Zhu}}]{Flambaum:2020xxo}%
  \BibitemOpen
  \bibfield  {author} {\bibinfo {author} {\bibfnamefont {Victor~V.}\
  \bibnamefont {Flambaum}}, \bibinfo {author} {\bibfnamefont {Liangliang}\
  \bibnamefont {Su}}, \bibinfo {author} {\bibfnamefont {Lei}\ \bibnamefont
  {Wu}}, \ and\ \bibinfo {author} {\bibfnamefont {Bin}\ \bibnamefont {Zhu}},\
  }\bibfield  {title} {\enquote {\bibinfo {title} {{Constraining sub-GeV dark
  matter from Migdal and Boosted effects}},}\ }\href@noop {} {\  (\bibinfo
  {year} {2020})},\ \Eprint {http://arxiv.org/abs/2012.09751} {arXiv:2012.09751
  [hep-ph]} \BibitemShut {NoStop}%
\bibitem [{\citenamefont {Wang}\ \emph {et~al.}(2021)\citenamefont {Wang},
  \citenamefont {Wu}, \citenamefont {Wu},\ and\ \citenamefont
  {Zhu}}]{Wang:2021oha}%
  \BibitemOpen
  \bibfield  {author} {\bibinfo {author} {\bibfnamefont {Wenyu}\ \bibnamefont
  {Wang}}, \bibinfo {author} {\bibfnamefont {Ke-Yun}\ \bibnamefont {Wu}},
  \bibinfo {author} {\bibfnamefont {Lei}\ \bibnamefont {Wu}}, \ and\ \bibinfo
  {author} {\bibfnamefont {Bin}\ \bibnamefont {Zhu}},\ }\bibfield  {title}
  {\enquote {\bibinfo {title} {{Direct Detection of Spin-Dependent Sub-GeV Dark
  Matter via Migdal Effect}},}\ }\href@noop {} {\  (\bibinfo {year} {2021})},\
  \Eprint {http://arxiv.org/abs/2112.06492} {arXiv:2112.06492 [hep-ph]}
  \BibitemShut {NoStop}%
\bibitem [{\citenamefont {An}\ \emph {et~al.}(2018)\citenamefont {An},
  \citenamefont {Pospelov}, \citenamefont {Pradler},\ and\ \citenamefont
  {Ritz}}]{An:2017ojc}%
  \BibitemOpen
  \bibfield  {author} {\bibinfo {author} {\bibfnamefont {Haipeng}\ \bibnamefont
  {An}}, \bibinfo {author} {\bibfnamefont {Maxim}\ \bibnamefont {Pospelov}},
  \bibinfo {author} {\bibfnamefont {Josef}\ \bibnamefont {Pradler}}, \ and\
  \bibinfo {author} {\bibfnamefont {Adam}\ \bibnamefont {Ritz}},\ }\bibfield
  {title} {\enquote {\bibinfo {title} {{Directly Detecting MeV-scale Dark
  Matter via Solar Reflection}},}\ }\href {\doibase
  10.1103/PhysRevLett.120.141801} {\bibfield  {journal} {\bibinfo  {journal}
  {Phys. Rev. Lett.}\ }\textbf {\bibinfo {volume} {120}},\ \bibinfo {pages}
  {141801} (\bibinfo {year} {2018})},\ \bibinfo {note} {[Erratum:
  Phys.Rev.Lett. 121, 259903 (2018)]},\ \Eprint
  {http://arxiv.org/abs/1708.03642} {arXiv:1708.03642 [hep-ph]} \BibitemShut
  {NoStop}%
\bibitem [{\citenamefont {Emken}\ \emph {et~al.}(2018)\citenamefont {Emken},
  \citenamefont {Kouvaris},\ and\ \citenamefont {Nielsen}}]{Emken:2017hnp}%
  \BibitemOpen
  \bibfield  {author} {\bibinfo {author} {\bibfnamefont {Timon}\ \bibnamefont
  {Emken}}, \bibinfo {author} {\bibfnamefont {Chris}\ \bibnamefont {Kouvaris}},
  \ and\ \bibinfo {author} {\bibfnamefont {Niklas~Gr\o{}nlund}\ \bibnamefont
  {Nielsen}},\ }\bibfield  {title} {\enquote {\bibinfo {title} {{The Sun as a
  sub-GeV Dark Matter Accelerator}},}\ }\href {\doibase
  10.1103/PhysRevD.97.063007} {\bibfield  {journal} {\bibinfo  {journal} {Phys.
  Rev. D}\ }\textbf {\bibinfo {volume} {97}},\ \bibinfo {pages} {063007}
  (\bibinfo {year} {2018})},\ \Eprint {http://arxiv.org/abs/1709.06573}
  {arXiv:1709.06573 [hep-ph]} \BibitemShut {NoStop}%
\bibitem [{\citenamefont {Chen}\ \emph {et~al.}(2021)\citenamefont {Chen},
  \citenamefont {Cui}, \citenamefont {Shu}, \citenamefont {Xue}, \citenamefont
  {Yuan},\ and\ \citenamefont {Yuan}}]{Chen:2020gcl}%
  \BibitemOpen
  \bibfield  {author} {\bibinfo {author} {\bibfnamefont {Yifan}\ \bibnamefont
  {Chen}}, \bibinfo {author} {\bibfnamefont {Ming-Yang}\ \bibnamefont {Cui}},
  \bibinfo {author} {\bibfnamefont {Jing}\ \bibnamefont {Shu}}, \bibinfo
  {author} {\bibfnamefont {Xiao}\ \bibnamefont {Xue}}, \bibinfo {author}
  {\bibfnamefont {Guan-Wen}\ \bibnamefont {Yuan}}, \ and\ \bibinfo {author}
  {\bibfnamefont {Qiang}\ \bibnamefont {Yuan}},\ }\bibfield  {title} {\enquote
  {\bibinfo {title} {{Sun heated MeV-scale dark matter and the XENON1T electron
  recoil excess}},}\ }\href {\doibase 10.1007/JHEP04(2021)282} {\bibfield
  {journal} {\bibinfo  {journal} {JHEP}\ }\textbf {\bibinfo {volume} {04}},\
  \bibinfo {pages} {282} (\bibinfo {year} {2021})},\ \Eprint
  {http://arxiv.org/abs/2006.12447} {arXiv:2006.12447 [hep-ph]} \BibitemShut
  {NoStop}%
\bibitem [{\citenamefont {Emken}(2021)}]{Emken:2021lgc}%
  \BibitemOpen
  \bibfield  {author} {\bibinfo {author} {\bibfnamefont {Timon}\ \bibnamefont
  {Emken}},\ }\bibfield  {title} {\enquote {\bibinfo {title} {{Solar reflection
  of light dark matter with heavy mediators}},}\ }\href@noop {} {\  (\bibinfo
  {year} {2021})},\ \Eprint {http://arxiv.org/abs/2102.12483} {arXiv:2102.12483
  [hep-ph]} \BibitemShut {NoStop}%
\bibitem [{\citenamefont {An}\ \emph {et~al.}(2021)\citenamefont {An},
  \citenamefont {Nie}, \citenamefont {Pospelov}, \citenamefont {Pradler},\ and\
  \citenamefont {Ritz}}]{An:2021qdl}%
  \BibitemOpen
  \bibfield  {author} {\bibinfo {author} {\bibfnamefont {Haipeng}\ \bibnamefont
  {An}}, \bibinfo {author} {\bibfnamefont {Haoming}\ \bibnamefont {Nie}},
  \bibinfo {author} {\bibfnamefont {Maxim}\ \bibnamefont {Pospelov}}, \bibinfo
  {author} {\bibfnamefont {Josef}\ \bibnamefont {Pradler}}, \ and\ \bibinfo
  {author} {\bibfnamefont {Adam}\ \bibnamefont {Ritz}},\ }\bibfield  {title}
  {\enquote {\bibinfo {title} {Solar reflection of dark matter},}\ }\href
  {\doibase 10.1103/PhysRevD.104.103026} {\bibfield  {journal} {\bibinfo
  {journal} {Phys. Rev. D}\ }\textbf {\bibinfo {volume} {104}},\ \bibinfo
  {pages} {103026} (\bibinfo {year} {2021})}\BibitemShut {NoStop}%
\bibitem [{\citenamefont {Bringmann}\ and\ \citenamefont
  {Pospelov}(2019)}]{Bringmann:2018cvk}%
  \BibitemOpen
  \bibfield  {author} {\bibinfo {author} {\bibfnamefont {Torsten}\ \bibnamefont
  {Bringmann}}\ and\ \bibinfo {author} {\bibfnamefont {Maxim}\ \bibnamefont
  {Pospelov}},\ }\bibfield  {title} {\enquote {\bibinfo {title} {{Novel direct
  detection constraints on light dark matter}},}\ }\href {\doibase
  10.1103/PhysRevLett.122.171801} {\bibfield  {journal} {\bibinfo  {journal}
  {Phys. Rev. Lett.}\ }\textbf {\bibinfo {volume} {122}},\ \bibinfo {pages}
  {171801} (\bibinfo {year} {2019})},\ \Eprint
  {http://arxiv.org/abs/1810.10543} {arXiv:1810.10543 [hep-ph]} \BibitemShut
  {NoStop}%
\bibitem [{\citenamefont {Ema}\ \emph {et~al.}(2019)\citenamefont {Ema},
  \citenamefont {Sala},\ and\ \citenamefont {Sato}}]{Ema:2018bih}%
  \BibitemOpen
  \bibfield  {author} {\bibinfo {author} {\bibfnamefont {Yohei}\ \bibnamefont
  {Ema}}, \bibinfo {author} {\bibfnamefont {Filippo}\ \bibnamefont {Sala}}, \
  and\ \bibinfo {author} {\bibfnamefont {Ryosuke}\ \bibnamefont {Sato}},\
  }\bibfield  {title} {\enquote {\bibinfo {title} {{Light Dark Matter at
  Neutrino Experiments}},}\ }\href {\doibase 10.1103/PhysRevLett.122.181802}
  {\bibfield  {journal} {\bibinfo  {journal} {Phys. Rev. Lett.}\ }\textbf
  {\bibinfo {volume} {122}},\ \bibinfo {pages} {181802} (\bibinfo {year}
  {2019})},\ \Eprint {http://arxiv.org/abs/1811.00520} {arXiv:1811.00520
  [hep-ph]} \BibitemShut {NoStop}%
\bibitem [{\citenamefont {Cappiello}\ and\ \citenamefont
  {Beacom}(2019)}]{Cappiello:2019qsw}%
  \BibitemOpen
  \bibfield  {author} {\bibinfo {author} {\bibfnamefont {Christopher~V.}\
  \bibnamefont {Cappiello}}\ and\ \bibinfo {author} {\bibfnamefont {John~F.}\
  \bibnamefont {Beacom}},\ }\bibfield  {title} {\enquote {\bibinfo {title}
  {{Strong New Limits on Light Dark Matter from Neutrino Experiments}},}\
  }\href {\doibase 10.1103/PhysRevD.104.069901} {\bibfield  {journal} {\bibinfo
   {journal} {Phys. Rev. D}\ }\textbf {\bibinfo {volume} {100}},\ \bibinfo
  {pages} {103011} (\bibinfo {year} {2019})},\ \bibinfo {note} {[Erratum:
  Phys.Rev.D 104, 069901 (2021)]},\ \Eprint {http://arxiv.org/abs/1906.11283}
  {arXiv:1906.11283 [hep-ph]} \BibitemShut {NoStop}%
\bibitem [{\citenamefont {Bondarenko}\ \emph {et~al.}(2020)\citenamefont
  {Bondarenko}, \citenamefont {Boyarsky}, \citenamefont {Bringmann},
  \citenamefont {Hufnagel}, \citenamefont {Schmidt-Hoberg},\ and\ \citenamefont
  {Sokolenko}}]{Bondarenko:2019vrb}%
  \BibitemOpen
  \bibfield  {author} {\bibinfo {author} {\bibfnamefont {Kyrylo}\ \bibnamefont
  {Bondarenko}}, \bibinfo {author} {\bibfnamefont {Alexey}\ \bibnamefont
  {Boyarsky}}, \bibinfo {author} {\bibfnamefont {Torsten}\ \bibnamefont
  {Bringmann}}, \bibinfo {author} {\bibfnamefont {Marco}\ \bibnamefont
  {Hufnagel}}, \bibinfo {author} {\bibfnamefont {Kai}\ \bibnamefont
  {Schmidt-Hoberg}}, \ and\ \bibinfo {author} {\bibfnamefont {Anastasia}\
  \bibnamefont {Sokolenko}},\ }\bibfield  {title} {\enquote {\bibinfo {title}
  {{Direct detection and complementary constraints for sub-GeV dark matter}},}\
  }\href {\doibase 10.1007/JHEP03(2020)118} {\bibfield  {journal} {\bibinfo
  {journal} {JHEP}\ }\textbf {\bibinfo {volume} {03}},\ \bibinfo {pages} {118}
  (\bibinfo {year} {2020})},\ \Eprint {http://arxiv.org/abs/1909.08632}
  {arXiv:1909.08632 [hep-ph]} \BibitemShut {NoStop}%
\bibitem [{\citenamefont {Roberts}\ \emph {et~al.}(2016)\citenamefont
  {Roberts}, \citenamefont {Dzuba}, \citenamefont {Flambaum}, \citenamefont
  {Pospelov},\ and\ \citenamefont {Stadnik}}]{Roberts:2016xfw}%
  \BibitemOpen
  \bibfield  {author} {\bibinfo {author} {\bibfnamefont {B.~M.}\ \bibnamefont
  {Roberts}}, \bibinfo {author} {\bibfnamefont {V.~A.}\ \bibnamefont {Dzuba}},
  \bibinfo {author} {\bibfnamefont {V.~V.}\ \bibnamefont {Flambaum}}, \bibinfo
  {author} {\bibfnamefont {M.}~\bibnamefont {Pospelov}}, \ and\ \bibinfo
  {author} {\bibfnamefont {Y.~V.}\ \bibnamefont {Stadnik}},\ }\bibfield
  {title} {\enquote {\bibinfo {title} {{Dark matter scattering on electrons:
  Accurate calculations of atomic excitations and implications for the DAMA
  signal}},}\ }\href {\doibase 10.1103/PhysRevD.93.115037} {\bibfield
  {journal} {\bibinfo  {journal} {Phys. Rev. D}\ }\textbf {\bibinfo {volume}
  {93}},\ \bibinfo {pages} {115037} (\bibinfo {year} {2016})},\ \Eprint
  {http://arxiv.org/abs/1604.04559} {arXiv:1604.04559 [hep-ph]} \BibitemShut
  {NoStop}%
\bibitem [{\citenamefont {Trickle}\ \emph {et~al.}(2020)\citenamefont
  {Trickle}, \citenamefont {Zhang}, \citenamefont {Zurek}, \citenamefont
  {Inzani},\ and\ \citenamefont {Griffin}}]{Trickle:2019nya}%
  \BibitemOpen
  \bibfield  {author} {\bibinfo {author} {\bibfnamefont {Tanner}\ \bibnamefont
  {Trickle}}, \bibinfo {author} {\bibfnamefont {Zhengkang}\ \bibnamefont
  {Zhang}}, \bibinfo {author} {\bibfnamefont {Kathryn~M.}\ \bibnamefont
  {Zurek}}, \bibinfo {author} {\bibfnamefont {Katherine}\ \bibnamefont
  {Inzani}}, \ and\ \bibinfo {author} {\bibfnamefont {Sin\'ead}\ \bibnamefont
  {Griffin}},\ }\bibfield  {title} {\enquote {\bibinfo {title} {{Multi-Channel
  Direct Detection of Light Dark Matter: Theoretical Framework}},}\ }\href
  {\doibase 10.1007/JHEP03(2020)036} {\bibfield  {journal} {\bibinfo  {journal}
  {JHEP}\ }\textbf {\bibinfo {volume} {03}},\ \bibinfo {pages} {036} (\bibinfo
  {year} {2020})},\ \Eprint {http://arxiv.org/abs/1910.08092} {arXiv:1910.08092
  [hep-ph]} \BibitemShut {NoStop}%
\bibitem [{\citenamefont {Catena}\ \emph {et~al.}(2020)\citenamefont {Catena},
  \citenamefont {Emken}, \citenamefont {Spaldin},\ and\ \citenamefont
  {Tarantino}}]{Catena:2019gfa}%
  \BibitemOpen
  \bibfield  {author} {\bibinfo {author} {\bibfnamefont {Riccardo}\
  \bibnamefont {Catena}}, \bibinfo {author} {\bibfnamefont {Timon}\
  \bibnamefont {Emken}}, \bibinfo {author} {\bibfnamefont {Nicola~A.}\
  \bibnamefont {Spaldin}}, \ and\ \bibinfo {author} {\bibfnamefont {Walter}\
  \bibnamefont {Tarantino}},\ }\bibfield  {title} {\enquote {\bibinfo {title}
  {{Atomic responses to general dark matter-electron interactions}},}\ }\href
  {\doibase 10.1103/PhysRevResearch.2.033195} {\bibfield  {journal} {\bibinfo
  {journal} {Phys. Rev. Res.}\ }\textbf {\bibinfo {volume} {2}},\ \bibinfo
  {pages} {033195} (\bibinfo {year} {2020})},\ \Eprint
  {http://arxiv.org/abs/1912.08204} {arXiv:1912.08204 [hep-ph]} \BibitemShut
  {NoStop}%
\bibitem [{\citenamefont {Trickle}\ \emph {et~al.}(2022)\citenamefont
  {Trickle}, \citenamefont {Zhang},\ and\ \citenamefont
  {Zurek}}]{Trickle:2020oki}%
  \BibitemOpen
  \bibfield  {author} {\bibinfo {author} {\bibfnamefont {Tanner}\ \bibnamefont
  {Trickle}}, \bibinfo {author} {\bibfnamefont {Zhengkang}\ \bibnamefont
  {Zhang}}, \ and\ \bibinfo {author} {\bibfnamefont {Kathryn~M.}\ \bibnamefont
  {Zurek}},\ }\bibfield  {title} {\enquote {\bibinfo {title} {Effective field
  theory of dark matter direct detection with collective excitations},}\ }\href
  {\doibase 10.1103/PhysRevD.105.015001} {\bibfield  {journal} {\bibinfo
  {journal} {Phys. Rev. D}\ }\textbf {\bibinfo {volume} {105}},\ \bibinfo
  {pages} {015001} (\bibinfo {year} {2022})}\BibitemShut {NoStop}%
\bibitem [{\citenamefont {Gelmini}\ \emph {et~al.}(2020)\citenamefont
  {Gelmini}, \citenamefont {Takhistov},\ and\ \citenamefont
  {Vitagliano}}]{Gelmini:2020xir}%
  \BibitemOpen
  \bibfield  {author} {\bibinfo {author} {\bibfnamefont {Graciela~B.}\
  \bibnamefont {Gelmini}}, \bibinfo {author} {\bibfnamefont {Volodymyr}\
  \bibnamefont {Takhistov}}, \ and\ \bibinfo {author} {\bibfnamefont {Edoardo}\
  \bibnamefont {Vitagliano}},\ }\bibfield  {title} {\enquote {\bibinfo {title}
  {{Scalar direct detection: In-medium effects}},}\ }\href {\doibase
  10.1016/j.physletb.2020.135779} {\bibfield  {journal} {\bibinfo  {journal}
  {Phys. Lett. B}\ }\textbf {\bibinfo {volume} {809}},\ \bibinfo {pages}
  {135779} (\bibinfo {year} {2020})},\ \Eprint
  {http://arxiv.org/abs/2006.13909} {arXiv:2006.13909 [hep-ph]} \BibitemShut
  {NoStop}%
\bibitem [{\citenamefont {Borah}\ \emph {et~al.}(2021)\citenamefont {Borah},
  \citenamefont {Mahapatra},\ and\ \citenamefont {Sahu}}]{Borah:2020smw}%
  \BibitemOpen
  \bibfield  {author} {\bibinfo {author} {\bibfnamefont {Debasish}\
  \bibnamefont {Borah}}, \bibinfo {author} {\bibfnamefont {Satyabrata}\
  \bibnamefont {Mahapatra}}, \ and\ \bibinfo {author} {\bibfnamefont
  {Narendra}\ \bibnamefont {Sahu}},\ }\bibfield  {title} {\enquote {\bibinfo
  {title} {{Connecting Low scale Seesaw for Neutrino Mass to Inelastic sub-GeV
  Dark Matter with Abelian Gauge Symmetry}},}\ }\href {\doibase
  10.1016/j.nuclphysb.2021.115407} {\bibfield  {journal} {\bibinfo  {journal}
  {Nucl. Phys. B}\ }\textbf {\bibinfo {volume} {968}},\ \bibinfo {pages}
  {115407} (\bibinfo {year} {2021})},\ \Eprint
  {http://arxiv.org/abs/2009.06294} {arXiv:2009.06294 [hep-ph]} \BibitemShut
  {NoStop}%
\bibitem [{\citenamefont {Griffin}\ \emph
  {et~al.}(2021{\natexlab{b}})\citenamefont {Griffin}, \citenamefont {Inzani},
  \citenamefont {Trickle}, \citenamefont {Zhang},\ and\ \citenamefont
  {Zurek}}]{Griffin:2021znd}%
  \BibitemOpen
  \bibfield  {author} {\bibinfo {author} {\bibfnamefont {Sin\'ead~M.}\
  \bibnamefont {Griffin}}, \bibinfo {author} {\bibfnamefont {Katherine}\
  \bibnamefont {Inzani}}, \bibinfo {author} {\bibfnamefont {Tanner}\
  \bibnamefont {Trickle}}, \bibinfo {author} {\bibfnamefont {Zhengkang}\
  \bibnamefont {Zhang}}, \ and\ \bibinfo {author} {\bibfnamefont {Kathryn~M.}\
  \bibnamefont {Zurek}},\ }\bibfield  {title} {\enquote {\bibinfo {title}
  {Extended calculation of dark matter-electron scattering in crystal
  targets},}\ }\href {\doibase 10.1103/PhysRevD.104.095015} {\bibfield
  {journal} {\bibinfo  {journal} {Phys. Rev. D}\ }\textbf {\bibinfo {volume}
  {104}},\ \bibinfo {pages} {095015} (\bibinfo {year}
  {2021}{\natexlab{b}})}\BibitemShut {NoStop}%
\bibitem [{\citenamefont {Catena}\ \emph {et~al.}(2021)\citenamefont {Catena},
  \citenamefont {Emken}, \citenamefont {Matas}, \citenamefont {Spaldin},\ and\
  \citenamefont {Urdshals}}]{Catena:2021qsr}%
  \BibitemOpen
  \bibfield  {author} {\bibinfo {author} {\bibfnamefont {Riccardo}\
  \bibnamefont {Catena}}, \bibinfo {author} {\bibfnamefont {Timon}\
  \bibnamefont {Emken}}, \bibinfo {author} {\bibfnamefont {Marek}\ \bibnamefont
  {Matas}}, \bibinfo {author} {\bibfnamefont {Nicola~A.}\ \bibnamefont
  {Spaldin}}, \ and\ \bibinfo {author} {\bibfnamefont {Einar}\ \bibnamefont
  {Urdshals}},\ }\bibfield  {title} {\enquote {\bibinfo {title} {{Crystal
  responses to general dark matter-electron interactions}},}\ }\href {\doibase
  10.1103/PhysRevResearch.3.033149} {\bibfield  {journal} {\bibinfo  {journal}
  {Phys. Rev. Res.}\ }\textbf {\bibinfo {volume} {3}},\ \bibinfo {pages}
  {033149} (\bibinfo {year} {2021})},\ \Eprint
  {http://arxiv.org/abs/2105.02233} {arXiv:2105.02233 [hep-ph]} \BibitemShut
  {NoStop}%
\bibitem [{\citenamefont {Hochberg}\ \emph
  {et~al.}(2021{\natexlab{c}})\citenamefont {Hochberg}, \citenamefont {Kahn},
  \citenamefont {Kurinsky}, \citenamefont {Lehmann}, \citenamefont {Yu},\ and\
  \citenamefont {Berggren}}]{Hochberg:2021pkt}%
  \BibitemOpen
  \bibfield  {author} {\bibinfo {author} {\bibfnamefont {Yonit}\ \bibnamefont
  {Hochberg}}, \bibinfo {author} {\bibfnamefont {Yonatan}\ \bibnamefont
  {Kahn}}, \bibinfo {author} {\bibfnamefont {Noah}\ \bibnamefont {Kurinsky}},
  \bibinfo {author} {\bibfnamefont {Benjamin~V.}\ \bibnamefont {Lehmann}},
  \bibinfo {author} {\bibfnamefont {To~Chin}\ \bibnamefont {Yu}}, \ and\
  \bibinfo {author} {\bibfnamefont {Karl~K.}\ \bibnamefont {Berggren}},\
  }\bibfield  {title} {\enquote {\bibinfo {title} {{Determining
  Dark-Matter\textendash{}Electron Scattering Rates from the Dielectric
  Function}},}\ }\href {\doibase 10.1103/PhysRevLett.127.151802} {\bibfield
  {journal} {\bibinfo  {journal} {Phys. Rev. Lett.}\ }\textbf {\bibinfo
  {volume} {127}},\ \bibinfo {pages} {151802} (\bibinfo {year}
  {2021}{\natexlab{c}})},\ \Eprint {http://arxiv.org/abs/2101.08263}
  {arXiv:2101.08263 [hep-ph]} \BibitemShut {NoStop}%
\bibitem [{\citenamefont {Knapen}\ \emph {et~al.}(2021)\citenamefont {Knapen},
  \citenamefont {Kozaczuk},\ and\ \citenamefont {Lin}}]{Knapen:2021run}%
  \BibitemOpen
  \bibfield  {author} {\bibinfo {author} {\bibfnamefont {Simon}\ \bibnamefont
  {Knapen}}, \bibinfo {author} {\bibfnamefont {Jonathan}\ \bibnamefont
  {Kozaczuk}}, \ and\ \bibinfo {author} {\bibfnamefont {Tongyan}\ \bibnamefont
  {Lin}},\ }\bibfield  {title} {\enquote {\bibinfo {title} {{Dark
  matter-electron scattering in dielectrics}},}\ }\href {\doibase
  10.1103/PhysRevD.104.015031} {\bibfield  {journal} {\bibinfo  {journal}
  {Phys. Rev. D}\ }\textbf {\bibinfo {volume} {104}},\ \bibinfo {pages}
  {015031} (\bibinfo {year} {2021})},\ \Eprint
  {http://arxiv.org/abs/2101.08275} {arXiv:2101.08275 [hep-ph]} \BibitemShut
  {NoStop}%
\bibitem [{\citenamefont {Kurinsky}\ \emph {et~al.}(2020)\citenamefont
  {Kurinsky}, \citenamefont {Baxter}, \citenamefont {Kahn},\ and\ \citenamefont
  {Krnjaic}}]{Kurinsky:2020dpb}%
  \BibitemOpen
  \bibfield  {author} {\bibinfo {author} {\bibfnamefont {Noah}\ \bibnamefont
  {Kurinsky}}, \bibinfo {author} {\bibfnamefont {Daniel}\ \bibnamefont
  {Baxter}}, \bibinfo {author} {\bibfnamefont {Yonatan}\ \bibnamefont {Kahn}},
  \ and\ \bibinfo {author} {\bibfnamefont {Gordan}\ \bibnamefont {Krnjaic}},\
  }\bibfield  {title} {\enquote {\bibinfo {title} {{Dark matter interpretation
  of excesses in multiple direct detection experiments}},}\ }\href {\doibase
  10.1103/PhysRevD.102.015017} {\bibfield  {journal} {\bibinfo  {journal}
  {Phys. Rev. D}\ }\textbf {\bibinfo {volume} {102}},\ \bibinfo {pages}
  {015017} (\bibinfo {year} {2020})},\ \Eprint
  {http://arxiv.org/abs/2002.06937} {arXiv:2002.06937 [hep-ph]} \BibitemShut
  {NoStop}%
\bibitem [{\citenamefont {Du}\ \emph {et~al.}(2022)\citenamefont {Du},
  \citenamefont {Egana-Ugrinovic}, \citenamefont {Essig},\ and\ \citenamefont
  {Sholapurkar}}]{Du:2020ldo}%
  \BibitemOpen
  \bibfield  {author} {\bibinfo {author} {\bibfnamefont {Peizhi}\ \bibnamefont
  {Du}}, \bibinfo {author} {\bibfnamefont {Daniel}\ \bibnamefont
  {Egana-Ugrinovic}}, \bibinfo {author} {\bibfnamefont {Rouven}\ \bibnamefont
  {Essig}}, \ and\ \bibinfo {author} {\bibfnamefont {Mukul}\ \bibnamefont
  {Sholapurkar}},\ }\bibfield  {title} {\enquote {\bibinfo {title} {Sources of
  low-energy events in low-threshold dark-matter and neutrino detectors},}\
  }\href {\doibase 10.1103/PhysRevX.12.011009} {\bibfield  {journal} {\bibinfo
  {journal} {Phys. Rev. X}\ }\textbf {\bibinfo {volume} {12}},\ \bibinfo
  {pages} {011009} (\bibinfo {year} {2022})}\BibitemShut {NoStop}%
\bibitem [{\citenamefont {Aprile}\ \emph {et~al.}(2020)\citenamefont {Aprile}
  \emph {et~al.}}]{XENON:2020rca}%
  \BibitemOpen
  \bibfield  {author} {\bibinfo {author} {\bibfnamefont {E.}~\bibnamefont
  {Aprile}} \emph {et~al.} (\bibinfo {collaboration} {XENON}),\ }\bibfield
  {title} {\enquote {\bibinfo {title} {{Excess electronic recoil events in
  XENON1T}},}\ }\href {\doibase 10.1103/PhysRevD.102.072004} {\bibfield
  {journal} {\bibinfo  {journal} {Phys. Rev. D}\ }\textbf {\bibinfo {volume}
  {102}},\ \bibinfo {pages} {072004} (\bibinfo {year} {2020})},\ \Eprint
  {http://arxiv.org/abs/2006.09721} {arXiv:2006.09721 [hep-ex]} \BibitemShut
  {NoStop}%
\bibitem [{\citenamefont {Tucker-Smith}\ and\ \citenamefont
  {Weiner}(2001)}]{Tucker-Smith:2001myb}%
  \BibitemOpen
  \bibfield  {author} {\bibinfo {author} {\bibfnamefont {David}\ \bibnamefont
  {Tucker-Smith}}\ and\ \bibinfo {author} {\bibfnamefont {Neal}\ \bibnamefont
  {Weiner}},\ }\bibfield  {title} {\enquote {\bibinfo {title} {{Inelastic dark
  matter}},}\ }\href {\doibase 10.1103/PhysRevD.64.043502} {\bibfield
  {journal} {\bibinfo  {journal} {Phys. Rev. D}\ }\textbf {\bibinfo {volume}
  {64}},\ \bibinfo {pages} {043502} (\bibinfo {year} {2001})},\ \Eprint
  {http://arxiv.org/abs/hep-ph/0101138} {arXiv:hep-ph/0101138} \BibitemShut
  {NoStop}%
\bibitem [{\citenamefont {Bell}\ \emph
  {et~al.}(2020{\natexlab{b}})\citenamefont {Bell}, \citenamefont {Dent},
  \citenamefont {Dutta}, \citenamefont {Ghosh}, \citenamefont {Kumar},\ and\
  \citenamefont {Newstead}}]{Bell:2020bes}%
  \BibitemOpen
  \bibfield  {author} {\bibinfo {author} {\bibfnamefont {Nicole~F.}\
  \bibnamefont {Bell}}, \bibinfo {author} {\bibfnamefont {James~B.}\
  \bibnamefont {Dent}}, \bibinfo {author} {\bibfnamefont {Bhaskar}\
  \bibnamefont {Dutta}}, \bibinfo {author} {\bibfnamefont {Sumit}\ \bibnamefont
  {Ghosh}}, \bibinfo {author} {\bibfnamefont {Jason}\ \bibnamefont {Kumar}}, \
  and\ \bibinfo {author} {\bibfnamefont {Jayden~L.}\ \bibnamefont {Newstead}},\
  }\bibfield  {title} {\enquote {\bibinfo {title} {{Explaining the XENON1T
  excess with Luminous Dark Matter}},}\ }\href {\doibase
  10.1103/PhysRevLett.125.161803} {\bibfield  {journal} {\bibinfo  {journal}
  {Phys. Rev. Lett.}\ }\textbf {\bibinfo {volume} {125}},\ \bibinfo {pages}
  {161803} (\bibinfo {year} {2020}{\natexlab{b}})},\ \Eprint
  {http://arxiv.org/abs/2006.12461} {arXiv:2006.12461 [hep-ph]} \BibitemShut
  {NoStop}%
\bibitem [{\citenamefont {Harigaya}\ \emph {et~al.}(2020)\citenamefont
  {Harigaya}, \citenamefont {Nakai},\ and\ \citenamefont
  {Suzuki}}]{Harigaya:2020ckz}%
  \BibitemOpen
  \bibfield  {author} {\bibinfo {author} {\bibfnamefont {Keisuke}\ \bibnamefont
  {Harigaya}}, \bibinfo {author} {\bibfnamefont {Yuichiro}\ \bibnamefont
  {Nakai}}, \ and\ \bibinfo {author} {\bibfnamefont {Motoo}\ \bibnamefont
  {Suzuki}},\ }\bibfield  {title} {\enquote {\bibinfo {title} {{Inelastic Dark
  Matter Electron Scattering and the XENON1T Excess}},}\ }\href {\doibase
  10.1016/j.physletb.2020.135729} {\bibfield  {journal} {\bibinfo  {journal}
  {Phys. Lett. B}\ }\textbf {\bibinfo {volume} {809}},\ \bibinfo {pages}
  {135729} (\bibinfo {year} {2020})},\ \Eprint
  {http://arxiv.org/abs/2006.11938} {arXiv:2006.11938 [hep-ph]} \BibitemShut
  {NoStop}%
\bibitem [{\citenamefont {Bloch}\ \emph {et~al.}(2021)\citenamefont {Bloch},
  \citenamefont {Caputo}, \citenamefont {Essig}, \citenamefont {Redigolo},
  \citenamefont {Sholapurkar},\ and\ \citenamefont {Volansky}}]{Bloch:2020uzh}%
  \BibitemOpen
  \bibfield  {author} {\bibinfo {author} {\bibfnamefont {Itay~M.}\ \bibnamefont
  {Bloch}}, \bibinfo {author} {\bibfnamefont {Andrea}\ \bibnamefont {Caputo}},
  \bibinfo {author} {\bibfnamefont {Rouven}\ \bibnamefont {Essig}}, \bibinfo
  {author} {\bibfnamefont {Diego}\ \bibnamefont {Redigolo}}, \bibinfo {author}
  {\bibfnamefont {Mukul}\ \bibnamefont {Sholapurkar}}, \ and\ \bibinfo {author}
  {\bibfnamefont {Tomer}\ \bibnamefont {Volansky}},\ }\bibfield  {title}
  {\enquote {\bibinfo {title} {{Exploring new physics with O(keV) electron
  recoils in direct detection experiments}},}\ }\href {\doibase
  10.1007/JHEP01(2021)178} {\bibfield  {journal} {\bibinfo  {journal} {JHEP}\
  }\textbf {\bibinfo {volume} {01}},\ \bibinfo {pages} {178} (\bibinfo {year}
  {2021})},\ \Eprint {http://arxiv.org/abs/2006.14521} {arXiv:2006.14521
  [hep-ph]} \BibitemShut {NoStop}%
\bibitem [{\citenamefont {Baryakhtar}\ \emph {et~al.}(2020)\citenamefont
  {Baryakhtar}, \citenamefont {Berlin}, \citenamefont {Liu},\ and\
  \citenamefont {Weiner}}]{Baryakhtar:2020rwy}%
  \BibitemOpen
  \bibfield  {author} {\bibinfo {author} {\bibfnamefont {Masha}\ \bibnamefont
  {Baryakhtar}}, \bibinfo {author} {\bibfnamefont {Asher}\ \bibnamefont
  {Berlin}}, \bibinfo {author} {\bibfnamefont {Hongwan}\ \bibnamefont {Liu}}, \
  and\ \bibinfo {author} {\bibfnamefont {Neal}\ \bibnamefont {Weiner}},\
  }\bibfield  {title} {\enquote {\bibinfo {title} {{Electromagnetic Signals of
  Inelastic Dark Matter Scattering}},}\ }\href@noop {} {\  (\bibinfo {year}
  {2020})},\ \Eprint {http://arxiv.org/abs/2006.13918} {arXiv:2006.13918
  [hep-ph]} \BibitemShut {NoStop}%
\bibitem [{\citenamefont {Choi}\ \emph {et~al.}(2021)\citenamefont {Choi},
  \citenamefont {Lee},\ and\ \citenamefont {Zhu}}]{Choi:2020ysq}%
  \BibitemOpen
  \bibfield  {author} {\bibinfo {author} {\bibfnamefont {Soo-Min}\ \bibnamefont
  {Choi}}, \bibinfo {author} {\bibfnamefont {Hyun~Min}\ \bibnamefont {Lee}}, \
  and\ \bibinfo {author} {\bibfnamefont {Bin}\ \bibnamefont {Zhu}},\ }\bibfield
   {title} {\enquote {\bibinfo {title} {{Exothermic dark mesons in light of
  electron recoil excess at XENON1T}},}\ }\href {\doibase
  10.1007/JHEP04(2021)251} {\bibfield  {journal} {\bibinfo  {journal} {JHEP}\
  }\textbf {\bibinfo {volume} {04}},\ \bibinfo {pages} {251} (\bibinfo {year}
  {2021})},\ \Eprint {http://arxiv.org/abs/2012.03713} {arXiv:2012.03713
  [hep-ph]} \BibitemShut {NoStop}%
\bibitem [{\citenamefont {An}\ and\ \citenamefont {Yang}(2021)}]{An:2020tcg}%
  \BibitemOpen
  \bibfield  {author} {\bibinfo {author} {\bibfnamefont {Haipeng}\ \bibnamefont
  {An}}\ and\ \bibinfo {author} {\bibfnamefont {Daneng}\ \bibnamefont {Yang}},\
  }\bibfield  {title} {\enquote {\bibinfo {title} {{Direct detection of
  freeze-in inelastic dark matter}},}\ }\href {\doibase
  10.1016/j.physletb.2021.136408} {\bibfield  {journal} {\bibinfo  {journal}
  {Phys. Lett. B}\ }\textbf {\bibinfo {volume} {818}},\ \bibinfo {pages}
  {136408} (\bibinfo {year} {2021})},\ \Eprint
  {http://arxiv.org/abs/2006.15672} {arXiv:2006.15672 [hep-ph]} \BibitemShut
  {NoStop}%
\bibitem [{\citenamefont {Chao}\ \emph {et~al.}(2020)\citenamefont {Chao},
  \citenamefont {Gao},\ and\ \citenamefont {Jin}}]{Chao:2020yro}%
  \BibitemOpen
  \bibfield  {author} {\bibinfo {author} {\bibfnamefont {Wei}\ \bibnamefont
  {Chao}}, \bibinfo {author} {\bibfnamefont {Yu}~\bibnamefont {Gao}}, \ and\
  \bibinfo {author} {\bibfnamefont {Ming~jie}\ \bibnamefont {Jin}},\ }\bibfield
   {title} {\enquote {\bibinfo {title} {{Pseudo-Dirac Dark Matter in
  XENON1T}},}\ }\href@noop {} {\  (\bibinfo {year} {2020})},\ \Eprint
  {http://arxiv.org/abs/2006.16145} {arXiv:2006.16145 [hep-ph]} \BibitemShut
  {NoStop}%
\bibitem [{\citenamefont {He}\ \emph {et~al.}(2021{\natexlab{a}})\citenamefont
  {He}, \citenamefont {Wang},\ and\ \citenamefont {Zheng}}]{He:2020wjs}%
  \BibitemOpen
  \bibfield  {author} {\bibinfo {author} {\bibfnamefont {Hong-Jian}\
  \bibnamefont {He}}, \bibinfo {author} {\bibfnamefont {Yu-Chen}\ \bibnamefont
  {Wang}}, \ and\ \bibinfo {author} {\bibfnamefont {Jiaming}\ \bibnamefont
  {Zheng}},\ }\bibfield  {title} {\enquote {\bibinfo {title} {{EFT Approach of
  Inelastic Dark Matter for Xenon Electron Recoil Detection}},}\ }\href
  {\doibase 10.1088/1475-7516/2021/01/042} {\bibfield  {journal} {\bibinfo
  {journal} {JCAP}\ }\textbf {\bibinfo {volume} {01}},\ \bibinfo {pages} {042}
  (\bibinfo {year} {2021}{\natexlab{a}})},\ \Eprint
  {http://arxiv.org/abs/2007.04963} {arXiv:2007.04963 [hep-ph]} \BibitemShut
  {NoStop}%
\bibitem [{\citenamefont {He}\ \emph {et~al.}(2021{\natexlab{b}})\citenamefont
  {He}, \citenamefont {Wang},\ and\ \citenamefont {Zheng}}]{He:2020sat}%
  \BibitemOpen
  \bibfield  {author} {\bibinfo {author} {\bibfnamefont {Hong-Jian}\
  \bibnamefont {He}}, \bibinfo {author} {\bibfnamefont {Yu-Chen}\ \bibnamefont
  {Wang}}, \ and\ \bibinfo {author} {\bibfnamefont {Jiaming}\ \bibnamefont
  {Zheng}},\ }\bibfield  {title} {\enquote {\bibinfo {title} {Gev-scale
  inelastic dark matter with dark photon mediator via direct detection and
  cosmological and laboratory constraints},}\ }\href {\doibase
  10.1103/PhysRevD.104.115033} {\bibfield  {journal} {\bibinfo  {journal}
  {Phys. Rev. D}\ }\textbf {\bibinfo {volume} {104}},\ \bibinfo {pages}
  {115033} (\bibinfo {year} {2021}{\natexlab{b}})}\BibitemShut {NoStop}%
\bibitem [{\citenamefont {Borah}\ \emph {et~al.}(2020)\citenamefont {Borah},
  \citenamefont {Mahapatra}, \citenamefont {Nanda},\ and\ \citenamefont
  {Sahu}}]{Borah:2020jzi}%
  \BibitemOpen
  \bibfield  {author} {\bibinfo {author} {\bibfnamefont {Debasish}\
  \bibnamefont {Borah}}, \bibinfo {author} {\bibfnamefont {Satyabrata}\
  \bibnamefont {Mahapatra}}, \bibinfo {author} {\bibfnamefont {Dibyendu}\
  \bibnamefont {Nanda}}, \ and\ \bibinfo {author} {\bibfnamefont {Narendra}\
  \bibnamefont {Sahu}},\ }\bibfield  {title} {\enquote {\bibinfo {title}
  {{Inelastic fermion dark matter origin of XENON1T excess with muon $(g - 2)$
  and light neutrino mass}},}\ }\href {\doibase 10.1016/j.physletb.2020.135933}
  {\bibfield  {journal} {\bibinfo  {journal} {Phys. Lett. B}\ }\textbf
  {\bibinfo {volume} {811}},\ \bibinfo {pages} {135933} (\bibinfo {year}
  {2020})},\ \Eprint {http://arxiv.org/abs/2007.10754} {arXiv:2007.10754
  [hep-ph]} \BibitemShut {NoStop}%
\bibitem [{\citenamefont {Baek}(2021)}]{Baek:2021yos}%
  \BibitemOpen
  \bibfield  {author} {\bibinfo {author} {\bibfnamefont {Seungwon}\
  \bibnamefont {Baek}},\ }\bibfield  {title} {\enquote {\bibinfo {title}
  {Inelastic dark matter, small scale problems, and the xenon1t excess},}\
  }\href {\doibase 10.1007/jhep10(2021)135} {\bibfield  {journal} {\bibinfo
  {journal} {Journal of High Energy Physics}\ }\textbf {\bibinfo {volume}
  {2021}} (\bibinfo {year} {2021}),\ 10.1007/jhep10(2021)135}\BibitemShut
  {NoStop}%
\bibitem [{\citenamefont {Feldstein}\ \emph {et~al.}(2010)\citenamefont
  {Feldstein}, \citenamefont {Graham},\ and\ \citenamefont
  {Rajendran}}]{Feldstein:2010su}%
  \BibitemOpen
  \bibfield  {author} {\bibinfo {author} {\bibfnamefont {Brian}\ \bibnamefont
  {Feldstein}}, \bibinfo {author} {\bibfnamefont {Peter~W.}\ \bibnamefont
  {Graham}}, \ and\ \bibinfo {author} {\bibfnamefont {Surjeet}\ \bibnamefont
  {Rajendran}},\ }\bibfield  {title} {\enquote {\bibinfo {title} {{Luminous
  Dark Matter}},}\ }\href {\doibase 10.1103/PhysRevD.82.075019} {\bibfield
  {journal} {\bibinfo  {journal} {Phys. Rev. D}\ }\textbf {\bibinfo {volume}
  {82}},\ \bibinfo {pages} {075019} (\bibinfo {year} {2010})},\ \Eprint
  {http://arxiv.org/abs/1008.1988} {arXiv:1008.1988 [hep-ph]} \BibitemShut
  {NoStop}%
\bibitem [{\citenamefont {Eby}\ \emph {et~al.}(2019)\citenamefont {Eby},
  \citenamefont {Fox}, \citenamefont {Harnik},\ and\ \citenamefont
  {Kribs}}]{Eby:2019mgs}%
  \BibitemOpen
  \bibfield  {author} {\bibinfo {author} {\bibfnamefont {Joshua}\ \bibnamefont
  {Eby}}, \bibinfo {author} {\bibfnamefont {Patrick~J.}\ \bibnamefont {Fox}},
  \bibinfo {author} {\bibfnamefont {Roni}\ \bibnamefont {Harnik}}, \ and\
  \bibinfo {author} {\bibfnamefont {Graham~D.}\ \bibnamefont {Kribs}},\
  }\bibfield  {title} {\enquote {\bibinfo {title} {{Luminous Signals of
  Inelastic Dark Matter in Large Detectors}},}\ }\href {\doibase
  10.1007/JHEP09(2019)115} {\bibfield  {journal} {\bibinfo  {journal} {JHEP}\
  }\textbf {\bibinfo {volume} {09}},\ \bibinfo {pages} {115} (\bibinfo {year}
  {2019})},\ \Eprint {http://arxiv.org/abs/1904.09994} {arXiv:1904.09994
  [hep-ph]} \BibitemShut {NoStop}%
\bibitem [{\citenamefont {Kavanagh}\ \emph {et~al.}(2017)\citenamefont
  {Kavanagh}, \citenamefont {Catena},\ and\ \citenamefont
  {Kouvaris}}]{Kavanagh:2016pyr}%
  \BibitemOpen
  \bibfield  {author} {\bibinfo {author} {\bibfnamefont {Bradley~J.}\
  \bibnamefont {Kavanagh}}, \bibinfo {author} {\bibfnamefont {Riccardo}\
  \bibnamefont {Catena}}, \ and\ \bibinfo {author} {\bibfnamefont {Chris}\
  \bibnamefont {Kouvaris}},\ }\bibfield  {title} {\enquote {\bibinfo {title}
  {{Signatures of Earth-scattering in the direct detection of Dark Matter}},}\
  }\href {\doibase 10.1088/1475-7516/2017/01/012} {\bibfield  {journal}
  {\bibinfo  {journal} {JCAP}\ }\textbf {\bibinfo {volume} {01}},\ \bibinfo
  {pages} {012} (\bibinfo {year} {2017})},\ \Eprint
  {http://arxiv.org/abs/1611.05453} {arXiv:1611.05453 [hep-ph]} \BibitemShut
  {NoStop}%
\bibitem [{\citenamefont {Pospelov}\ \emph {et~al.}(2020)\citenamefont
  {Pospelov}, \citenamefont {Rajendran},\ and\ \citenamefont
  {Ramani}}]{pospelov:2020}%
  \BibitemOpen
  \bibfield  {author} {\bibinfo {author} {\bibfnamefont {Maxim}\ \bibnamefont
  {Pospelov}}, \bibinfo {author} {\bibfnamefont {Surjeet}\ \bibnamefont
  {Rajendran}}, \ and\ \bibinfo {author} {\bibfnamefont {Harikrishnan}\
  \bibnamefont {Ramani}},\ }\bibfield  {title} {\enquote {\bibinfo {title}
  {Metastable nuclear isomers as dark matter accelerators},}\ }\href {\doibase
  10.1103/physrevd.101.055001} {\bibfield  {journal} {\bibinfo  {journal}
  {Physical Review D}\ }\textbf {\bibinfo {volume} {101}} (\bibinfo {year}
  {2020}),\ 10.1103/physrevd.101.055001}\BibitemShut {NoStop}%
\bibitem [{\citenamefont {Emken}\ and\ \citenamefont
  {Kouvaris}(2017)}]{Emken:2017qmp}%
  \BibitemOpen
  \bibfield  {author} {\bibinfo {author} {\bibfnamefont {Timon}\ \bibnamefont
  {Emken}}\ and\ \bibinfo {author} {\bibfnamefont {Chris}\ \bibnamefont
  {Kouvaris}},\ }\bibfield  {title} {\enquote {\bibinfo {title} {{DaMaSCUS: The
  Impact of Underground Scatterings on Direct Detection of Light Dark
  Matter}},}\ }\href {\doibase 10.1088/1475-7516/2017/10/031} {\bibfield
  {journal} {\bibinfo  {journal} {JCAP}\ }\textbf {\bibinfo {volume} {10}},\
  \bibinfo {pages} {031} (\bibinfo {year} {2017})},\ \Eprint
  {http://arxiv.org/abs/1706.02249} {arXiv:1706.02249 [hep-ph]} \BibitemShut
  {NoStop}%
\bibitem [{\citenamefont {Tucker-Smith}\ and\ \citenamefont
  {Weiner}(2005)}]{Tucker-Smith:2004mxa}%
  \BibitemOpen
  \bibfield  {author} {\bibinfo {author} {\bibfnamefont {David}\ \bibnamefont
  {Tucker-Smith}}\ and\ \bibinfo {author} {\bibfnamefont {Neal}\ \bibnamefont
  {Weiner}},\ }\bibfield  {title} {\enquote {\bibinfo {title} {{The Status of
  inelastic dark matter}},}\ }\href {\doibase 10.1103/PhysRevD.72.063509}
  {\bibfield  {journal} {\bibinfo  {journal} {Phys. Rev. D}\ }\textbf {\bibinfo
  {volume} {72}},\ \bibinfo {pages} {063509} (\bibinfo {year} {2005})},\
  \Eprint {http://arxiv.org/abs/hep-ph/0402065} {arXiv:hep-ph/0402065}
  \BibitemShut {NoStop}%
\bibitem [{\citenamefont {Catena}\ and\ \citenamefont
  {Ullio}(2010)}]{Catena:2009mf}%
  \BibitemOpen
  \bibfield  {author} {\bibinfo {author} {\bibfnamefont {Riccardo}\
  \bibnamefont {Catena}}\ and\ \bibinfo {author} {\bibfnamefont {Piero}\
  \bibnamefont {Ullio}},\ }\bibfield  {title} {\enquote {\bibinfo {title} {{A
  novel determination of the local dark matter density}},}\ }\href {\doibase
  10.1088/1475-7516/2010/08/004} {\bibfield  {journal} {\bibinfo  {journal}
  {JCAP}\ }\textbf {\bibinfo {volume} {08}},\ \bibinfo {pages} {004} (\bibinfo
  {year} {2010})},\ \Eprint {http://arxiv.org/abs/0907.0018} {arXiv:0907.0018
  [astro-ph.CO]} \BibitemShut {NoStop}%
\bibitem [{\citenamefont {Evans}\ \emph {et~al.}(2019)\citenamefont {Evans},
  \citenamefont {O'Hare},\ and\ \citenamefont {McCabe}}]{Evans:2018bqy}%
  \BibitemOpen
  \bibfield  {author} {\bibinfo {author} {\bibfnamefont {N.~Wyn}\ \bibnamefont
  {Evans}}, \bibinfo {author} {\bibfnamefont {Ciaran A.~J.}\ \bibnamefont
  {O'Hare}}, \ and\ \bibinfo {author} {\bibfnamefont {Christopher}\
  \bibnamefont {McCabe}},\ }\bibfield  {title} {\enquote {\bibinfo {title}
  {{Refinement of the standard halo model for dark matter searches in light of
  the Gaia Sausage}},}\ }\href {\doibase 10.1103/PhysRevD.99.023012} {\bibfield
   {journal} {\bibinfo  {journal} {Phys. Rev. D}\ }\textbf {\bibinfo {volume}
  {99}},\ \bibinfo {pages} {023012} (\bibinfo {year} {2019})},\ \Eprint
  {http://arxiv.org/abs/1810.11468} {arXiv:1810.11468 [astro-ph.GA]}
  \BibitemShut {NoStop}%
\bibitem [{\citenamefont {Abdelhameed}\ \emph {et~al.}(2019)\citenamefont
  {Abdelhameed} \emph {et~al.}}]{CRESST:2019jnq}%
  \BibitemOpen
  \bibfield  {author} {\bibinfo {author} {\bibfnamefont {A.~H.}\ \bibnamefont
  {Abdelhameed}} \emph {et~al.} (\bibinfo {collaboration} {CRESST}),\
  }\bibfield  {title} {\enquote {\bibinfo {title} {{First results from the
  CRESST-III low-mass dark matter program}},}\ }\href {\doibase
  10.1103/PhysRevD.100.102002} {\bibfield  {journal} {\bibinfo  {journal}
  {Phys. Rev. D}\ }\textbf {\bibinfo {volume} {100}},\ \bibinfo {pages}
  {102002} (\bibinfo {year} {2019})},\ \Eprint
  {http://arxiv.org/abs/1904.00498} {arXiv:1904.00498 [astro-ph.CO]}
  \BibitemShut {NoStop}%
\bibitem [{\citenamefont {Agnese}\ \emph {et~al.}(2016)\citenamefont {Agnese}
  \emph {et~al.}}]{SuperCDMS:2015eex}%
  \BibitemOpen
  \bibfield  {author} {\bibinfo {author} {\bibfnamefont {R.}~\bibnamefont
  {Agnese}} \emph {et~al.} (\bibinfo {collaboration} {SuperCDMS}),\ }\bibfield
  {title} {\enquote {\bibinfo {title} {{New Results from the Search for
  Low-Mass Weakly Interacting Massive Particles with the CDMS Low Ionization
  Threshold Experiment}},}\ }\href {\doibase 10.1103/PhysRevLett.116.071301}
  {\bibfield  {journal} {\bibinfo  {journal} {Phys. Rev. Lett.}\ }\textbf
  {\bibinfo {volume} {116}},\ \bibinfo {pages} {071301} (\bibinfo {year}
  {2016})},\ \Eprint {http://arxiv.org/abs/1509.02448} {arXiv:1509.02448
  [astro-ph.CO]} \BibitemShut {NoStop}%
\bibitem [{\citenamefont {Aprile}\ \emph {et~al.}(2018)\citenamefont {Aprile}
  \emph {et~al.}}]{XENON:2018voc}%
  \BibitemOpen
  \bibfield  {author} {\bibinfo {author} {\bibfnamefont {E.}~\bibnamefont
  {Aprile}} \emph {et~al.} (\bibinfo {collaboration} {XENON}),\ }\bibfield
  {title} {\enquote {\bibinfo {title} {{Dark Matter Search Results from a One
  Ton-Year Exposure of XENON1T}},}\ }\href {\doibase
  10.1103/PhysRevLett.121.111302} {\bibfield  {journal} {\bibinfo  {journal}
  {Phys. Rev. Lett.}\ }\textbf {\bibinfo {volume} {121}},\ \bibinfo {pages}
  {111302} (\bibinfo {year} {2018})},\ \Eprint
  {http://arxiv.org/abs/1805.12562} {arXiv:1805.12562 [astro-ph.CO]}
  \BibitemShut {NoStop}%
\bibitem [{\citenamefont {Bringmann}\ \emph {et~al.}(2017)\citenamefont
  {Bringmann} \emph {et~al.}}]{GAMBITDarkMatterWorkgroup:2017fax}%
  \BibitemOpen
  \bibfield  {author} {\bibinfo {author} {\bibfnamefont {Torsten}\ \bibnamefont
  {Bringmann}} \emph {et~al.} (\bibinfo {collaboration} {GAMBIT Dark Matter
  Workgroup}),\ }\bibfield  {title} {\enquote {\bibinfo {title} {{DarkBit: A
  GAMBIT module for computing dark matter observables and likelihoods}},}\
  }\href {\doibase 10.1140/epjc/s10052-017-5155-4} {\bibfield  {journal}
  {\bibinfo  {journal} {Eur. Phys. J. C}\ }\textbf {\bibinfo {volume} {77}},\
  \bibinfo {pages} {831} (\bibinfo {year} {2017})},\ \Eprint
  {http://arxiv.org/abs/1705.07920} {arXiv:1705.07920 [hep-ph]} \BibitemShut
  {NoStop}%
\bibitem [{\citenamefont {Athron}\ \emph {et~al.}(2019)\citenamefont {Athron}
  \emph {et~al.}}]{GAMBIT:2018eea}%
  \BibitemOpen
  \bibfield  {author} {\bibinfo {author} {\bibfnamefont {Peter}\ \bibnamefont
  {Athron}} \emph {et~al.} (\bibinfo {collaboration} {GAMBIT}),\ }\bibfield
  {title} {\enquote {\bibinfo {title} {{Global analyses of Higgs portal singlet
  dark matter models using GAMBIT}},}\ }\href {\doibase
  10.1140/epjc/s10052-018-6513-6} {\bibfield  {journal} {\bibinfo  {journal}
  {Eur. Phys. J. C}\ }\textbf {\bibinfo {volume} {79}},\ \bibinfo {pages} {38}
  (\bibinfo {year} {2019})},\ \Eprint {http://arxiv.org/abs/1808.10465}
  {arXiv:1808.10465 [hep-ph]} \BibitemShut {NoStop}%
\bibitem [{\citenamefont {Bunge}\ \emph {et~al.}(1993)\citenamefont {Bunge},
  \citenamefont {Barrientos},\ and\ \citenamefont {Bunge}}]{BUNGE1993113}%
  \BibitemOpen
  \bibfield  {author} {\bibinfo {author} {\bibfnamefont {C.F.}\ \bibnamefont
  {Bunge}}, \bibinfo {author} {\bibfnamefont {J.A.}\ \bibnamefont
  {Barrientos}}, \ and\ \bibinfo {author} {\bibfnamefont {A.V.}\ \bibnamefont
  {Bunge}},\ }\bibfield  {title} {\enquote {\bibinfo {title}
  {Roothaan-hartree-fock ground-state atomic wave functions: Slater-type
  orbital expansions and expectation values for z = 2-54},}\ }\href {\doibase
  https://doi.org/10.1006/adnd.1993.1003} {\bibfield  {journal} {\bibinfo
  {journal} {Atomic Data and Nuclear Data Tables}\ }\textbf {\bibinfo {volume}
  {53}},\ \bibinfo {pages} {113--162} (\bibinfo {year} {1993})}\BibitemShut
  {NoStop}%
\bibitem [{\citenamefont {Bethe}\ and\ \citenamefont
  {Salpeter}(1957)}]{bethe2012quantum}%
  \BibitemOpen
  \bibfield  {author} {\bibinfo {author} {\bibfnamefont {"H.~A.}\ \bibnamefont
  {Bethe}}\ and\ \bibinfo {author} {\bibfnamefont {E.~E.}\ \bibnamefont
  {Salpeter}},\ }\href@noop {} {\emph {\bibinfo {title} {Quantum Mechanics of
  One-and Two-electron Atoms}}}\ (\bibinfo  {publisher} {Springer, Berlin},\
  \bibinfo {year} {1957})\BibitemShut {NoStop}%
\bibitem [{\citenamefont {Carrillo~Gonz\'alez}\ and\ \citenamefont
  {Toro}(2021)}]{CarrilloGonzalez:2021lxm}%
  \BibitemOpen
  \bibfield  {author} {\bibinfo {author} {\bibfnamefont {Mariana}\ \bibnamefont
  {Carrillo~Gonz\'alez}}\ and\ \bibinfo {author} {\bibfnamefont {Natalia}\
  \bibnamefont {Toro}},\ }\bibfield  {title} {\enquote {\bibinfo {title}
  {{Cosmology and Signals of Light Pseudo-Dirac Dark Matter}},}\ }\href@noop {}
  {\  (\bibinfo {year} {2021})},\ \Eprint {http://arxiv.org/abs/2108.13422}
  {arXiv:2108.13422 [hep-ph]} \BibitemShut {NoStop}%
\bibitem [{\citenamefont {Fitzpatrick}\ \emph {et~al.}(2021)\citenamefont
  {Fitzpatrick}, \citenamefont {Liu}, \citenamefont {Slatyer},\ and\
  \citenamefont {Tsai}}]{Fitzpatrick:2021cij}%
  \BibitemOpen
  \bibfield  {author} {\bibinfo {author} {\bibfnamefont {Patrick~J.}\
  \bibnamefont {Fitzpatrick}}, \bibinfo {author} {\bibfnamefont {Hongwan}\
  \bibnamefont {Liu}}, \bibinfo {author} {\bibfnamefont {Tracy~R.}\
  \bibnamefont {Slatyer}}, \ and\ \bibinfo {author} {\bibfnamefont {Yu-Dai}\
  \bibnamefont {Tsai}},\ }\bibfield  {title} {\enquote {\bibinfo {title} {{New
  Thermal Relic Targets for Inelastic Vector-Portal Dark Matter}},}\
  }\href@noop {} {\  (\bibinfo {year} {2021})},\ \Eprint
  {http://arxiv.org/abs/2105.05255} {arXiv:2105.05255 [hep-ph]} \BibitemShut
  {NoStop}%
\bibitem [{\citenamefont {Finkbeiner}\ \emph {et~al.}(2008)\citenamefont
  {Finkbeiner}, \citenamefont {Padmanabhan},\ and\ \citenamefont
  {Weiner}}]{Finkbeiner:2008gw}%
  \BibitemOpen
  \bibfield  {author} {\bibinfo {author} {\bibfnamefont {Douglas~P.}\
  \bibnamefont {Finkbeiner}}, \bibinfo {author} {\bibfnamefont {Nikhil}\
  \bibnamefont {Padmanabhan}}, \ and\ \bibinfo {author} {\bibfnamefont {Neal}\
  \bibnamefont {Weiner}},\ }\bibfield  {title} {\enquote {\bibinfo {title}
  {{CMB and 21-cm Signals for Dark Matter with a Long-Lived Excited State}},}\
  }\href {\doibase 10.1103/PhysRevD.78.063530} {\bibfield  {journal} {\bibinfo
  {journal} {Phys. Rev. D}\ }\textbf {\bibinfo {volume} {78}},\ \bibinfo
  {pages} {063530} (\bibinfo {year} {2008})},\ \Eprint
  {http://arxiv.org/abs/0805.3531} {arXiv:0805.3531 [astro-ph]} \BibitemShut
  {NoStop}%
\bibitem [{\citenamefont {Collar}\ and\ \citenamefont
  {Avignone}(1992)}]{Collar:1992qc}%
  \BibitemOpen
  \bibfield  {author} {\bibinfo {author} {\bibfnamefont {J.~I.}\ \bibnamefont
  {Collar}}\ and\ \bibinfo {author} {\bibfnamefont {F.~T.}\ \bibnamefont
  {Avignone}},\ }\bibfield  {title} {\enquote {\bibinfo {title} {{Diurnal
  modulation effects in cold dark matter experiments}},}\ }\href {\doibase
  10.1016/0370-2693(92)90873-3} {\bibfield  {journal} {\bibinfo  {journal}
  {Phys. Lett. B}\ }\textbf {\bibinfo {volume} {275}},\ \bibinfo {pages}
  {181--185} (\bibinfo {year} {1992})}\BibitemShut {NoStop}%
\bibitem [{\citenamefont {Collar}\ and\ \citenamefont
  {Avignone}(1993)}]{Collar:1993ss}%
  \BibitemOpen
  \bibfield  {author} {\bibinfo {author} {\bibfnamefont {J.~I.}\ \bibnamefont
  {Collar}}\ and\ \bibinfo {author} {\bibfnamefont {F.~T.}\ \bibnamefont
  {Avignone}, \bibfnamefont {III}},\ }\bibfield  {title} {\enquote {\bibinfo
  {title} {{The Effect of elastic scattering in the Earth on cold dark matter
  experiments}},}\ }\href {\doibase 10.1103/PhysRevD.47.5238} {\bibfield
  {journal} {\bibinfo  {journal} {Phys. Rev. D}\ }\textbf {\bibinfo {volume}
  {47}},\ \bibinfo {pages} {5238--5246} (\bibinfo {year} {1993})}\BibitemShut
  {NoStop}%
\bibitem [{\citenamefont {Hasenbalg}\ \emph {et~al.}(1997)\citenamefont
  {Hasenbalg}, \citenamefont {Abriola}, \citenamefont {Avignone}, \citenamefont
  {Collar}, \citenamefont {Di~Gregorio}, \citenamefont {Gattone}, \citenamefont
  {Huck}, \citenamefont {Tomasi},\ and\ \citenamefont
  {Urteaga}}]{Hasenbalg:1997hs}%
  \BibitemOpen
  \bibfield  {author} {\bibinfo {author} {\bibfnamefont {F.}~\bibnamefont
  {Hasenbalg}}, \bibinfo {author} {\bibfnamefont {D.}~\bibnamefont {Abriola}},
  \bibinfo {author} {\bibfnamefont {F.~T.}\ \bibnamefont {Avignone}}, \bibinfo
  {author} {\bibfnamefont {J.~I.}\ \bibnamefont {Collar}}, \bibinfo {author}
  {\bibfnamefont {D.~E.}\ \bibnamefont {Di~Gregorio}}, \bibinfo {author}
  {\bibfnamefont {A.~O.}\ \bibnamefont {Gattone}}, \bibinfo {author}
  {\bibfnamefont {H.}~\bibnamefont {Huck}}, \bibinfo {author} {\bibfnamefont
  {D.}~\bibnamefont {Tomasi}}, \ and\ \bibinfo {author} {\bibfnamefont
  {I.}~\bibnamefont {Urteaga}},\ }\bibfield  {title} {\enquote {\bibinfo
  {title} {{Cold dark matter identification: Diurnal modulation revisited}},}\
  }\href {\doibase 10.1103/PhysRevD.55.7350} {\bibfield  {journal} {\bibinfo
  {journal} {Phys. Rev. D}\ }\textbf {\bibinfo {volume} {55}},\ \bibinfo
  {pages} {7350--7355} (\bibinfo {year} {1997})},\ \Eprint
  {http://arxiv.org/abs/astro-ph/9702165} {arXiv:astro-ph/9702165} \BibitemShut
  {NoStop}%
\bibitem [{\citenamefont {Kavanagh}\ \emph {et~al.}(2021)\citenamefont
  {Kavanagh}, \citenamefont {Emken},\ and\ \citenamefont
  {Catena}}]{Kavanagh:2020cvn}%
  \BibitemOpen
  \bibfield  {author} {\bibinfo {author} {\bibfnamefont {Bradley~J.}\
  \bibnamefont {Kavanagh}}, \bibinfo {author} {\bibfnamefont {Timon}\
  \bibnamefont {Emken}}, \ and\ \bibinfo {author} {\bibfnamefont {Riccardo}\
  \bibnamefont {Catena}},\ }\bibfield  {title} {\enquote {\bibinfo {title}
  {Measuring the local dark matter density in the laboratory},}\ }\href
  {\doibase 10.1103/PhysRevD.104.083023} {\bibfield  {journal} {\bibinfo
  {journal} {Phys. Rev. D}\ }\textbf {\bibinfo {volume} {104}},\ \bibinfo
  {pages} {083023} (\bibinfo {year} {2021})}\BibitemShut {NoStop}%
\bibitem [{\citenamefont {Lundberg}\ and\ \citenamefont
  {Edsjö}(2004)}]{Lundberg_2004}%
  \BibitemOpen
  \bibfield  {author} {\bibinfo {author} {\bibfnamefont {Johan}\ \bibnamefont
  {Lundberg}}\ and\ \bibinfo {author} {\bibfnamefont {Joakim}\ \bibnamefont
  {Edsjö}},\ }\bibfield  {title} {\enquote {\bibinfo {title} {Weakly
  interacting massive particle diffusion in the solar system including solar
  depletion and its effect on earth capture rates},}\ }\href {\doibase
  10.1103/physrevd.69.123505} {\bibfield  {journal} {\bibinfo  {journal}
  {Physical Review D}\ }\textbf {\bibinfo {volume} {69}} (\bibinfo {year}
  {2004}),\ 10.1103/physrevd.69.123505}\BibitemShut {NoStop}%
\bibitem [{\citenamefont {Takahashi}\ \emph {et~al.}(2020)\citenamefont
  {Takahashi}, \citenamefont {Yamada},\ and\ \citenamefont
  {Yin}}]{Takahashi:2020bpq}%
  \BibitemOpen
  \bibfield  {author} {\bibinfo {author} {\bibfnamefont {Fuminobu}\
  \bibnamefont {Takahashi}}, \bibinfo {author} {\bibfnamefont {Masaki}\
  \bibnamefont {Yamada}}, \ and\ \bibinfo {author} {\bibfnamefont {Wen}\
  \bibnamefont {Yin}},\ }\bibfield  {title} {\enquote {\bibinfo {title}
  {{XENON1T Excess from Anomaly-Free Axionlike Dark Matter and Its Implications
  for Stellar Cooling Anomaly}},}\ }\href {\doibase
  10.1103/PhysRevLett.125.161801} {\bibfield  {journal} {\bibinfo  {journal}
  {Phys. Rev. Lett.}\ }\textbf {\bibinfo {volume} {125}},\ \bibinfo {pages}
  {161801} (\bibinfo {year} {2020})},\ \Eprint
  {http://arxiv.org/abs/2006.10035} {arXiv:2006.10035 [hep-ph]} \BibitemShut
  {NoStop}%
\bibitem [{\citenamefont {Alonso-\'Alvarez}\ \emph {et~al.}(2020)\citenamefont
  {Alonso-\'Alvarez}, \citenamefont {Ertas}, \citenamefont {Jaeckel},
  \citenamefont {Kahlhoefer},\ and\ \citenamefont
  {Thormaehlen}}]{Alonso-Alvarez:2020cdv}%
  \BibitemOpen
  \bibfield  {author} {\bibinfo {author} {\bibfnamefont {Gonzalo}\ \bibnamefont
  {Alonso-\'Alvarez}}, \bibinfo {author} {\bibfnamefont {Fatih}\ \bibnamefont
  {Ertas}}, \bibinfo {author} {\bibfnamefont {Joerg}\ \bibnamefont {Jaeckel}},
  \bibinfo {author} {\bibfnamefont {Felix}\ \bibnamefont {Kahlhoefer}}, \ and\
  \bibinfo {author} {\bibfnamefont {Lennert~J.}\ \bibnamefont {Thormaehlen}},\
  }\bibfield  {title} {\enquote {\bibinfo {title} {{Hidden Photon Dark Matter
  in the Light of XENON1T and Stellar Cooling}},}\ }\href {\doibase
  10.1088/1475-7516/2020/11/029} {\bibfield  {journal} {\bibinfo  {journal}
  {JCAP}\ }\textbf {\bibinfo {volume} {11}},\ \bibinfo {pages} {029} (\bibinfo
  {year} {2020})},\ \Eprint {http://arxiv.org/abs/2006.11243} {arXiv:2006.11243
  [hep-ph]} \BibitemShut {NoStop}%
\bibitem [{\citenamefont {Athron}\ \emph {et~al.}(2021)\citenamefont {Athron}
  \emph {et~al.}}]{Athron:2020maw}%
  \BibitemOpen
  \bibfield  {author} {\bibinfo {author} {\bibfnamefont {Peter}\ \bibnamefont
  {Athron}} \emph {et~al.},\ }\bibfield  {title} {\enquote {\bibinfo {title}
  {{Global fits of axion-like particles to XENON1T and astrophysical data}},}\
  }\href {\doibase 10.1007/JHEP05(2021)159} {\bibfield  {journal} {\bibinfo
  {journal} {JHEP}\ }\textbf {\bibinfo {volume} {05}},\ \bibinfo {pages} {159}
  (\bibinfo {year} {2021})},\ \Eprint {http://arxiv.org/abs/2007.05517}
  {arXiv:2007.05517 [astro-ph.CO]} \BibitemShut {NoStop}%
\bibitem [{\citenamefont {Bozorgnia}\ \emph {et~al.}(2011)\citenamefont
  {Bozorgnia}, \citenamefont {Gelmini},\ and\ \citenamefont
  {Gondolo}}]{Bozorgnia:2011tk}%
  \BibitemOpen
  \bibfield  {author} {\bibinfo {author} {\bibfnamefont {Nassim}\ \bibnamefont
  {Bozorgnia}}, \bibinfo {author} {\bibfnamefont {Graciela~B.}\ \bibnamefont
  {Gelmini}}, \ and\ \bibinfo {author} {\bibfnamefont {Paolo}\ \bibnamefont
  {Gondolo}},\ }\bibfield  {title} {\enquote {\bibinfo {title} {{Daily
  modulation due to channeling in direct dark matter crystalline detectors}},}\
  }\href {\doibase 10.1103/PhysRevD.84.023516} {\bibfield  {journal} {\bibinfo
  {journal} {Phys. Rev. D}\ }\textbf {\bibinfo {volume} {84}},\ \bibinfo
  {pages} {023516} (\bibinfo {year} {2011})},\ \Eprint
  {http://arxiv.org/abs/1101.2876} {arXiv:1101.2876 [astro-ph.CO]} \BibitemShut
  {NoStop}%
\bibitem [{\citenamefont {Griffin}\ \emph {et~al.}(2018)\citenamefont
  {Griffin}, \citenamefont {Knapen}, \citenamefont {Lin},\ and\ \citenamefont
  {Zurek}}]{Griffin:2018bjn}%
  \BibitemOpen
  \bibfield  {author} {\bibinfo {author} {\bibfnamefont {Sinead}\ \bibnamefont
  {Griffin}}, \bibinfo {author} {\bibfnamefont {Simon}\ \bibnamefont {Knapen}},
  \bibinfo {author} {\bibfnamefont {Tongyan}\ \bibnamefont {Lin}}, \ and\
  \bibinfo {author} {\bibfnamefont {Kathryn~M.}\ \bibnamefont {Zurek}},\
  }\bibfield  {title} {\enquote {\bibinfo {title} {{Directional Detection of
  Light Dark Matter with Polar Materials}},}\ }\href {\doibase
  10.1103/PhysRevD.98.115034} {\bibfield  {journal} {\bibinfo  {journal} {Phys.
  Rev. D}\ }\textbf {\bibinfo {volume} {98}},\ \bibinfo {pages} {115034}
  (\bibinfo {year} {2018})},\ \Eprint {http://arxiv.org/abs/1807.10291}
  {arXiv:1807.10291 [hep-ph]} \BibitemShut {NoStop}%
\bibitem [{\citenamefont {Blanco}\ \emph {et~al.}(2021)\citenamefont {Blanco},
  \citenamefont {Kahn}, \citenamefont {Lillard},\ and\ \citenamefont
  {McDermott}}]{Blanco:2021hlm}%
  \BibitemOpen
  \bibfield  {author} {\bibinfo {author} {\bibfnamefont {Carlos}\ \bibnamefont
  {Blanco}}, \bibinfo {author} {\bibfnamefont {Yonatan}\ \bibnamefont {Kahn}},
  \bibinfo {author} {\bibfnamefont {Benjamin}\ \bibnamefont {Lillard}}, \ and\
  \bibinfo {author} {\bibfnamefont {Samuel~D.}\ \bibnamefont {McDermott}},\
  }\bibfield  {title} {\enquote {\bibinfo {title} {{Dark Matter Daily
  Modulation With Anisotropic Organic Crystals}},}\ }\href {\doibase
  10.1103/PhysRevD.104.036011} {\bibfield  {journal} {\bibinfo  {journal}
  {Phys. Rev. D}\ }\textbf {\bibinfo {volume} {104}},\ \bibinfo {pages}
  {036011} (\bibinfo {year} {2021})},\ \Eprint
  {http://arxiv.org/abs/2103.08601} {arXiv:2103.08601 [hep-ph]} \BibitemShut
  {NoStop}%
\bibitem [{\citenamefont {Kouvaris}\ and\ \citenamefont
  {Nielsen}(2015)}]{Kouvaris:2015xga}%
  \BibitemOpen
  \bibfield  {author} {\bibinfo {author} {\bibfnamefont {Chris}\ \bibnamefont
  {Kouvaris}}\ and\ \bibinfo {author} {\bibfnamefont {Niklas~Gr\o{}nlund}\
  \bibnamefont {Nielsen}},\ }\bibfield  {title} {\enquote {\bibinfo {title}
  {{Daily modulation and gravitational focusing in direct dark matter search
  experiments}},}\ }\href {\doibase 10.1103/PhysRevD.92.075016} {\bibfield
  {journal} {\bibinfo  {journal} {Phys. Rev. D}\ }\textbf {\bibinfo {volume}
  {92}},\ \bibinfo {pages} {075016} (\bibinfo {year} {2015})},\ \Eprint
  {http://arxiv.org/abs/1505.02615} {arXiv:1505.02615 [hep-ph]} \BibitemShut
  {NoStop}%
\bibitem [{\citenamefont {Chang}\ \emph {et~al.}(2010)\citenamefont {Chang},
  \citenamefont {Weiner},\ and\ \citenamefont {Yavin}}]{Chang:2010en}%
  \BibitemOpen
  \bibfield  {author} {\bibinfo {author} {\bibfnamefont {Spencer}\ \bibnamefont
  {Chang}}, \bibinfo {author} {\bibfnamefont {Neal}\ \bibnamefont {Weiner}}, \
  and\ \bibinfo {author} {\bibfnamefont {Itay}\ \bibnamefont {Yavin}},\
  }\bibfield  {title} {\enquote {\bibinfo {title} {{Magnetic Inelastic Dark
  Matter}},}\ }\href {\doibase 10.1103/PhysRevD.82.125011} {\bibfield
  {journal} {\bibinfo  {journal} {Phys. Rev. D}\ }\textbf {\bibinfo {volume}
  {82}},\ \bibinfo {pages} {125011} (\bibinfo {year} {2010})},\ \Eprint
  {http://arxiv.org/abs/1007.4200} {arXiv:1007.4200 [hep-ph]} \BibitemShut
  {NoStop}%
\bibitem [{\citenamefont {Finkbeiner}\ and\ \citenamefont
  {Weiner}(2016)}]{Finkbeiner:2014sja}%
  \BibitemOpen
  \bibfield  {author} {\bibinfo {author} {\bibfnamefont {Douglas~P.}\
  \bibnamefont {Finkbeiner}}\ and\ \bibinfo {author} {\bibfnamefont {Neal}\
  \bibnamefont {Weiner}},\ }\bibfield  {title} {\enquote {\bibinfo {title}
  {{X-ray line from exciting dark matter}},}\ }\href {\doibase
  10.1103/PhysRevD.94.083002} {\bibfield  {journal} {\bibinfo  {journal} {Phys.
  Rev. D}\ }\textbf {\bibinfo {volume} {94}},\ \bibinfo {pages} {083002}
  (\bibinfo {year} {2016})},\ \Eprint {http://arxiv.org/abs/1402.6671}
  {arXiv:1402.6671 [hep-ph]} \BibitemShut {NoStop}%
\bibitem [{\citenamefont {D'Eramo}\ \emph
  {et~al.}(2016{\natexlab{a}})\citenamefont {D'Eramo}, \citenamefont
  {Hambleton}, \citenamefont {Profumo},\ and\ \citenamefont
  {Stefaniak}}]{DEramo:2016gqz}%
  \BibitemOpen
  \bibfield  {author} {\bibinfo {author} {\bibfnamefont {Francesco}\
  \bibnamefont {D'Eramo}}, \bibinfo {author} {\bibfnamefont {Kevin}\
  \bibnamefont {Hambleton}}, \bibinfo {author} {\bibfnamefont {Stefano}\
  \bibnamefont {Profumo}}, \ and\ \bibinfo {author} {\bibfnamefont {Tim}\
  \bibnamefont {Stefaniak}},\ }\bibfield  {title} {\enquote {\bibinfo {title}
  {{Dark matter inelastic up-scattering with the interstellar plasma: A new
  source of x-ray lines, including at 3.5 keV}},}\ }\href {\doibase
  10.1103/PhysRevD.93.103011} {\bibfield  {journal} {\bibinfo  {journal} {Phys.
  Rev. D}\ }\textbf {\bibinfo {volume} {93}},\ \bibinfo {pages} {103011}
  (\bibinfo {year} {2016}{\natexlab{a}})},\ \Eprint
  {http://arxiv.org/abs/1603.04859} {arXiv:1603.04859 [hep-ph]} \BibitemShut
  {NoStop}%
\bibitem [{\citenamefont {Duerr}\ \emph {et~al.}(2021)\citenamefont {Duerr},
  \citenamefont {Ferber}, \citenamefont {Garcia-Cely}, \citenamefont {Hearty},\
  and\ \citenamefont {Schmidt-Hoberg}}]{Duerr:2020muu}%
  \BibitemOpen
  \bibfield  {author} {\bibinfo {author} {\bibfnamefont {Michael}\ \bibnamefont
  {Duerr}}, \bibinfo {author} {\bibfnamefont {Torben}\ \bibnamefont {Ferber}},
  \bibinfo {author} {\bibfnamefont {Camilo}\ \bibnamefont {Garcia-Cely}},
  \bibinfo {author} {\bibfnamefont {Christopher}\ \bibnamefont {Hearty}}, \
  and\ \bibinfo {author} {\bibfnamefont {Kai}\ \bibnamefont {Schmidt-Hoberg}},\
  }\bibfield  {title} {\enquote {\bibinfo {title} {{Long-lived Dark Higgs and
  Inelastic Dark Matter at Belle II}},}\ }\href {\doibase
  10.1007/JHEP04(2021)146} {\bibfield  {journal} {\bibinfo  {journal} {JHEP}\
  }\textbf {\bibinfo {volume} {04}},\ \bibinfo {pages} {146} (\bibinfo {year}
  {2021})},\ \Eprint {http://arxiv.org/abs/2012.08595} {arXiv:2012.08595
  [hep-ph]} \BibitemShut {NoStop}%
\bibitem [{\citenamefont {D'Eramo}\ \emph
  {et~al.}(2016{\natexlab{b}})\citenamefont {D'Eramo}, \citenamefont
  {Kavanagh},\ and\ \citenamefont {Panci}}]{DEramo:2016gos}%
  \BibitemOpen
  \bibfield  {author} {\bibinfo {author} {\bibfnamefont {Francesco}\
  \bibnamefont {D'Eramo}}, \bibinfo {author} {\bibfnamefont {Bradley~J.}\
  \bibnamefont {Kavanagh}}, \ and\ \bibinfo {author} {\bibfnamefont {Paolo}\
  \bibnamefont {Panci}},\ }\bibfield  {title} {\enquote {\bibinfo {title} {{You
  can hide but you have to run: direct detection with vector mediators}},}\
  }\href {\doibase 10.1007/JHEP08(2016)111} {\bibfield  {journal} {\bibinfo
  {journal} {JHEP}\ }\textbf {\bibinfo {volume} {08}},\ \bibinfo {pages} {111}
  (\bibinfo {year} {2016}{\natexlab{b}})},\ \Eprint
  {http://arxiv.org/abs/1605.04917} {arXiv:1605.04917 [hep-ph]} \BibitemShut
  {NoStop}%
\bibitem [{\citenamefont {D'Eramo}\ \emph {et~al.}(2017)\citenamefont
  {D'Eramo}, \citenamefont {Kavanagh},\ and\ \citenamefont
  {Panci}}]{DEramo:2017zqw}%
  \BibitemOpen
  \bibfield  {author} {\bibinfo {author} {\bibfnamefont {Francesco}\
  \bibnamefont {D'Eramo}}, \bibinfo {author} {\bibfnamefont {Bradley~J.}\
  \bibnamefont {Kavanagh}}, \ and\ \bibinfo {author} {\bibfnamefont {Paolo}\
  \bibnamefont {Panci}},\ }\bibfield  {title} {\enquote {\bibinfo {title}
  {{Probing Leptophilic Dark Sectors with Hadronic Processes}},}\ }\href
  {\doibase 10.1016/j.physletb.2017.05.063} {\bibfield  {journal} {\bibinfo
  {journal} {Phys. Lett. B}\ }\textbf {\bibinfo {volume} {771}},\ \bibinfo
  {pages} {339--348} (\bibinfo {year} {2017})},\ \Eprint
  {http://arxiv.org/abs/1702.00016} {arXiv:1702.00016 [hep-ph]} \BibitemShut
  {NoStop}%
\bibitem [{\citenamefont {Kopp}\ \emph {et~al.}(2009)\citenamefont {Kopp},
  \citenamefont {Niro}, \citenamefont {Schwetz},\ and\ \citenamefont
  {Zupan}}]{Kopp:2009et}%
  \BibitemOpen
  \bibfield  {author} {\bibinfo {author} {\bibfnamefont {Joachim}\ \bibnamefont
  {Kopp}}, \bibinfo {author} {\bibfnamefont {Viviana}\ \bibnamefont {Niro}},
  \bibinfo {author} {\bibfnamefont {Thomas}\ \bibnamefont {Schwetz}}, \ and\
  \bibinfo {author} {\bibfnamefont {Jure}\ \bibnamefont {Zupan}},\ }\bibfield
  {title} {\enquote {\bibinfo {title} {{DAMA/LIBRA and leptonically interacting
  Dark Matter}},}\ }\href {\doibase 10.1103/PhysRevD.80.083502} {\bibfield
  {journal} {\bibinfo  {journal} {Phys. Rev. D}\ }\textbf {\bibinfo {volume}
  {80}},\ \bibinfo {pages} {083502} (\bibinfo {year} {2009})},\ \Eprint
  {http://arxiv.org/abs/0907.3159} {arXiv:0907.3159 [hep-ph]} \BibitemShut
  {NoStop}%
\bibitem [{\citenamefont {Agnese}\ \emph {et~al.}(2017)\citenamefont {Agnese}
  \emph {et~al.}}]{SuperCDMS:2016wui}%
  \BibitemOpen
  \bibfield  {author} {\bibinfo {author} {\bibfnamefont {R.}~\bibnamefont
  {Agnese}} \emph {et~al.} (\bibinfo {collaboration} {SuperCDMS}),\ }\bibfield
  {title} {\enquote {\bibinfo {title} {{Projected Sensitivity of the SuperCDMS
  SNOLAB experiment}},}\ }\href {\doibase 10.1103/PhysRevD.95.082002}
  {\bibfield  {journal} {\bibinfo  {journal} {Phys. Rev. D}\ }\textbf {\bibinfo
  {volume} {95}},\ \bibinfo {pages} {082002} (\bibinfo {year} {2017})},\
  \Eprint {http://arxiv.org/abs/1610.00006} {arXiv:1610.00006
  [physics.ins-det]} \BibitemShut {NoStop}%
\end{thebibliography}%

\end{document}